\documentclass[aps,prl,twocolumn,amsmath,amssymb,showpacs,superscriptaddress,notitlepage,longbibliography]{revtex4-1}
\usepackage[colorlinks=true,linkcolor=blue,anchorcolor=red,citecolor=blue, urlcolor=blue]{hyperref}
\usepackage{bm} 
\usepackage{graphicx}
\usepackage{adjustbox}
\usepackage{color}
\usepackage{xcolor}
\usepackage{comment}
\definecolor{myRed}{HTML}{E69F00}   
\definecolor{myGreen}{HTML}{009E73} 
\definecolor{myBlue}{HTML}{0072B2}  

\usepackage{varwidth}
\usepackage{mathtools}

\usepackage{soul}
\usepackage{enumerate}
\usepackage{mathrsfs}
\usepackage{tikz}
\usetikzlibrary{automata}
\usepackage{circuitikz}
\usetikzlibrary{patterns}
\usetikzlibrary{through,hobby}
\usetikzlibrary{decorations.markings}
\usetikzlibrary{fadings,shadings}
\usetikzlibrary{plotmarks}
\usetikzlibrary{positioning}
\usetikzlibrary{decorations,arrows}
\usetikzlibrary{decorations.pathreplacing}
\usetikzlibrary{chains, scopes, positioning, backgrounds, shapes, fit, shadows, calc, arrows.meta}
\usepackage{braket}

\usepackage{amsthm}

\usepackage{CJK}
\usepackage[utf8]{inputenc}

\begin{document}

\title{Anyon Permutations in Quantum Double Models through Constant-depth Circuits}
\author{Yabo Li (\begin{CJK}{UTF8}{gbsn}李雅博\end{CJK})}
    \email{liyb.poiuy@gmail.com}
	\affiliation{Center for Quantum Phenomena, Department of Physics, New York University, 726 Broadway, New York, New York 10003, USA}
\author{Zijian Song (\begin{CJK}{UTF8}{gbsn}宋子健\end{CJK})}
    \email{zjsong.physics@gmail.com}
    \affiliation{C. N. Yang Institute for Theoretical Physics, State University of New York at Stony Brook, New York 11794-3840, USA}

\begin{abstract}
    We provide explicit constant-depth local unitary circuits that realize general anyon permutations in Kitaev's quantum double models. This construction can be naturally understood through a correspondence between anyon permutation symmetries of two-dimensional topological orders and self-dualities in one-dimensional systems, where local gates implement self-duality transformations on the boundaries of microscopic regions. From this holographic perspective, general anyon permutations in the $D(G)$ quantum double correspond to compositions of three classes of one-dimensional self-dualities, including gauging of certain subgroups of $G$, stacking with $G$ symmetry-protected topological phases, and outer automorphisms of the group $G$. We construct  circuits realizing the first class by employing self-dual unitary gauging maps, and present transversal circuits for the latter two classes.
\end{abstract}

\maketitle

\textit{Introduction} - Topological orders are of broad interest for both fundamental and practical reasons. While significant progress has been made in their classification~\cite{wen1991non,kitaev2003fault,levin2005string,chen2013symmetry,gu2015classification,etingof2015tensor,wen2017colloquium}, recent developments have also revealed a holographic correspondence between topological orders and generalized symmetries in one lower dimension~\cite{kong2020algebraic,kong2020classification,kong2022one,chatterjee2023symmetry,ji2021unified,moradi2023topological,bhardwaj2024categorical,bhardwaj2025gapped,huang2025topological}. From a practical perspective, topological orders provide a natural platform for quantum memory and error correction due to their intrinsic robustness against local noise~\cite{freedman2002modular,kitaev2003fault,bravyi1998quantum,dennis2002topological,nayak2008non,terhal2015quantum}. In topological codes, quantum information is encoded in the topologically protected ground state subspace, and logical operations act within this subspace. A particularly important class of such operations consists of transformations that permute the anyonic excitations (anyons) while preserving their fusion and braiding structures. Remarkably, it has recently been shown that implementing anyon permutations in certain topological codes enables non-Clifford logical operations, which are a key ingredient for universal fault-tolerant quantum computation~\cite{kobayashi2025clifford,warman2025transversal}.

Since the allowed anyon permutations are always associated with invertible domain walls in topological order~\cite{kitaev2012models}, they can, in principle be realized by sweeping the associated domain wall across the system. The creation of such domain walls in lattice models have been studied extensively~\cite{kitaev2012models,yoshida2015topological,yoshida2016topological,yoshida2017gapped,kesselring2018boundaries,barkeshli2023codimension,barkeshli2024higher,tantivasadakarn2024string,li2024domain,zhu2022topological,Song2025magic}. In parallel, logical operations implementing anyon permutations via automorphism or sequences of local measurements have also been explored~\cite{hsin2025automorphism,hastings2021dynamically,aasen2022adiabatic,davydova2023quantum,kobayashi2024cross}. Existing works, however, typically focus on a specific class of anyon permutations in quantum double models for particular groups. 

\begin{figure*}
    \centering
    \includegraphics[width=0.9\linewidth]{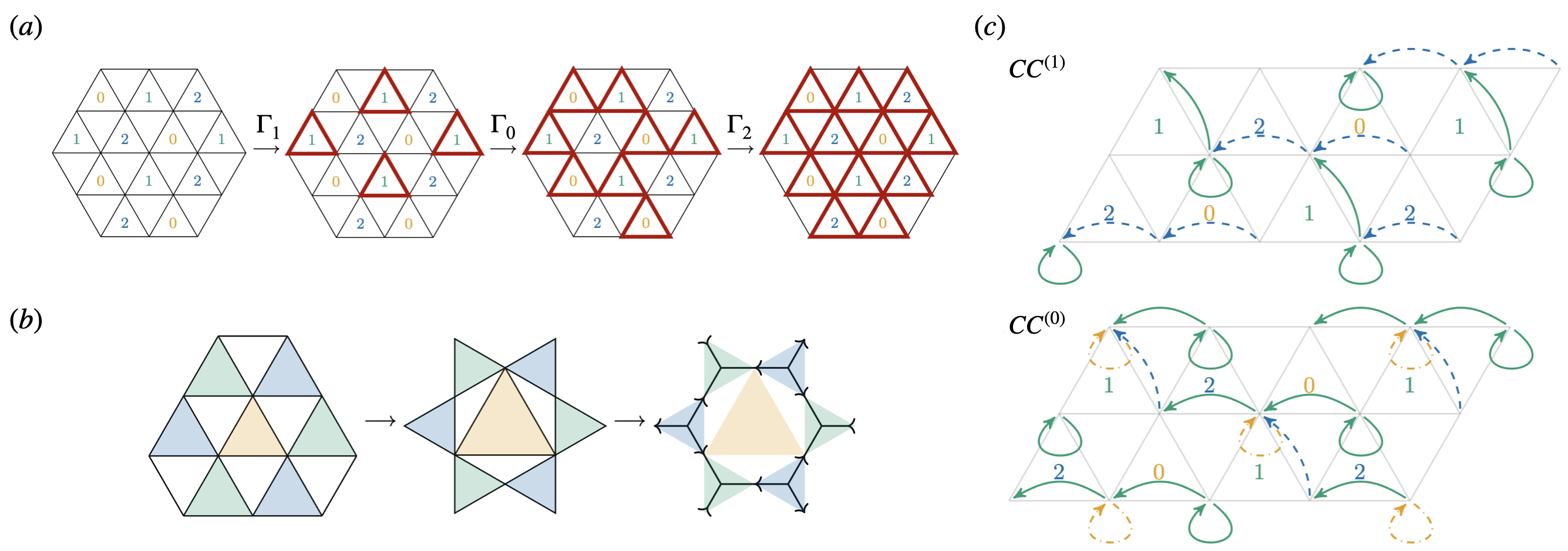}
    \caption{(a) Illustration of the electric--magnetic duality circuit. After applying the gauging maps on all three colors of triangles, the corresponding domain wall is effectively swept across the entire system. (b) The lattice deformation from the original triangular lattice to a hexagonal lattice. (c) The global controlled-conjugation circuit applied after $\Gamma_1$ and $\Gamma_0$ for the $D(G)$ quantum double. A single arrow denotes a controlled-conjugation gate. The local gates are applied in the order: green (solid), blue (dashed), and yellow (dash-dotted). The $CC^{(2)}$ circuit after $\Gamma_2$ is given by on-site controlled-conjugation gates.}
    \label{fig:figure_1}
\end{figure*}

In this work, we present a general finite-depth local unitary circuit construction that realizes arbitrary anyon permutations in the quantum double models. The local gates in our circuits implement duality transformations along the one-dimensional boundaries of microscopic regions. As illustrated in Fig.~\ref{fig:figure_1}(a), each duality transformation inserts a domain wall along one microscopic boundary corresponding to a given anyon permutation. By applying these duality transformations throughout the system, one effectively sweeps the corresponding domain walls, thereby realizing anyon automorphisms in the entire system. This construction provides a lattice level demonstration of the holographic correspondence between anyon permutations in two-dimensional topological orders and duality transformations of generalized symmetries in one dimension

\textit{Anyon permutation in the $\mathbb{Z}_2$ toric code} - We begin with the electric-magnetic duality of the $\mathbb{Z}_2$ toric code model, defined on a triangular lattice as shown in Fig.~\ref{fig:figure_1}(a). There is one qubit at each vertex, the Pauli matrices acting on which are denoted by $X$, $Y$, and $Z$. We assign three colors to the up-triangles, which we label by $0$, $1$, and $2$, such that neighboring up-triangles carry different labels. The stabilizers of the toric code model are generated by
\begin{equation}
\begin{tikzpicture}[scale=0.6, baseline=(current bounding box.center)]
\tikzset{
  hexbond/.style={line width=0.6pt},
  vdot/.style={circle, fill=black, inner sep=1.2pt}
}
\coordinate (A_1) at ( 0, 2*0.8660254);
\coordinate (A_2) at ( -1, 0);
\coordinate (A_3) at ( 1, 0);
\draw[hexbond] (A_1)--(A_2)--(A_3)--cycle;

\coordinate (A_4) at ( -2, 2*0.8660254);
\coordinate (A_5) at ( 0, -2*0.8660254);
\coordinate (A_6) at ( 2, 2*0.8660254);
\draw[hexbond] (A_4)--(A_5)--(A_6)--cycle;

\node at ($(A_1)+(0,0.3)$) {$Z$};
\node at ($(A_2)+(-0.4,0)$) {$Z$};
\node at ($(A_3)+(0.4,0)$) {$Z$};
\node at ($(A_4)+(0,0.3)$) {$Z$};
\node at ($(A_5)+(0,-0.2)$) {$Z$};
\node at ($(A_6)+(0,0.3)$) {$Z$};
\node[myRed] at (0, 2*0.8660254/3) {$0$}; 

\node at (0+2.5, 0) {,};
\coordinate (A_7) at ( 0+4.5, 0.8660254);
\coordinate (A_8) at ( -1+4.5, -0.8660254);
\coordinate (A_9) at ( 1+4.5, -0.8660254);
\draw[hexbond] (A_7)--(A_8)--(A_9)--cycle;

\node at ($(A_7)+(0,0.3)$) {$X$};
\node at ($(A_8)+(-0.4,0)$) {$X$};
\node at ($(A_9)+(0.4,0)$) {$X$};
\node[myGreen] at (-0.2+4.5, 2*0.8660254/3-0.8660254) {$1$};
\node at (0+4.5, 2*0.8660254/3-0.1-0.8660254) {$,$};
\node[myBlue] at (0.2+4.5, 2*0.8660254/3-0.8660254) {$2$};
\end{tikzpicture}. \label{eq:toric_code}
\end{equation}
The toric code ground states are the simultaneous $+1$-eigenstates for all the stabilizers. After a lattice deformation shown in Fig.~\ref{fig:figure_1}(b), the stabilizers above constitute the Hamiltonian of the $D(\mathbb{Z}_2)$ quantum double on the hexagonal lattice, where the first term corresponds to the plaquette term, and the last two terms correspond to the star terms~\cite{kitaev2003fault}. In the $D(\mathbb{Z}_2)$ quantum double model, a localized $e$ excitation is the violation of a star term, while a localized $m$ excitation is the violation of a plaquette term. They are created by strings of Pauli $Z$ and $X$ operators respectively. Non-contractible string operators, on the other hand, commute with all the stabilizers and act nontrivially on the ground state subspace. The electric-magnetic duality preserves the ground state subspace while exchange the non-contractible $X$ string operators with the $Z$ string operators. A finite-depth circuit implementing this $e$-$m$ exchange is presented in Ref.~\cite{aasen2022adiabatic}, which we review below for later convenience.

On a one-dimensional qubit chain with $N$ sites, the basis states are given by $\ket{\{s_i\}}=\ket{s_1, s_2, \ldots, s_N}$ with $s_i=0,1$. The $\mathbb{Z}_2$ symmetry acting on the system is defined as $\eta \equiv \prod_j X_j$. The gauging map realizing a $\mathbb{Z}_2$ self-duality is defined as follows:
\begin{equation}
    \begin{aligned}
        |\{s_i\}\rangle\ \overset{\Gamma}{\longmapsto}\ \sum_{\{t_i\}} \bigg(\prod_j\chi(t_j, s_{j-1}+s_j)\bigg)
        \ |\{t_i\} \rangle,
    \end{aligned}
    \label{eq:KW}
\end{equation}
where $\chi(t, s) := (-1)^{t\cdot s}$. When the product over site index $j$ in Eq.~\eqref{eq:KW} is taken from $1$ to $N$ with periodic boundary condition $s_{0}= s_{N}$, this gauging map realizes the Kramers-Wannier duality~\cite{seiberg2024non}, which satisfies $\eta\cdot \Gamma = \Gamma = \Gamma\cdot \eta$ and implements a locality preserving map on the $\mathbb{Z}_2$ symmetric local operators:
\begin{align}
    \Gamma: X_i \mapsto Z_i Z_{i+1}, \quad Z_i Z_{i+1} \mapsto X_{i+1}.
    \label{eq:Z_2_operator_map}
\end{align}

When the product over site index $j$ in Eq.~\eqref{eq:KW} is taken from $2$ to $N$, with two extra $s$ gate at site $1$ and $N$, we can define a {\it unitary gauging map} as $U_{\Gamma} = S_1 H_1 CZ_{1,2}H_2 \cdots CZ_{N-1,N} H_N S_N$. It is known as the sequential unitary realization of the Kramers-Wannier duality since $\Gamma = U_{\Gamma} (1+\eta)$~\cite{ho2019efficient,chen2024sequential}.

The realization of the electric-magnetic duality in the $\mathbb{Z}_2$ toric code is schematically shown in Fig.~\ref{fig:figure_1}(a). We treat each 1-triangle as a one-dimensional chain of three sites. The stabilizer, given by the product of Pauli $X$ operators on its vertices, then defines a $\mathbb{Z}_2$ symmetry.
The corresponding unitary gauging map acts on the three sites of the chain. It can be decomposed into single-qubit Hadamard gates, single-qubit $S$ gates, and two-qubit controlled-$Z$ gates as we discussed above, and it implements the operator maps in Eq.~\eqref{eq:Z_2_operator_map} within the $\eta = 1$ sector. We denote the set of toric code stabilizers as $H^{(0)}$, in which the superscript $0$ emphasizes that stabilizers centered on the 0-triangles are the plaquette terms. After applying the unitary gauging maps on all 1-triangles, followed by those on the 0-triangles and then on the 2-triangles, the set of stabilizers is eventually mapped back to $H^{(0)}$:
\begin{align*}
    H^{(0)} \xrightarrow[]{\Gamma_1} H^{(1)} \xrightarrow[]{\Gamma_0} H^{(2)} \xrightarrow[]{\Gamma_2} H^{(0)}.
\end{align*}
While the above sequence of gauging maps preserves the ground state subspace, it realizes the electric-magnetic duality, as non-contractible $X$ and $Z$ string operators are exchanged. We refer the reader to Ref.~\cite{aasen2022adiabatic} and to the supplementary material (see Appendix~\ref{sec:toric_code}) for further details.

\textit{Anyon permutations and self-dualities} - It is not a coincidence that the anyon permutation in the above model can be realized by performing one-dimensional self-duality transformations. At the lattice level, a $G$ self-duality is a locality preserving automorphism, or quantum cellular automaton (QCA), on the algebra of $G$-symmetric local operators~\cite{jones2024dhr,ma2024quantum}. In the above $\mathbb{Z}_2$ example, the Kramers-Wannier duality implements a QCA on the $\mathbb{Z}_2$ symmetric algebra generated by $X_i$ and $Z_i Z_{i+1}$ as shown in Eq.~\eqref{eq:Z_2_operator_map}. For Abelian symmetry $G$, it is proven that a QCA on the $G$ symmetric algebra always corresponds to an anyon permutation in two-dimensional $D(G)$ quantum double, up to local symmetric unitaries and translations~\cite{ma2024quantum}.

In the continuum, it is believed that $G$ self-dualities of a quantum field theory is mathematically described by invertible $\mathrm{Vec}_G$-bimodule categories~\cite{diatlyk2024gauging}. Their tensor products form a group denoted by $\mathcal{G}(G)$, commonly refereed to as the \textit{Brauer-Picard group} or the duality group~\cite{moradi2023topological}. An important theorem in the category theory states that this group is isomorphic to the group of anyon permutations, also known as the \textit{braided autoequivalences}, of the $D(G)$ quantum double~\cite{etingof2010fusion}. See Appendix~\ref{sec:mathematics} for further details.

This holographic correspondence between anyon permutation in 2D quantum doubles and 1D self-dualities can also be heuristically understood via the SymTFT sandwich construction, where a quasi-1D system is constructed from a thin slab of topological order sandwiched between a symmetry boundary and a physical boundary. When the bulk topological order is $D(G)$, and the symmetry boundary is chosen to be the electric boundary where all charge anyons are condensed, the quasi-1D system is equipped with a $G$ symmetry. Since an anyon permutation corresponds to an invertible domain wall in the bulk, fusing this domain wall to the physical boundary implements the associated self-duality on the quasi-1D system.

The duality group $\mathcal{G}(G)$ is generated by three classes of 1D self-dualities~\cite{nikshych2014categorical}. The first class consists of gauging dualities. The second class is given by stacking 1D SPT phases classified by $H^2(G, U(1))$. The last class corresponds to the outer automorphism group of $G$. In the remainder of this paper, we will define the duality maps for all three classes, and demonstrate how the holographic correspondence manifests at the lattice level, in the construction of the anyon-permutation circuit.

\textit{$D(G)$ quantum double model} - The Hamiltonian of quantum double model for finite group $G$ on the same triangular lattice composed of terms centering on the up-triangles:
\begin{equation}
\centering{\begin{tikzpicture}[scale=0.6, baseline=(current bounding box.center)]
\tikzset{
  hexbond/.style={line width=0.6pt},
  vdot/.style={circle, fill=black, inner sep=1.2pt}
}

\coordinate (A_1) at ( 0, 2*0.8660254);
\coordinate (A_2) at ( -1, 0);
\coordinate (A_3) at ( 1, 0);
\draw[hexbond] (A_1)--(A_2)--(A_3)--cycle;

\node at ($(A_1)+(0,0.2)$) {$L^{g}$};
\node at ($(A_2)+(-0.5,0)$) {$L^{g}$};
\node at ($(A_3)+(0.5,0)$) {$L^{g}$};
\node[myGreen] at (0, 2*0.8660254/3) {$1$};
\node at ($(A_3)+(1,-.1)$) {,};

\coordinate (B_1) at ( 0+4, 2*0.8660254);
\coordinate (B_2) at ( -1+4, 0);
\coordinate (B_3) at ( 1+4, 0);
\draw[hexbond] (B_1)--(B_2)--(B_3)--cycle;

\node at ($(B_1)+(0,0.2)$) {$R^{g}$};
\node at ($(B_2)+(-0.5,0)$) {$R^{g}$};
\node at ($(B_3)+(0.5,0)$) {$R^{g}$};
\node[myBlue] at (0+4, 2*0.8660254/3) {$2$};

\coordinate (C_1) at ( 0+2, 2*0.8660254-2.7);
\coordinate (C_2) at ( -1+2, 0-2.7);
\coordinate (C_3) at ( 1+2, 0-2.7);
\draw[hexbond] (C_1)--(C_2)--(C_3)--cycle;

\coordinate (C_4) at ( -2+2, 2*0.8660254-2.7);
\coordinate (C_5) at ( 0+2, -2*0.8660254-2.7);
\coordinate (C_6) at ( 2+2, 2*0.8660254-2.7);
\draw[hexbond] (C_4)--(C_5)--(C_6)--cycle;

\node at ($(C_4)+(0,-0.4)$) {\footnotesize $1$};
\node at ($(C_1)+(0,-0.4)$) {\footnotesize $2$};
\node at ($(C_6)+(0,-0.4)$) {\footnotesize $3$};
\node at ($(C_2)+(0,-0.4)$) {\footnotesize $6$};
\node at ($(C_3)+(0,-0.4)$) {\footnotesize $4$};
\node at ($(C_5)+(0,0.4)$) {\footnotesize $5$};

\node at (0+2, -2.5-2.7) {$\frac{1}{d_{\rho}}\mathrm{Tr}(Z_1^{\rho \dagger}Z^{\rho}_2 Z_3^{\rho \dagger} Z^{\rho}_4 Z_5^{\rho \dagger}Z^{\rho}_6$)};
\node[myRed] at (0+2, 2*0.8660254/3-2.7) {$0$}; 
\end{tikzpicture}}
\label{eq:QD_terms}
\end{equation}
The left and right multiplication operators $L^{g}$ and $R^{g}$, as well as the matrix-valued operator $Z^{\rho}$ in the above terms, are defined as
\begin{equation}
    \begin{aligned}
        L^{g} &= \sum_{h}\ket{gh}\bra{h},\ R^{g} = \sum_{h} \ket{hg}\bra{h},\\
        Z^{\rho} &= \sum_h \rho(h)\ket{h}\bra{h},\quad  \rho\in \mathrm{Rep}(G),
    \end{aligned}
\end{equation}
and $d_{\rho}$ is the dimension of representation $\rho$. The term $\mathrm{Tr}(Z_1^{\rho\dagger}\cdots Z_6^{\rho})$ in Eq.~\eqref{eq:QD_terms} is a matrix-product operator, with the trace taking over virtual degrees of freedom in the representation space. The ground states of the $D(G)$ quantum double model are the simultaneous $+1$-eigenstates of all terms defined above. Anyonic excitations are labeled by pairs $(C_u,\mu)$, in which $C_u=\{gu\bar{g}: g\in G\}$ is the conjugacy class of an element $u\in G$, $\bar{g} = g^{-1}$, and $\mu$ is an irreducible representation of the centralizer group $E=\{g\in G: ug=gu\}$~\cite{dijkgraaf1991quasi,kitaev2003fault}. When the conjugacy class $C$ in the pair is the identity class, i.e. $C=C_e$, the corresponding anyon is a pure charge. On the other hand, when $\mu$ is the trivial representation of the centralizer, the corresponding anyon is a pure flux. In the following, we present circuits that realize anyon permutations for all three classes of generators of the duality group $\mathcal{G}(G)$.

\textit{Class I: gauging dualities} - The first class of generators of $\mathcal{G}(G)$ consists of the 1D gauging self-dualities, which correspond to the partial electric-magnetic dualities in the 2d quantum double~\cite{buerschaper2013electric,hu2020electric,li2024non}. This is a direct generalization of the $e$-$m$ duality in the $\mathbb{Z}_2$ toric code. As in the $\mathbb{Z}_2$ example, our one-dimensional setup is a qudit chain with $N$ sites. The basis states are given by $\ket{\{g_i\}}=\ket{g_1, g_2, \ldots, g_N}$ with $g_i\in G$. The $G$ symmetry acting on the system is generated by $\eta_g \equiv \prod_j L^g_j$ for $g\in G$. 

For a gauging self-duality to exist, the group $G$ must contain an Abelian normal subgroup $N \subseteq G$. The group $G$ is then an extension of the quotient group $Q=G/N$ by the normal subgroup $N$. This group extension is generally specified by two defining data: a conjugation action on the $N$ elements $n\mapsto \sigma^q(n)$ for $q\in Q$, and a 2-cocycle $\omega \in Z^2(Q, N)$. Elements of $G$ can be written in the form $g = (n, q)$, with the multiplication rule:
\begin{align}
    (n_1, q_1)(n_2, q_2) = \bigl(n_1\, \sigma^{q_1}(n_2)\, \omega(q_1, q_2),\, q_1 q_2 \bigr).
\end{align}
Hence, the local degree of freedom is naturally decomposed into an $N$ qudit and a $Q$ qudit. When the extension split, namely when the 2-cocycle $\omega$ is trivial, a self-dual gauging map can be defined with a bicharacter $\chi$ of $N$ that is invariant under the conjugation of $Q$ elements. Explicitly, if $\chi(n_1, n_2) = \chi(\sigma^q(n_1), \sigma^q(n_2))$ for any $n_1,n_2\in N$ and $q\in Q$, then the gauging map
\begin{align}
    |\{n_i, q_i\} \rangle \ \overset{\Gamma}{\longmapsto}\ \sum_{\{m_i\}} \left(\prod_j \chi(m_j, \bar{n}_{j-1} n_j)\right) |\{m_i, q_i\} \rangle
\end{align}
realizes a self-duality of the $G$ symmetry. Any $G$-symmetric state will be mapped to another $G$-symmetric state as $\eta_{(n,1)}\cdot \Gamma = \Gamma = \Gamma\cdot \eta_{(n,1)}$ and $\eta_{(1,q)}\cdot \Gamma = \Gamma\cdot \eta_{(1,q)}$. We refer the reader to Appendix~\ref{sec:general_gauging}--\ref{sec:twisted_gauging} for more details, where we also discuss self-dual twisted gauging maps when the group extension does not split.

As a QCA on the $G$-symmetric algebra, $\Gamma$ implements a locality-preserving transformation on the symmetric local operators. Here we list several such mappings for later reference:
\begin{equation}
    \begin{aligned}
        &R^{(n,1)}_i \overset{\Gamma}{\longmapsto} Z^{\widehat{\sigma^{q_i}(n)}}_{i} Z^{\widehat{\sigma^{q_{i}}(n)} \dagger}_{i+1},\quad Z^{\hat{n}}_{i} Z^{\hat{n} \dagger}_{i+1} \overset{\Gamma}{\longmapsto}  L^{(n,1)}_{i+1},\\
        &R^{(1,q)}_i \overset{\Gamma}{\longmapsto} R^{(1,q)}_i,\quad T^q_i \overset{\Gamma}{\longmapsto} T^q_i.
    \end{aligned}
    \label{eq:Operator_maps}
\end{equation}
The diagonal operators $Z^{\hat{n}}_i$ and $T^q_i$ are defined as
\begin{equation}
    \begin{aligned}
    Z^{\hat{n}} &= \sum_{\substack{m\in N\\ q\in Q}} \chi(m, n)\, |m, q\rangle \langle m, q|,\\
    T^q &= \sum_{m\in N}\ket{m,q}\bra{m,q},
    \end{aligned}
\end{equation}
on the $i$-th site. We use $Z^{\widehat{\sigma^{q_i}(n)}}_{j}$ to denote the following controlled-phase gate: 
\begin{equation}
    \begin{aligned}
    Z^{\widehat{\sigma^{q_i}(n)}}_{j} = \sum_{\substack{m_i,m_j\in N\\ q_i,q_j\in Q}} &\chi( m_j, \sigma^{q_i}(n))\, |m_i, q_i\rangle \langle m_i, q_i|\\
    &\otimes|m_j, q_j\rangle \langle m_j, q_j|.
    \end{aligned}
\end{equation}

The star terms centered on a 1-triangle (2-triangle) in Eq.~\eqref{eq:QD_terms} are generated by a product of the $L^{(n,1)}$ ($R^{(n,1)}$) operators and a product of the $L^{(1,q)}$ ($R^{(1,q)}$) operators. The plaquette terms centered on a 0-triangle can be decomposed into a flux-free projector on the $Q$ degrees of freedom and a flux-free projector on the $N$ degrees of freedom. Due to the conjugation action from $q\in Q$ on $N$, the $N$ flux-free projector has dependence explicitly on the $Q$ degrees of freedom, which can be written as:
\begin{equation}
\begin{tikzpicture}[scale=0.8, baseline=(current bounding box.center)]
\tikzset{
  hexbond/.style={line width=0.6pt},
  vdot/.style={circle, fill=black, inner sep=1.2pt}
}

\coordinate (A_1) at ( 0, 2*0.8660254);
\coordinate (A_2) at ( -1, 0);
\coordinate (A_3) at ( 1, 0);
\draw[hexbond] (A_1)--(A_2)--(A_3)--cycle;

\coordinate (A_4) at ( -2, 2*0.8660254);
\coordinate (A_5) at ( 0, -2*0.8660254);
\coordinate (A_6) at ( 2, 2*0.8660254);
\draw[hexbond] (A_4)--(A_5)--(A_6)--cycle;

\node at ($(A_4)+(0,-0.3)$) {\footnotesize $1$};
\node at ($(A_1)+(0,-0.3)$) {\footnotesize $2$};
\node at ($(A_6)+(0,-0.3)$) {\footnotesize $3$};
\node at ($(A_2)+(0,-0.3)$) {\footnotesize $6$};
\node at ($(A_3)+(0,-0.3)$) {\footnotesize $4$};
\node at ($(A_5)+(0,0.3)$) {\footnotesize $5$};

\node at ($(A_4)+(0,0.4)$) {$Z_1^{\widehat{\sigma^{q_2 \bar{q}_3}(n)} \dagger}$};
\node at ($(A_1)+(0,0.4)$) {$Z^{\widehat{\sigma^{q_2 \bar{q}_3}(n)}}_2$};
\node at ($(A_6)+(0,0.4)$) {$Z_3^{\hat{n} \dagger}$};
\node at ($(A_2)+(-0.9,0)$) {$Z_6^{\widehat{\sigma^{q_5 \bar{q}_4}(n)}}$};
\node at ($(A_3)+(0.5,0)$) {$Z_4^{\hat{n}}$};
\node at ($(A_5)+(0,-0.4)$) {$Z_5^{\widehat{\sigma^{q_5 \bar{q}_4}(n)} \dagger}$};

\node[myRed] at (0, 2*0.8660254/3) {$0$}; 
\end{tikzpicture}.
\end{equation}

To realize the associated anyon permutation, we first apply the unitary gauging maps on all the 1-triangles. According to Eq.~\eqref{eq:Operator_maps}, the above flux-free term is not mapped to the symmetry operator $\eta_{(n,1)}$ on the 0-triangle due to their dependence on the $Q$ degrees of freedom, in contrast to the $\mathbb{Z}_2$ case. This can be fixed by a further action of local controlled-conjugation gates shown in Fig.~\ref{fig:figure_1}(c). Once the flux-free terms are mapped to $\eta_{(n,1)}$ on all 0-triangles, we can further apply the unitary gauging maps on them. Finally, it can be shown that after a sequence of such operations, the set $H^{(0)}$ of local Hamiltonian terms and their products are mapped back to itself:
\begin{align*}
    H^{(0)} \xrightarrow[]{CC^{(1)} \circ \Gamma_1} H^{(1)} \xrightarrow[]{CC^{(0)} \circ \Gamma_0} H^{(2)} \xrightarrow[]{CC^{(2)} \circ \Gamma_2} H^{(0)}.
\end{align*}

While preserving the ground state subspace, the above circuit realizes a partial $e$-$m$ duality. In the $D(G)$ quantum double, there are pure flux anyons labeled by $([n],1)$, where $[n]$ is the conjugacy class of an element $n\in N\subseteq G$. There are also pure charge anyons labeled by $([e], \pi_n)$, in which $\pi_n$ is an irreducible representation of $G$, that corresponds to the orbit of an $N$-representation under the conjugation of $Q$. By examining the transformation of ribbon operators, one can show that these two types of anyons are exchanged under the action of the above circuit. See Appendix~\ref{sec:general_ribbon_gauging}.

When $G = D_4 = \langle r,s: r^4 = s^2 = rsrs = e\rangle$, choosing the normal subgroup to be $N = \langle r^2, s\rangle$, we construct a circuit in Appendix \ref{sec:D4_gauging} implementing the following anyon permutation:
\begin{equation}
    \begin{aligned}
        m_r &\mapsto m_{rg}, \quad m_g \mapsto m_g, \quad m_b \mapsto m_{gb}, \\
        e_r &\mapsto e_r, \quad e_g \mapsto e_{rgb}, \quad e_b \mapsto e_b. \label{eq:D_4_anyon_permutation}
    \end{aligned}
\end{equation}
See Table \ref{tab:dictionary} for the anyon notations for $D_4$ quantum double. As shown in Ref.~\cite{kobayashi2025clifford}, with an appropriate boundary, the above anyon permutation realizes the logical $T$-gate.

\textit{Class II: stacking of SPT states} - The second class of $\mathcal{G}(G)$ generators are the elements in group $H^2(G, U(1))$, which classifies the 1D SPT phases. The corresponding anyon permutations in the 2d quantum double are always flux-preserving~\cite{barkeshli2023codimension,li2024domain}. These anyon permutations can be realized by a circuit composed of local commuting gates, as recently pointed out in Ref.~\cite{warman2025transversal}.

To be explicit, on a 1D qudit chain with even number $N$ sites, a 2-cocycle representative $\alpha$ in $H^2(G, U(1))$ gives an SPT state. Its associate SPT entangler defines a class II $G$ self-duality since it commutes with the $G$ symmetry operator $\eta_g = \prod_j L^g_j$,
\begin{equation}
    \begin{aligned}
        E_{\alpha} = \sum_{\{g_i\}} \bigg(\prod_j\frac{\alpha(g_{2j-1}, \bar{g}_{2j-1} g_{2j})}{\alpha(g_{2j+1}, \bar{g}_{2j+1}g_{2j})}\bigg) \ket{\{g_i\}}\bra{\{g_i\}}.
    \end{aligned}
    \label{eq:entangler}
\end{equation}
On the other hand, the matrix product operators $\mathrm{Tr}(Z_1^{\rho \dagger} Z^{\rho}_2 \cdots Z_{N-1}^{\rho \dagger} Z^{\rho}_N)$ for $\rho \in \mathrm{Rep}(G)$ generate $\mathrm{Rep}(G)$ symmetry. This symmetry emerges from the above $G$ symmetry via a gauging map $\Lambda$~\cite{yoshida2017gapped}:
\begin{equation}
    \begin{aligned}
        |\{g_i\}\rangle\ \overset{\Lambda}{\longmapsto}\ 
        \ |\bar{g}_{1}g_{N},\bar{g}_{1}g_{2}, \bar{g}_3 g_2, \cdots\rangle.
    \end{aligned}
    \label{eq:Lambda}
\end{equation}

In the $D(G)$ quantum double model, the flux-free term in Eq.~\eqref{eq:QD_terms} defines such a $\mathrm{Rep}(G)$ symmetry on each hexagonal plaquette, on which qudits are located on the edges. We label the qudits from 1 to 6 as in Eq.~\eqref{eq:QD_terms}, and the corresponding states are given by $|g_1\rangle, \cdots, |g_6\rangle$. The unitary self-duality associated with this hexagon is given by $M_{\alpha} \equiv \Lambda E_{\alpha} \Lambda^{\dagger}$:
\begin{equation}
\begin{aligned}
M_{\alpha} = \sum_{\{g_i\}}&\frac{\alpha(g_*,g_2)\alpha(g_* g_2 \bar{g}_3, g_4)\alpha(g_* g_1 \bar{g}_6, g_6)}{ \alpha(g_*, g_1)\alpha(g_* g_2 \bar{g}_3, g_3) \alpha(g_* g_1 \bar{g}_6, g_5)} |\{g_i\} \rangle \langle \{g_i\}|.
\end{aligned}
\label{eq:phase_gate}
\end{equation}
The ancillary element $g_*$ can be taken as identity, since this gate does not depend on $g_*$ despite its explicit appearance in the above definition. The phase gate $M_{\alpha}$ inserts a gauged-SPT defect on the boundary of this plaquette~\cite{barkeshli2023codimension,li2024domain}. Taking a product of $M_{\alpha}$ over all the plaquettes effectively sweeps the gauged-SPT defect across the entire system, implementing a flux-preserving anyon permutations.

When $G = \mathbb{Z}_2\times \mathbb{Z}_2$, there is one nontrivial class of 2-cocycle. The circuit implements the following anyon permutations in 2 copies of $\mathbb{Z}_2$ toric code:
\begin{align}
    m_1 \to m_1 e_2,\quad m_2 \to m_2 e_1,\quad e_1 \to e_1,\quad e_2 \to e_2,
\end{align}
where $e_i,m_i$ are the excitations in the $i$-th copy. This circuit is transversal, since it is composed of non-overlapping four-qubit gates. See more details in Appendix \ref{app:class_II}. When $G$ is the dihedral group of size $8N$, Ref.~\cite{warman2025transversal} constructs the explicit circuit in the quantum double model with specific choices of boundary, and suggests that it implements a logical $T^{1/N}$ gate. When $N=1$, this class-II anyon permutation in $D_4$ quantum double is
\begin{equation}
    \begin{aligned}
        m_r &\mapsto m_{rb}, \quad m_b \mapsto m_b, \quad m_g \mapsto m_{gb}, \\
        e_b &\mapsto e_r, \quad e_b \mapsto e_{rgb}, \quad e_g \mapsto e_g. \label{eq:D_4_anyon_permutation_2}
    \end{aligned}
\end{equation}

\textit{Class III: group outer automorphisms} - The third class of $\mathcal{G}(G)$ generators are the outer $G$ automorphisms. The corresponding anyon permutations in the 2D quantum double are the permutations of the pure charges and pure fluxes respectively. The outer automorphism group $\mathsf{Out}(G)$ contains automorphisms $\phi$ of $G$, up to the inner automorphisms given by conjugations of $G$ elements. A transversal circuit composed of on-site gates $U_{\phi}=\sum_{g}\ket{\phi(g)}\bra{g}$ on all qudits permutes the star terms while keeping the flux-free terms invariant, thus preserves the ground state subspace. 

When $G=D_4$, an outer automorphism of $D_4$ is given by exchanging $s$ with $rs$. Under the corresponding transversal circuit, anyons in the quantum double transform as:
\begin{equation}
\begin{aligned}
    m_g &\mapsto m_r,\quad e_g \mapsto e_r,\quad m_r \mapsto m_g,\\
    e_r &\mapsto e_g,\quad m_b \mapsto m_b,\quad e_b \mapsto e_b.
\end{aligned}
\label{eq:D_4_anyon_permutation_3}
\end{equation}
With appropriate boundary, this anyon permutation can realize a logical $T$-gate~\cite{davydova2025universal}. In fact, as pointed out in Ref.~\cite{kobayashi2025clifford}, it is in the same conjugacy class of the anyon permutation in Eq.~\eqref{eq:D_4_anyon_permutation}.

Near the completion of this work, we learned of an upcoming work that studies the logical gates implemented via group automorphisms~\cite{tyler2026universal}.

\textit{Discussion and outlook} - We presented a family of finite-depth local unitary circuits for general anyon permutations in $D(G)$ quantum double model on the hexagonal lattice. The three classes of generators of the group of anyon permutations $\mathcal{G}(G)$ are respectively obtained from a sequence of unitary gauging maps, insertion of gauged-SPT defects, and transversal action of group outer automorphisms, all of which are associated with certain self-dualities of $G$ symmetry in 1D.

Among these three classes, class-I circuits consist of single-qudit Hadamard gates, single-qudit $S$ gates, two-qudit controlled-$Z$ gates, and two-qudit controlled-conjugation gates, depending on the group structure. Class-II circuits consist of multi-qudit phase gates $M_{\alpha}$, whose form depends on the explicit choice of the $2$-cocycle $\alpha$. Class-III circuits consist solely of single-qudit automorphism gates $U_{\phi}$.

When $G$ is Abelian, the class-I elements are fundamental in $\mathcal{G}(G)$, since class-II and class-III elements only generate a proper subgroup, while class-I elements together with class-II (or class-III) elements generate $\mathcal{G}(G)$. For the general non-Abelian group $G$, no such clear distinction between the three classes of elements in $\mathcal{G}(G)$ is known to the best of our knowledge. For example, in the $D_4$ quantum double, the three classes of anyon permutations shown in Eq.~(\ref{eq:D_4_anyon_permutation}, \ref{eq:D_4_anyon_permutation_2}, \ref{eq:D_4_anyon_permutation_3}) lie in the same conjugacy class of $\mathcal{G}(D_4)=S_4$~\cite{nikshych2014categorical}.

Along similar lines, Ref.~\cite{lootens2022mapping} constructs constant-depth circuits that map between states in Morita-equivalent extended string-net models, although the precise relation between those circuits and the circuits presented in this work remains unclear. Moreover, we expect that the lattice-level holographic relation presented here persists in twisted quantum double models~\cite{hu2013twisted}, as well as in generalized quantum double models based on Hopf algebras~\cite{buerschaper2013hierarchy}. Another interesting question is whether, from our construction of unitary circuits, we can construct a Floquet code that implements anyon permutation in the $D(G)$ quantum double model via a sequence of local measurements, generalizing the construction in Ref.~\cite{hastings2021dynamically,aasen2022adiabatic,davydova2023quantum}.

Recently, Refs.~\cite{kobayashi2025clifford,warman2025transversal} showed that certain anyon permutations in non-Abelian topological codes, when combined with specific boundary conditions, can realize logical operations at the third or higher levels of the Clifford hierarchy. Our construction provides general recipes to realizing the corresponding permutations. In a related direction, we note that although group inner automorphisms do not induce any nontrivial anyon permutation, their transversal actions could potentially implement non-Clifford logical gates in the presence of appropriate boundaries~\cite{sajith2025non}. The precise relation between anyon permutations and logical operations in topological code with topological boundaries remains an open question.

From an implementation perspective, universal quantum computation can be achieved by combining the preparation and error correction of non-Abelian topological codes~\cite{iqbal2024non,minev2025realizing,gacs1983reliable,harringtonPhD,wootton2016active,dauphinais2017fault,jing2025intrinsic}, the non-Clifford logical operations implemented via the finite-depth circuits presented here, and code switching through anyon condensation. A detailed discussion of the fault-tolerant properties will be presented in a separate work~\cite{workinprogress2026}.

\textit{Acknowledgment} - We thank Isaac Kim, Matteo Dell'Acqua, and Yifan Wang for many insightful conversations. We thank Tyler Ellison and Vieri Mattei for discussions on anyon permutations in $D_4$ quantum double. ZS is supported by the National Science Foundation under Grant No. DRL I-TEST 2148467. YL is supported by the U.S. National Science Foundation under Grant No. NSF DMR-2316598.

\bibliography{reference}

\appendix 
\onecolumngrid
\setcounter{secnumdepth}{2}

\newpage

\begin{center}
{\large\bfseries Supplemental Material for ``Anyon Permutations in Quantum Double Models through Constant-depth Circuits"}
\end{center}

\tableofcontents

\section{Structure of the anyon permutation group} \label{sec:mathematics}

For a unitary fusion category $\mathcal{C}$, its Drinfel'd center describes a two-dimensional topological order $\mathcal{Z}(\mathcal{C})$. Anyon permutations in this topological order that preserve anyon fusion and braiding are classified by the \textit{braided autoequivalences} of $\mathcal{Z}(\mathcal{C})$~\cite{kitaev2012models}, denoted by $\mathsf{Aut}^{br}(\mathcal{Z}(\mathcal{C}))$. Ref.~\cite{etingof2010fusion} establishes the following relation:
\begin{align}
    \mathsf{BrPic}(\mathcal{C}) \cong \mathsf{Aut}^{br}(\mathcal{Z}(\mathcal{C})).
\end{align}
Here, $\mathsf{BrPic}(\mathcal{C})$ denotes the Brauer-Picard group of $\mathcal{C}$, defined as the group of invertible $\mathcal{C}$-$\mathcal{C}$ bimodule categories under the relative tensor product. In the main text, we denote this group by $\mathcal{G}(\mathcal{C})$. In particular, the Kitaev's quantum double models~\cite{kitaev2003fault} are described by the Drinfel'd center of fusion category $\mathrm{Vec}_G$ for finite group $G$.

The structure of the Brauer-Picard group $\mathrm{Vec}_G$ for a finite group $G$ has been studied extensively. Here we will review some of the results in Ref.~\cite{nikshych2014categorical}. We denote by $\mathbb{L}(G)$ the {\it categorical Lagrangian Grassmannian}, i.e. the set of pairs $(N,\alpha)$, where $N$ is a normal Abelian subgroup of $G$ and $\alpha$ is a $G$-invariant cohomology class in $H^2(N,U(1))$. Physically, $\mathbb{L}(G)$ is the set of all $G$ (twisted-)gauging maps, such that the dual symmetry group $G_{(N,\alpha)}$ has the same order as $G$. We further denote by $\mathbb{L}_0(G)\subset \mathbb{L}(G)$ the set of {\it self-dual} $G$ (twisted-)gauging maps that have $G_{(N,\alpha)}\cong G$. In Proposition 7.6 of Ref.~\cite{nikshych2014categorical}, it is shown that the action of $\mathcal{G}(G)$ on $\mathbb{L}_0(G)$ is transitive, and the stabilizer group of the canonical element $\mathcal{L}_0$ associated with pair $(\{e\},1)$ is given by
\begin{equation}
    \mathsf{Stab} \cong H^2(G,U(1)) \rtimes \mathsf{Out}(G),
    \label{eq:stab}
\end{equation}
which is a subgroup of $\mathcal{G}(G)$ generated by the classes-II and III elements discussed in the main text. Following this result, we can assign an element $\Gamma_{\mathcal{L}} \in \mathcal{G}(G)$ for $\mathcal{L}\in \mathbb{L}_0(G)$, such that its action maps $\mathcal{L}_0$ to $\mathcal{L}$. These elements $\{\Gamma_{\mathcal{L}}\}$ are the class-I generators discussed in the main text.

For an arbitrary element $\mathbf{g}\in\mathcal{G}(G)$, its action maps $\mathcal{L}_0$ to some element $\mathcal{L}\in\mathbb{L}_0(G)$, then the composition $\Gamma^{-1}_{\mathcal{L}} \cdot \mathbf{g}$ stabilizes $\mathcal{L}_0$. As a result, $\Gamma^{-1}_{\mathcal{L}} \cdot \mathbf{g}$ belongs to the stabilizer group in Eq.~\eqref{eq:stab}, and can therefore be written as $M_{\alpha}\cdot U_{\phi}$ for some 2-cocycle $[\alpha]\in H^2(G,U(1))$, and outer automorphism $\phi \in \mathsf{Out}(G)$. In other words, any element $\mathbf{g}\in\mathcal{G}(G)$ can be written as
\begin{equation}
    \mathbf{g} = \Gamma_{\mathcal{L}}\cdot M_{\alpha}\cdot U_{\phi}.
\end{equation}
Furthermore, the size of $\mathsf{BrPic}(\mathrm{Vec}_G)$ can also be computed:
\begin{align}
    |\mathsf{BrPic}(\mathrm{Vec}_G)| = |H^2(G, U(1))| \cdot |\mathsf{Out}(G)| \cdot |\mathbb{L}_0(G)|.
    \label{eq:BrPic_order}
\end{align}

In the example of $G = \mathbb{Z}_2$, both $H^2(\mathbb{Z}_2,U(1))$ and $\mathsf{Out}(\mathbb{Z}_2)$ are trivial. There is only one nontrivial element in $\mathcal{G}(\mathbb{Z}_2)$, which corresponds to the 1D Kramers-Wannier gauging map or the electric-magnetic duality in the 2D toric code, as explained in the main text. In the example of $G = D_4$, both $H^2(D_4,U(1))$ and $\mathsf{Out}(D_4)$ are $\mathbb{Z}_2$, while $|\mathbb{L}_0(G)|=6$ corresponding to the following pairs~\cite{nikshych2014categorical}:
\begin{equation}
    (1,1),\ (\mathbb{Z}_4 = \langle r\rangle, 1),\ (\mathbb{Z}_2^2 = \langle s, r^2\rangle, 1),\ (\mathbb{Z}_2^2 = \langle sr, r^2\rangle, 1),\ (\mathbb{Z}_2^2 = \langle s, r^2\rangle, \mu_1),\ (\mathbb{Z}_2^2 = \langle sr, r^2\rangle, \mu_2),
\end{equation}
where $\mu_1$ and $\mu_2$ denote the nontrivial cohomology classes. In fact, using the above results, it can be shown that $\mathcal{G}(D_4) = S_4$~\cite{nikshych2014categorical}.
In the following sections, we will demonstrate the explicit gauging map and the associated anyon permutation for the third element above.

\section{Review of the electric-magnetic duality in the toric code model} \label{sec:toric_code}

In this section, we review the electric-magnetic duality circuit discussed in Ref.~\cite{aasen2022adiabatic}. Consider the triangular lattice in Fig.~\ref{fig:figure_1}(a). We assign qubit degrees of freedom at each vertex. The Hamiltonian stabilizers are given as follows:

\raisebox{-0.5\height}{
\begin{minipage}{0.95\textwidth}
\centering
\begin{tikzpicture}[scale=0.9, baseline=(current bounding box.center)]
\tikzset{
  hexbond/.style={line width=0.6pt},
  vdot/.style={circle, fill=black, inner sep=1.2pt}
}

\coordinate (A_1) at ( 0, 2*0.8660254);
\coordinate (A_2) at ( -1, 0);
\coordinate (A_3) at ( 1, 0);
\draw[hexbond] (A_1)--(A_2)--(A_3)--cycle;

\node at ($(A_1)+(0,0.2)$) {$X$};
\node at ($(A_2)+(-0.4,0)$) {$X$};
\node at ($(A_3)+(0.4,0)$) {$X$};
\node[myGreen] at (0, 2*0.8660254/3) {$1$};
\end{tikzpicture},
\begin{tikzpicture}[scale=0.9, baseline=(current bounding box.center)]
\tikzset{
  hexbond/.style={line width=0.6pt},
  vdot/.style={circle, fill=black, inner sep=1.2pt}
}

\coordinate (A_1) at ( 0, 2*0.8660254);
\coordinate (A_2) at ( -1, 0);
\coordinate (A_3) at ( 1, 0);
\draw[hexbond] (A_1)--(A_2)--(A_3)--cycle;

\node at ($(A_1)+(0,0.2)$) {$X$};
\node at ($(A_2)+(-0.4,0)$) {$X$};
\node at ($(A_3)+(0.4,0)$) {$X$};
\node[myBlue] at (0, 2*0.8660254/3) {$2$};
\end{tikzpicture},
\begin{tikzpicture}[scale=0.9, baseline=(current bounding box.center)]
\tikzset{
  hexbond/.style={line width=0.6pt},
  vdot/.style={circle, fill=black, inner sep=1.2pt}
}

\coordinate (A_1) at ( 0, 2*0.8660254);
\coordinate (A_2) at ( -1, 0);
\coordinate (A_3) at ( 1, 0);
\draw[hexbond] (A_1)--(A_2)--(A_3)--cycle;

\coordinate (A_4) at ( -2, 2*0.8660254);
\coordinate (A_5) at ( 0, -2*0.8660254);
\coordinate (A_6) at ( 2, 2*0.8660254);
\draw[hexbond] (A_4)--(A_5)--(A_6)--cycle;

\node at ($(A_1)+(0,0.2)$) {$Z$};
\node at ($(A_2)+(-0.4,0)$) {$Z$};
\node at ($(A_3)+(0.4,0)$) {$Z$};
\node at ($(A_4)+(0,0.2)$) {$Z$};
\node at ($(A_5)+(0,-0.2)$) {$Z$};
\node at ($(A_6)+(0,0.2)$) {$Z$};
\node[myRed] at (0, 2*0.8660254/3) {$0$}; 
\end{tikzpicture}.
\end{minipage}
}

Consider the following state $|\{n_1, n_2, n_3, \ldots, n_L\}\rangle$, where $n_i \in \{0,1\}$, the $\mathbb{Z}_2$ gauging map is defined as follows:
\begin{align}
    \Gamma: |\{n_1, n_2, \ldots, n_L\} \mapsto \sum_{\{m_i\}} (-1)^{m_i (n_i - n_{i-1})} |\{m_1, m_2, \ldots, m_L\}\rangle.
\end{align}
This map can be equivalently realized in the $\mathbb{Z}_2$ symmetric sector by a depth $\mathcal{O}(L)$ local unitary circuit. It implements a locality preserving map on the symmetric local operators:
\begin{align}
    \Gamma: X_i \mapsto Z_i Z_{i+1}, \quad Z_i Z_{i+1} \mapsto X_{i+1}.
\end{align}

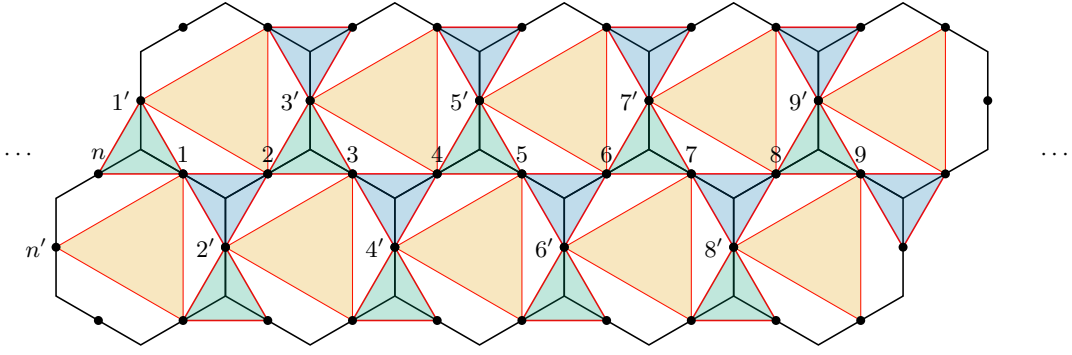
\begin{figure}
    \centering
    \begin{tikzpicture}[scale=1.3]
\tikzset{
  hexbond/.style={line width=0.6pt},
  vdot/.style={circle, fill=black, inner sep=1.2pt}
}



\coordinate (A_U)  at ( 0.0000000,  1.0000000);
\coordinate (A_UR) at ( 0.8660254,  0.5000000);
\coordinate (A_LR) at ( 0.8660254, -0.5000000);
\coordinate (A_D)  at ( 0.0000000, -1.0000000);
\coordinate (A_LL) at (-0.8660254, -0.5000000);
\coordinate (A_UL) at (-0.8660254,  0.5000000);
\draw[hexbond] (A_U)--(A_UR)--(A_LR)--(A_D)--(A_LL)--(A_UL)--cycle;

\draw[red] ($(A_U)!0.5!(A_UR)$)--($(A_LR)!0.5!(A_D)$)--($(A_LL)!0.5!(A_UL)$)--($(A_U)!0.5!(A_UR)$);
\fill[myRed, nearly transparent] ($(A_U)!0.5!(A_UR)$)--($(A_LR)!0.5!(A_D)$)--($(A_LL)!0.5!(A_UL)$)--($(A_U)!0.5!(A_UR)$);

\coordinate (B_U)  at ( 1.7320508+0.0000000,  0.0000000+1.0000000);
\coordinate (B_UR) at ( 1.7320508+0.8660254,  0.0000000+0.5000000);
\coordinate (B_LR) at ( 1.7320508+0.8660254,  0.0000000-0.5000000);
\coordinate (B_D)  at ( 1.7320508+0.0000000,  0.0000000-1.0000000);
\coordinate (B_LL) at ( 1.7320508-0.8660254,  0.0000000-0.5000000);
\coordinate (B_UL) at ( 1.7320508-0.8660254,  0.0000000+0.5000000);
\draw[hexbond] (B_U)--(B_UR)--(B_LR)--(B_D)--(B_LL)--(B_UL)--cycle;

\draw[red] ($(B_U)!0.5!(B_UR)$)--($(B_LR)!0.5!(B_D)$)--($(B_LL)!0.5!(B_UL)$)--($(B_U)!0.5!(B_UR)$);
\fill[myRed, nearly transparent] ($(B_U)!0.5!(B_UR)$)--($(B_LR)!0.5!(B_D)$)--($(B_LL)!0.5!(B_UL)$)--($(B_U)!0.5!(B_UR)$);

\coordinate (C_U)  at ( 3.4641016+0.0000000,  0.0000000+1.0000000);
\coordinate (C_UR) at ( 3.4641016+0.8660254,  0.0000000+0.5000000);
\coordinate (C_LR) at ( 3.4641016+0.8660254,  0.0000000-0.5000000);
\coordinate (C_D)  at ( 3.4641016+0.0000000,  0.0000000-1.0000000);
\coordinate (C_LL) at ( 3.4641016-0.8660254,  0.0000000-0.5000000);
\coordinate (C_UL) at ( 3.4641016-0.8660254,  0.0000000+0.5000000);
\draw[hexbond] (C_U)--(C_UR)--(C_LR)--(C_D)--(C_LL)--(C_UL)--cycle;

\draw[red] ($(C_U)!0.5!(C_UR)$)--($(C_LR)!0.5!(C_D)$)--($(C_LL)!0.5!(C_UL)$)--($(C_U)!0.5!(C_UR)$);
\fill[myRed, nearly transparent] ($(C_U)!0.5!(C_UR)$)--($(C_LR)!0.5!(C_D)$)--($(C_LL)!0.5!(C_UL)$)--($(C_U)!0.5!(C_UR)$);

\coordinate (D_U)  at ( 5.1961524+0.0000000,  0.0000000+1.0000000);
\coordinate (D_UR) at ( 5.1961524+0.8660254,  0.0000000+0.5000000);
\coordinate (D_LR) at ( 5.1961524+0.8660254,  0.0000000-0.5000000);
\coordinate (D_D)  at ( 5.1961524+0.0000000,  0.0000000-1.0000000);
\coordinate (D_LL) at ( 5.1961524-0.8660254,  0.0000000-0.5000000);
\coordinate (D_UL) at ( 5.1961524-0.8660254,  0.0000000+0.5000000);
\draw[hexbond] (D_U)--(D_UR)--(D_LR)--(D_D)--(D_LL)--(D_UL)--cycle;

\draw[red] ($(D_U)!0.5!(D_UR)$)--($(D_LR)!0.5!(D_D)$)--($(D_LL)!0.5!(D_UL)$)--($(D_U)!0.5!(D_UR)$);
\fill[myRed, nearly transparent] ($(D_U)!0.5!(D_UR)$)--($(D_LR)!0.5!(D_D)$)--($(D_LL)!0.5!(D_UL)$)--($(D_U)!0.5!(D_UR)$);

\coordinate (E_U)  at ( 6.9282032+0.0000000,  0.0000000+1.0000000);
\coordinate (E_UR) at ( 6.9282032+0.8660254,  0.0000000+0.5000000);
\coordinate (E_LR) at ( 6.9282032+0.8660254,  0.0000000-0.5000000);
\coordinate (E_D)  at ( 6.9282032+0.0000000,  0.0000000-1.0000000);
\coordinate (E_LL) at ( 6.9282032-0.8660254,  0.0000000-0.5000000);
\coordinate (E_UL) at ( 6.9282032-0.8660254,  0.0000000+0.5000000);
\draw[hexbond] (E_U)--(E_UR)--(E_LR)--(E_D)--(E_LL)--(E_UL)--cycle;

\draw[red] ($(E_U)!0.5!(E_UR)$)--($(E_LR)!0.5!(E_D)$)--($(E_LL)!0.5!(E_UL)$)--($(E_U)!0.5!(E_UR)$);
\fill[myRed, nearly transparent] ($(E_U)!0.5!(E_UR)$)--($(E_LR)!0.5!(E_D)$)--($(E_LL)!0.5!(E_UL)$)--($(E_U)!0.5!(E_UR)$);


\coordinate (F_U)  at ( 0.8660254+0.0000000,  1.5+1.0000000);
\coordinate (F_UR) at ( 0.8660254+0.8660254,  1.5+0.5000000);
\coordinate (F_LR) at ( 0.8660254+0.8660254,  1.5-0.5000000);
\coordinate (F_D)  at ( 0.8660254+0.0000000,  1.5-1.0000000);
\coordinate (F_LL) at ( 0.8660254-0.8660254,  1.5-0.5000000);
\coordinate (F_UL) at ( 0.8660254-0.8660254,  1.5+0.5000000);
\draw[hexbond] (F_U)--(F_UR)--(F_LR)--(F_D)--(F_LL)--(F_UL)--cycle;

\draw[red] ($(F_U)!0.5!(F_UR)$)--($(F_LR)!0.5!(F_D)$)--($(F_LL)!0.5!(F_UL)$)--($(F_U)!0.5!(F_UR)$);
\fill[myRed, nearly transparent] ($(F_U)!0.5!(F_UR)$)--($(F_LR)!0.5!(F_D)$)--($(F_LL)!0.5!(F_UL)$)--($(F_U)!0.5!(F_UR)$);

\coordinate (G_U)  at ( 2.5980762+0.0000000,  1.5+1.0000000);
\coordinate (G_UR) at ( 2.5980762+0.8660254,  1.5+0.5000000);
\coordinate (G_LR) at ( 2.5980762+0.8660254,  1.5-0.5000000);
\coordinate (G_D)  at ( 2.5980762+0.0000000,  1.5-1.0000000);
\coordinate (G_LL) at ( 2.5980762-0.8660254,  1.5-0.5000000);
\coordinate (G_UL) at ( 2.5980762-0.8660254,  1.5+0.5000000);
\draw[hexbond] (G_U)--(G_UR)--(G_LR)--(G_D)--(G_LL)--(G_UL)--cycle;

\draw[red] ($(G_U)!0.5!(G_UR)$)--($(G_LR)!0.5!(G_D)$)--($(G_LL)!0.5!(G_UL)$)--($(G_U)!0.5!(G_UR)$);
\fill[myRed, nearly transparent] ($(G_U)!0.5!(G_UR)$)--($(G_LR)!0.5!(G_D)$)--($(G_LL)!0.5!(G_UL)$)--($(G_U)!0.5!(G_UR)$);

\coordinate (H_U)  at ( 4.3301270+0.0000000,  1.5+1.0000000);
\coordinate (H_UR) at ( 4.3301270+0.8660254,  1.5+0.5000000);
\coordinate (H_LR) at ( 4.3301270+0.8660254,  1.5-0.5000000);
\coordinate (H_D)  at ( 4.3301270+0.0000000,  1.5-1.0000000);
\coordinate (H_LL) at ( 4.3301270-0.8660254,  1.5-0.5000000);
\coordinate (H_UL) at ( 4.3301270-0.8660254,  1.5+0.5000000);
\draw[hexbond] (H_U)--(H_UR)--(H_LR)--(H_D)--(H_LL)--(H_UL)--cycle;

\draw[red] ($(H_U)!0.5!(H_UR)$)--($(H_LR)!0.5!(H_D)$)--($(H_LL)!0.5!(H_UL)$)--($(H_U)!0.5!(H_UR)$);
\fill[myRed, nearly transparent] ($(H_U)!0.5!(H_UR)$)--($(H_LR)!0.5!(H_D)$)--($(H_LL)!0.5!(H_UL)$)--($(H_U)!0.5!(H_UR)$);

\coordinate (I_U)  at ( 6.0621778+0.0000000,  1.5+1.0000000);
\coordinate (I_UR) at ( 6.0621778+0.8660254,  1.5+0.5000000);
\coordinate (I_LR) at ( 6.0621778+0.8660254,  1.5-0.5000000);
\coordinate (I_D)  at ( 6.0621778+0.0000000,  1.5-1.0000000);
\coordinate (I_LL) at ( 6.0621778-0.8660254,  1.5-0.5000000);
\coordinate (I_UL) at ( 6.0621778-0.8660254,  1.5+0.5000000);
\draw[hexbond] (I_U)--(I_UR)--(I_LR)--(I_D)--(I_LL)--(I_UL)--cycle;

\draw[red] ($(I_U)!0.5!(I_UR)$)--($(I_LR)!0.5!(I_D)$)--($(I_LL)!0.5!(I_UL)$)--($(I_U)!0.5!(I_UR)$);
\fill[myRed, nearly transparent] ($(I_U)!0.5!(I_UR)$)--($(I_LR)!0.5!(I_D)$)--($(I_LL)!0.5!(I_UL)$)--($(I_U)!0.5!(I_UR)$);

\coordinate (J_U)  at ( 7.7942286+0.0000000,  1.5+1.0000000);
\coordinate (J_UR) at ( 7.7942286+0.8660254,  1.5+0.5000000);
\coordinate (J_LR) at ( 7.7942286+0.8660254,  1.5-0.5000000);
\coordinate (J_D)  at ( 7.7942286+0.0000000,  1.5-1.0000000);
\coordinate (J_LL) at ( 7.7942286-0.8660254,  1.5-0.5000000);
\coordinate (J_UL) at ( 7.7942286-0.8660254,  1.5+0.5000000);
\draw[hexbond] (J_U)--(J_UR)--(J_LR)--(J_D)--(J_LL)--(J_UL)--cycle;

\draw[red] ($(J_U)!0.5!(J_UR)$)--($(J_LR)!0.5!(J_D)$)--($(J_LL)!0.5!(J_UL)$)--($(J_U)!0.5!(J_UR)$);
\fill[myRed, nearly transparent] ($(J_U)!0.5!(J_UR)$)--($(J_LR)!0.5!(J_D)$)--($(J_LL)!0.5!(J_UL)$)--($(J_U)!0.5!(J_UR)$);

\tikzset{vtri/.style={line width=0.6pt}} 

\draw[vtri,red]
  ($(A_U)!0.5!(A_UR)$) -- ($(A_UR)!0.5!(A_LR)$) -- ($(B_UL)!0.5!(B_U)$) -- cycle;
\fill[myBlue, nearly transparent] ($(A_U)!0.5!(A_UR)$) -- ($(A_UR)!0.5!(A_LR)$) -- ($(B_UL)!0.5!(B_U)$) -- cycle;

\draw[vtri,red]
  ($(B_U)!0.5!(B_UR)$) -- ($(B_UR)!0.5!(B_LR)$) -- ($(C_UL)!0.5!(C_U)$) -- cycle;
\fill[myBlue, nearly transparent] ($(B_U)!0.5!(B_UR)$) -- ($(B_UR)!0.5!(B_LR)$) -- ($(C_UL)!0.5!(C_U)$) -- cycle;

\draw[vtri,red]
  ($(C_U)!0.5!(C_UR)$) -- ($(C_UR)!0.5!(C_LR)$) -- ($(D_UL)!0.5!(D_U)$) -- cycle;
\fill[myBlue, nearly transparent] ($(C_U)!0.5!(C_UR)$) -- ($(C_UR)!0.5!(C_LR)$) -- ($(D_UL)!0.5!(D_U)$) -- cycle;

\draw[vtri,red]
  ($(D_U)!0.5!(D_UR)$) -- ($(D_UR)!0.5!(D_LR)$) -- ($(E_UL)!0.5!(E_U)$) -- cycle;
\fill[myBlue, nearly transparent] ($(D_U)!0.5!(D_UR)$) -- ($(D_UR)!0.5!(D_LR)$) -- ($(E_UL)!0.5!(E_U)$) -- cycle;

\draw[vtri,red]
  ($(E_U)!0.5!(E_UR)$) -- ($(E_UR)!0.5!(E_LR)$) -- ($(J_D)!0.5!(J_LR)$) -- cycle;
\fill[myBlue, nearly transparent] ($(E_U)!0.5!(E_UR)$) -- ($(E_UR)!0.5!(E_LR)$) -- ($(J_D)!0.5!(J_LR)$) -- cycle;

\draw[vtri,red]
  ($(A_U)!0.5!(A_UL)$) -- ($(A_U)!0.5!(A_UR)$) -- ($(F_UL)!0.5!(F_LL)$) -- cycle;
\fill[myGreen, nearly transparent] ($(A_U)!0.5!(A_UL)$) -- ($(A_U)!0.5!(A_UR)$) -- ($(F_UL)!0.5!(F_LL)$) -- cycle;

\draw[vtri,red]
  ($(B_U)!0.5!(B_UL)$) -- ($(B_U)!0.5!(B_UR)$) -- ($(G_UL)!0.5!(G_LL)$) -- cycle;
\fill[myGreen, nearly transparent] ($(B_U)!0.5!(B_UL)$) -- ($(B_U)!0.5!(B_UR)$) -- ($(G_UL)!0.5!(G_LL)$) -- cycle;

\draw[vtri,red]
  ($(C_U)!0.5!(C_UL)$) -- ($(C_U)!0.5!(C_UR)$) -- ($(H_UL)!0.5!(H_LL)$) -- cycle;
\fill[myGreen, nearly transparent] ($(C_U)!0.5!(C_UL)$) -- ($(C_U)!0.5!(C_UR)$) -- ($(H_UL)!0.5!(H_LL)$) -- cycle;

\draw[vtri,red]
  ($(D_U)!0.5!(D_UL)$) -- ($(D_U)!0.5!(D_UR)$) -- ($(I_UL)!0.5!(I_LL)$) -- cycle;
\fill[myGreen, nearly transparent]  ($(D_U)!0.5!(D_UL)$) -- ($(D_U)!0.5!(D_UR)$) -- ($(I_UL)!0.5!(I_LL)$) -- cycle;

\draw[vtri,red]
  ($(E_U)!0.5!(E_UL)$) -- ($(E_U)!0.5!(E_UR)$) -- ($(J_UL)!0.5!(J_LL)$) -- cycle;
\fill[myGreen, nearly transparent] ($(E_U)!0.5!(E_UL)$) -- ($(E_U)!0.5!(E_UR)$) -- ($(J_UL)!0.5!(J_LL)$) -- cycle;

\draw[vtri,red]
  ($(A_UR)!0.5!(A_LR)$) -- ($(A_LR)!0.5!(A_D)$) -- ($(B_LL)!0.5!(B_D)$) -- cycle;
\fill[myGreen, nearly transparent] ($(A_UR)!0.5!(A_LR)$) -- ($(A_LR)!0.5!(A_D)$) -- ($(B_LL)!0.5!(B_D)$) -- cycle;

\draw[vtri,red]
  ($(B_UR)!0.5!(B_LR)$) -- ($(B_LR)!0.5!(B_D)$) -- ($(C_LL)!0.5!(C_D)$) -- cycle;
\fill[myGreen, nearly transparent] ($(B_UR)!0.5!(B_LR)$) -- ($(B_LR)!0.5!(B_D)$) -- ($(C_LL)!0.5!(C_D)$) -- cycle;

\draw[vtri,red]
  ($(C_UR)!0.5!(C_LR)$) -- ($(C_LR)!0.5!(C_D)$) -- ($(D_LL)!0.5!(D_D)$) -- cycle;
\fill[myGreen, nearly transparent] ($(C_UR)!0.5!(C_LR)$) -- ($(C_LR)!0.5!(C_D)$) -- ($(D_LL)!0.5!(D_D)$) -- cycle;

\draw[vtri,red]
  ($(D_UR)!0.5!(D_LR)$) -- ($(D_LR)!0.5!(D_D)$) -- ($(E_LL)!0.5!(E_D)$) -- cycle;
\fill[myGreen, nearly transparent] ($(D_UR)!0.5!(D_LR)$) -- ($(D_LR)!0.5!(D_D)$) -- ($(E_LL)!0.5!(E_D)$) -- cycle;

\draw[vtri,red]
  ($(F_U)!0.5!(F_UR)$) -- ($(F_UR)!0.5!(F_LR)$) -- ($(G_U)!0.5!(G_UL)$) -- cycle;
\fill[myBlue, nearly transparent] ($(F_U)!0.5!(F_UR)$) -- ($(F_UR)!0.5!(F_LR)$) -- ($(G_U)!0.5!(G_UL)$) -- cycle;

\draw[vtri,red]
  ($(G_U)!0.5!(G_UR)$) -- ($(G_UR)!0.5!(G_LR)$) -- ($(H_U)!0.5!(H_UL)$) -- cycle;
\fill[myBlue, nearly transparent] ($(G_U)!0.5!(G_UR)$) -- ($(G_UR)!0.5!(G_LR)$) -- ($(H_U)!0.5!(H_UL)$) -- cycle;

\draw[vtri,red]
  ($(H_U)!0.5!(H_UR)$) -- ($(H_UR)!0.5!(H_LR)$) -- ($(I_U)!0.5!(I_UL)$) -- cycle;
\fill[myBlue, nearly transparent] ($(H_U)!0.5!(H_UR)$) -- ($(H_UR)!0.5!(H_LR)$) -- ($(I_U)!0.5!(I_UL)$) -- cycle;

\draw[vtri,red]
  ($(I_U)!0.5!(I_UR)$) -- ($(I_UR)!0.5!(I_LR)$) -- ($(J_U)!0.5!(J_UL)$) -- cycle;
\fill[myBlue, nearly transparent] ($(I_U)!0.5!(I_UR)$) -- ($(I_UR)!0.5!(I_LR)$) -- ($(J_U)!0.5!(J_UL)$) -- cycle;

\node[vdot] at ($(A_U)!0.5!(A_UR)$) {};
\node[vdot] at ($(A_UR)!0.5!(A_LR)$) {};
\node[vdot] at ($(A_LR)!0.5!(A_D)$) {};
\node[vdot] at ($(A_D)!0.5!(A_LL)$) {};
\node[vdot] at ($(A_LL)!0.5!(A_UL)$) {};
\node[vdot] at ($(A_UL)!0.5!(A_U)$) {};
\node[vdot] at ($(B_U)!0.5!(B_UR)$) {};
\node[vdot] at ($(B_UR)!0.5!(B_LR)$) {};
\node[vdot] at ($(B_LR)!0.5!(B_D)$) {};
\node[vdot] at ($(B_D)!0.5!(B_LL)$) {};
\node[vdot] at ($(B_LL)!0.5!(B_UL)$) {};
\node[vdot] at ($(B_UL)!0.5!(B_U)$) {};
\node[vdot] at ($(C_U)!0.5!(C_UR)$) {};
\node[vdot] at ($(C_UR)!0.5!(C_LR)$) {};
\node[vdot] at ($(C_LR)!0.5!(C_D)$) {};
\node[vdot] at ($(C_D)!0.5!(C_LL)$) {};
\node[vdot] at ($(C_LL)!0.5!(C_UL)$) {};
\node[vdot] at ($(C_UL)!0.5!(C_U)$) {};
\node[vdot] at ($(D_U)!0.5!(D_UR)$) {};
\node[vdot] at ($(D_UR)!0.5!(D_LR)$) {};
\node[vdot] at ($(D_LR)!0.5!(D_D)$) {};
\node[vdot] at ($(D_D)!0.5!(D_LL)$) {};
\node[vdot] at ($(D_LL)!0.5!(D_UL)$) {};
\node[vdot] at ($(D_UL)!0.5!(D_U)$) {};
\node[vdot] at ($(E_U)!0.5!(E_UR)$) {};
\node[vdot] at ($(E_UR)!0.5!(E_LR)$) {};
\node[vdot] at ($(E_LR)!0.5!(E_D)$) {};
\node[vdot] at ($(E_D)!0.5!(E_LL)$) {};
\node[vdot] at ($(E_LL)!0.5!(E_UL)$) {};
\node[vdot] at ($(E_UL)!0.5!(E_U)$) {};
\node[vdot] at ($(F_U)!0.5!(F_UR)$) {};
\node[vdot] at ($(F_UR)!0.5!(F_LR)$) {};
\node[vdot] at ($(F_LR)!0.5!(F_D)$) {};
\node[vdot] at ($(F_D)!0.5!(F_LL)$) {};
\node[vdot] at ($(F_LL)!0.5!(F_UL)$) {};
\node[vdot] at ($(F_UL)!0.5!(F_U)$) {};
\node[vdot] at ($(G_U)!0.5!(G_UR)$) {};
\node[vdot] at ($(G_UR)!0.5!(G_LR)$) {};
\node[vdot] at ($(G_LR)!0.5!(G_D)$) {};
\node[vdot] at ($(G_D)!0.5!(G_LL)$) {};
\node[vdot] at ($(G_LL)!0.5!(G_UL)$) {};
\node[vdot] at ($(G_UL)!0.5!(G_U)$) {};
\node[vdot] at ($(H_U)!0.5!(H_UR)$) {};
\node[vdot] at ($(H_UR)!0.5!(H_LR)$) {};
\node[vdot] at ($(H_LR)!0.5!(H_D)$) {};
\node[vdot] at ($(H_D)!0.5!(H_LL)$) {};
\node[vdot] at ($(H_LL)!0.5!(H_UL)$) {};
\node[vdot] at ($(H_UL)!0.5!(H_U)$) {};
\node[vdot] at ($(I_U)!0.5!(I_UR)$) {};
\node[vdot] at ($(I_UR)!0.5!(I_LR)$) {};
\node[vdot] at ($(I_LR)!0.5!(I_D)$) {};
\node[vdot] at ($(I_D)!0.5!(I_LL)$) {};
\node[vdot] at ($(I_LL)!0.5!(I_UL)$) {};
\node[vdot] at ($(I_UL)!0.5!(I_U)$) {};
\node[vdot] at ($(J_U)!0.5!(J_UR)$) {};
\node[vdot] at ($(J_UR)!0.5!(J_LR)$) {};
\node[vdot] at ($(J_LR)!0.5!(J_D)$) {};
\node[vdot] at ($(J_D)!0.5!(J_LL)$) {};
\node[vdot] at ($(J_LL)!0.5!(J_UL)$) {};
\node[vdot] at ($(J_UL)!0.5!(J_U)$) {};

\node at ($(A_U)!0.5!(A_UL)+(-0.8,0.2)$) {$\cdots$};
\node at ($(A_U)!0.5!(A_UL)+(9.8,0.2)$) {$\cdots$};
\node at ($(A_U)!0.5!(A_UL)+(0,0.2)$) {$n$};
\node at ($(A_U)!0.5!(A_UR)+(0,0.2)$) {1};
\node at ($(B_U)!0.5!(B_UL)+(0,0.2)$) {2};
\node at ($(B_U)!0.5!(B_UR)+(0,0.2)$) {3};
\node at ($(C_U)!0.5!(C_UL)+(0,0.2)$) {4};
\node at ($(C_U)!0.5!(C_UR)+(0,0.2)$) {5};
\node at ($(D_U)!0.5!(D_UL)+(0,0.2)$) {6};
\node at ($(D_U)!0.5!(D_UR)+(0,0.2)$) {7};
\node at ($(E_U)!0.5!(E_UL)+(0,0.2)$) {8};
\node at ($(E_U)!0.5!(E_UR)+(0,0.2)$) {9};

\node at ($(A_UL)!0.5!(A_LL)+(-0.2,0)$) {$n'$};
\node at ($(B_UL)!0.5!(B_LL)+(-0.2,0)$) {$2'$};
\node at ($(C_UL)!0.5!(C_LL)+(-0.2,0)$) {$4'$};
\node at ($(D_UL)!0.5!(D_LL)+(-0.2,0)$) {$6'$};
\node at ($(E_UL)!0.5!(E_LL)+(-0.2,0)$) {$8'$};

\node at ($(F_LL)!0.5!(F_UL)+(-0.2,0)$) {$1'$};
\node at ($(G_LL)!0.5!(G_UL)+(-0.2,0)$) {$3'$};
\node at ($(H_LL)!0.5!(H_UL)+(-0.2,0)$) {$5'$};
\node at ($(I_LL)!0.5!(I_UL)+(-0.2,0)$) {$7'$};
\node at ($(J_LL)!0.5!(J_UL)+(-0.2,0)$) {$9'$};

\end{tikzpicture}
    \caption{The deformed 3-colorable triangular lattice with periodic boundary condition. Qubits (Qudits) are placed on the edges of this hexagonal lattice. Generalized $X$ terms are defined on each vertex, while generalized $Z$ terms are defined on each plaquette.}
    \label{fig:deformed_lattice}
\end{figure}

We now implement the unitary gauging maps in the manner illustrated in Fig.~\ref{fig:figure_1}(a). We first apply the gauging map on all the $1$-triangles, then on all the $0$-triangles, and finally on all the $2$-triangles. It is straightforward to check that the set of stabilizers returns to itself after the above process. However, anyon ribbon operators transform nontrivially. Consider the lattice shown in Fig.~\ref{fig:deformed_lattice}. A closed electric anyon ribbon operator can be written as
\begin{align}
    Z_1 Z_2 \cdots Z_n.
\end{align}
After applying the gauging map on all the $1$-triangles, this operator transforms to
\begin{align}
    X_1 X_3 \cdots X_{n-1}.
\end{align}
We then apply the gauging map on all the $0$-triangles, which gives
\begin{align}
    Z_1 Z_{2'} Z_3 Z_{4'} \cdots Z_{n'}.
\end{align}
Finally, applying the gauging map on all the $2$-triangles yields
\begin{align}
    X_{2'} X_{4'} \cdots X_{n'},
\end{align}
which corresponds to the magnetic anyonic ribbon operator. One can also check that the inverse map is realized by applying the above process once again. Therefore, we conclude that the above procedure realizes the $e$-$m$ anyon permutation. This $e$-$m$ duality circuit is in constant depth, since each unitary gauging map is implemented by a constant-depth circuit.

\section{Anyon permutation: class I}
\label{sec:general_gauging}

In this section, we discuss in detail how to realize the class I anyon permutation using constant-depth local unitary circuits.

\subsection{Gauging map and self-duality}

Consider a 1-dimensional spin chain with local Hilbert space spanned by $\mathbb{C}[G= N \rtimes Q]$, where $N$ is an abelian normal subgroup and $Q$ is the quotient group $G/N$. This is represented by the following group extension:
\begin{align}
    1 \to N \to G \to Q = G/N \to 1,
\end{align}
together with a conjugation action $\sigma: Q \to \mathsf{Aut}(N)$ and a $2$-cocycle $[\omega] \in H^2(Q, N)$. Group elements are labeled by pairs $(n,q)$, with $n \in N$ and $q \in Q$. The group multiplication is given as follows:
\begin{align}
    (n_1, q_1)(n_2, q_2) = \bigl(n_1\, \sigma^{q_1}(n_2)\, \omega(q_1, q_2),\, q_1 q_2 \bigr).
\end{align}

Gauging the $N$ symmetry maps it to its Pontryagin dual $\hat{N}=\mathrm{Hom}(N,U(1))$, which is isomorphic to $N$ when $N$ is abelian. We may therefore choose a group isomorphism $f: \hat{N} \to N$, so that after gauging and applying $f$, the local Hilbert space is identified with the original Hilbert space. This is realized with the help of a $G$-invariant bicharacter of $N$. The explicit (left)-gauging maps is given as follows: 
\begin{equation}
    \begin{aligned}
        \Gamma: | \{n_i, q_i\} \rangle \to \sum_{\{m_i\}} \prod_{s} \chi(m_s, n_{s-1}^{-1} n_s) |\{m_i, q_i\}\rangle. \label{eq:general_gauging}
    \end{aligned}
\end{equation}
In the above definition, $\chi(\cdot, \cdot)$ is a bicharacter, $\chi: N \times N \to U(1)$, satisfying the following conditions:
\begin{equation}
    \begin{aligned}
        &\chi(m, n) \chi(m', n) = \chi(m m', n),\quad \chi(m, n) \chi(m, n') = \chi(m, n n'),\quad \chi(\sigma^q({m}), \sigma^q(n)) = \chi(m, n).\label{eq:bicharacter_identities}
    \end{aligned}
\end{equation}

We define the following local operators:
\begin{equation}
    \begin{aligned}
        L^n_i |n_i, q_i\rangle = |n n_i, q_i\rangle,&\quad R^n_i |n_i, q_i\rangle = |n_i \prescript{q_i}{}{n}, q_i\rangle,\quad Z^{\hat{n}}_i = \sum_{m_i} \chi(m_i, n) |m_i, q_i\rangle \langle m_i, q_i|, \\
        L^q_i |n_i, q_i\rangle = |\sigma^q({n_i}), q q_i\rangle,&\quad R^q_i |n_i, q_i\rangle = |n_i, q_iq\rangle,\quad T_i^{q} = \sum_{m_i\in N} |n_i, q \rangle \langle n_i, q |
    \end{aligned}
\end{equation}

Consider the $G$-symmetric state 
\begin{equation}
    \begin{aligned}
        |\Psi\rangle = \frac{1}{|G|^L} \sum_{\{n_i, q_i\}} f(\{n_i, q_i\}) | \{n_i, q_i\}\rangle.
    \end{aligned}
\end{equation}
that is invariant under the action of $\prod_i L^{(n,q)}_i$. After gauging, the state is mapped to $|\Gamma\rangle = \Gamma (|\Psi\rangle)$, the emergent $G$ symmetry can be checked as follows.
\begin{equation}
    \begin{aligned}
        \prod_i L^{n}_i |\Gamma\rangle &= \frac{1}{|G|^{L}} \sum_{\{m_i, n_i, q_i\}} \prod_s \chi(m_s, n_{s-1}^{-1} n_s) f(\{n_i, q_i\}) |\{n m_i, q_i\}\rangle \\
        &= \frac{1}{|G|^{L}} \sum_{\{m_i, n_i, q_i\}} \prod_s \chi(n^{-1} m_s, n_{s-1}^{-1} n_s) f(\{n_i, q_i\})|\{m_i, q_i\}\rangle \\
        &= \frac{1}{|G|^{L}} \sum_{\{m_i, n_i, q_1\}} \left(\prod_{s'} \chi(n^{-1}, n_{s'-1}^{-1} n_{s'}) \right)\prod_s \chi(m_s, n_{s-1}^{-1} n_s)f(\{n_i, q_i\}) |\{m_i, q_i\}\rangle \\
        &= \frac{1}{|G|^{L}} \sum_{\{m_i, n_i, q_i\}} \prod_s \chi(m_s, n_{s-1}^{-1} n_s)f(\{n_i, q_i\}) |\{m_i, q_i\}\rangle \\
        &= |\Gamma\rangle,
    \end{aligned}
\end{equation}
From line 3 to line 4 we use the fact that $\prod_i \chi(n^{-1}, n_{i-1}^{-1} n_i) = \chi(n^{-1},1) = 1$. We also have,
\begin{equation}
    \begin{aligned}
         \prod_i L^{q}_i |\Gamma\rangle &= \frac{1}{|G|^{L}} \sum_{\{m_i, n_i, q_i\}} \prod_s \chi(m_s, n_{s-1}^{-1} n_s) f(\{n_i, q_i\})|\{\prescript{q}{}{m_i}, q q_i\}\rangle \\
         &= \frac{1}{|G|^{L}} \sum_{\{m_i, n_i, q_i\}} \prod_s \chi(\sigma^{q^{-1}}({m_s}), n_{s-1}^{-1} n_s) f(\{n_i, q_i\})|m_i, q q_i\}\rangle \\
         &= \frac{1}{|G|^{L}} \sum_{\{m_i, n_i, q_i\}} \prod_s \chi(m_s, \sigma^{q}{(n_{s-1}^{-1} n_s)}) f(\{n_i, q_i\})|\{m_i, q q_i\}\rangle \\
         &= \Gamma \left( \prod_i L^{q}_i| \Psi\rangle\right) \\
         &= |\Gamma\rangle.
    \end{aligned}
\end{equation}
From the line 2 to line 3, we apply the identity in Eq.~\eqref{eq:bicharacter_identities}. From line 4 to line 5, we use the fact that $|\Psi\rangle$ is symmetric under the $\prod_i L^{q}_i$ operator. Therefore, after gauging the $N$ symmetry, the resulting state exhibits an emergent $G$ symmetry, and the gauging map defined above realizes a self-duality.

\subsection{$G = N \rtimes Q$-quantum double model}

When $G = N \rtimes Q$, the quantum double Hamiltonian defined in Eq.~\eqref{eq:QD_terms} can be written as follows. On each $0$-triangle, we define the following terms,

\raisebox{-0.5\height}{
\begin{minipage}{\textwidth}
\centering
$B^{(0)}_n$ = \begin{tikzpicture}[scale=1, baseline=(current bounding box.center)]
\tikzset{
  hexbond/.style={line width=0.6pt},
  vdot/.style={circle, fill=black, inner sep=1.2pt}
}

\coordinate (A_1) at ( 0, 2*0.8660254);
\coordinate (A_2) at ( -1, 0);
\coordinate (A_3) at ( 1, 0);
\draw[hexbond] (A_1)--(A_2)--(A_3)--cycle;

\coordinate (A_4) at ( -2, 2*0.8660254);
\coordinate (A_5) at ( 0, -2*0.8660254);
\coordinate (A_6) at ( 2, 2*0.8660254);
\draw[hexbond] (A_4)--(A_5)--(A_6)--cycle;

\node at ($(A_4)+(0,-0.3)$) {\footnotesize $1$};
\node at ($(A_1)+(0,-0.3)$) {\footnotesize $2$};
\node at ($(A_6)+(0,-0.3)$) {\footnotesize $3$};
\node at ($(A_2)+(0,-0.3)$) {\footnotesize $6$};
\node at ($(A_3)+(0,-0.3)$) {\footnotesize $4$};
\node at ($(A_5)+(0,0.3)$) {\footnotesize $5$};

\node at ($(A_4)+(0,0.3)$) {$Z_1^{\widehat{\sigma^{q_2 \bar{q}_3}(n) \dagger}}$};
\node at ($(A_1)+(0,0.3)$) {$Z^{\widehat{\sigma^{q_2 \bar{q}_3}(n)}}_2$};
\node at ($(A_6)+(0,0.3)$) {$Z_3^{\hat{n} \dagger}$};
\node at ($(A_2)+(-0.8,0)$) {$Z_6^{\widehat{\sigma^{q_5 \bar{q}_4}(n)}}$};
\node at ($(A_3)+(0.5,0)$) {$Z_4^{\hat{n}}$};
\node at ($(A_5)+(0,-0.3)$) {$Z_5^{\widehat{\sigma^{q_5 \bar{q}_4}(n) \dagger}}$};

\node[myRed] at (0, 2*0.8660254/3) {$0$}; 
\end{tikzpicture}, \quad $B^{(0)}_q = \delta_{\prod_i q_i, 1}|\{q_i\}\rangle \langle \{q_i\}|$.
\end{minipage}
}

These are the flux-free conditions, and we use subscripts $n$ and $q$ to distinguish operators acting on the $n$ and $q$ degrees of freedom. On each 1-triangle and 2-triangle, we define the following terms.

\raisebox{-0.5\height}{
\begin{minipage}{0.95\textwidth}
\centering
$A^{(1)}_{n,q}=$\begin{tikzpicture}[scale=1, baseline=(current bounding box.center)]
\tikzset{
  hexbond/.style={line width=0.6pt},
  vdot/.style={circle, fill=black, inner sep=1.2pt}
}

\coordinate (A_1) at ( 0, 2*0.8660254);
\coordinate (A_2) at ( -1, 0);
\coordinate (A_3) at ( 1, 0);
\draw[hexbond] (A_1)--(A_2)--(A_3)--cycle;

\node at ($(A_1)+(0,0.2)$) {$L^{nq}$};
\node at ($(A_2)+(-0.4,0)$) {$L^{nq}$};
\node at ($(A_3)+(0.4,0)$) {$L^{nq}$};
\node[myGreen] at (0, 2*0.8660254/3) {$1$};
\end{tikzpicture},
\quad
$A^{(2)}_{n,q}=$\begin{tikzpicture}[scale=1, baseline=(current bounding box.center)]
\tikzset{
  hexbond/.style={line width=0.6pt},
  vdot/.style={circle, fill=black, inner sep=1.2pt}
}

\coordinate (A_1) at ( 0, 2*0.8660254);
\coordinate (A_2) at ( -1, 0);
\coordinate (A_3) at ( 1, 0);
\draw[hexbond] (A_1)--(A_2)--(A_3)--cycle;

\node at ($(A_1)+(0,0.2)$) {$R^{nq}$};
\node at ($(A_2)+(-0.4,0)$) {$R^{nq}$};
\node at ($(A_3)+(0.4,0)$) {$R^{nq}$};
\node[myBlue] at (0, 2*0.8660254/3) {$2$};
\end{tikzpicture}.
\end{minipage}
}

The ground state of a $G$-quantum double model is a simultaneous $+1$-eigenstate of the above terms. To relate these terms with the original Hamiltonian in Ref.~\cite{kitaev2003fault}, we can apply a lattice deformation shown in Fig.~\ref{fig:figure_1}(b). It is straightforward to see the above Hamiltonian defines the $\mathcal{D}(G)$-quantum double model on a hexagonal lattice.

\subsection{Operator maps}

In this subsection, we show how operators transform under the gauging map. Consider the local $L^{n}_s$ operator. Before gauging, we have
\begin{equation}
    \begin{aligned}
        L^{n}_s | \cdots, n_{s-1}, q_{s-1}, n_s, q_s, n_{s+1}, q_{s+1}, \cdots\rangle = | \cdots, n_{s-1}, q_{s-1}, n n_s, q_s, n_{s+1}, q_{s+1}, \cdots\rangle.
    \end{aligned}
\end{equation}
After the gauging map Eq.~\eqref{eq:general_gauging}, the above state becomes,
\begin{equation}
    \begin{aligned}
        &\quad\ \chi(m_s, n) \chi(m_{s+1}, n^{-1})\prod_{i} \chi(m_i, n_{i-1}^{-1} n_i) | \cdots, m_{s-1}, q_{s-1}, m_s, q_s, m_{s+1}, q_{s+1}, \cdots\rangle \\
        &=Z^{\hat{n}}_{s} Z^{\hat{n} \dagger}_{s+1} \prod_{i} \chi(m_i, n_{i-1}^{-1} n_i) | \cdots, m_{s-1}, q_{s-1}, m_s, q_s, m_{s+1}, q_{s+1}, \cdots\rangle.
    \end{aligned}
\end{equation}

Consider the local operator $R_{i}^{n}$. Before gauging, we have,
\begin{equation}
    \begin{aligned}
         R^{n}_s | \cdots, n_{s-1}, q_{s-1}, n_s, q_s, n_{s+1}, q_{s+1}, \cdots\rangle = | \cdots, n_{s-1}, q_{s-1}, \sigma^{q_s}({n}) n_s, q_s, n_{s+1}, q_{s+1}, \cdots\rangle.
    \end{aligned}
\end{equation}
After gauging, the above state becomes,
\begin{equation}
    \begin{aligned}
         &\quad\ \chi(m_s, \sigma^{q_s}({n})) \chi(m_{s+1}, \sigma^{q_{s}}({n^{-1}}))\prod_{i} \chi(m_i, n_{i-1}^{-1} n_i) | \cdots, m_{s-1}, q_{s-1}, m_s, q_s, m_{s+1}, q_{s+1}, \cdots\rangle \\
        &=Z^{\widehat{\sigma^{q_s}(n)}}_{s} Z^{\widehat{\sigma^{q_{s}}(n) \dagger}}_{s+1} \prod_{i} \chi(m_i, n_{i-1}^{-1} n_i) | \cdots, m_{s-1}, q_{s-1}, m_s, q_s, m_{s+1}, q_{s+1}, \cdots\rangle.
    \end{aligned}
\end{equation}

It's also straightforward to check $\prod_i L^{q}$, $T_i^q$, and $R^{q}_i$ are invariant under the gauging map. The maps of operators under gauging are summarized below,
\begin{equation}
    \begin{aligned}
    &L^{n}_i \to Z^{\hat{n}}_{i} Z^{\hat{n} \dagger}_{i+1} \to  L^{n}_{i+1}, \quad R^{n}_s \to Z^{\widehat{\sigma^{q_s}(n)}}_{s} Z^{\widehat{\sigma^{q_{s}}(n) \dagger}}_{s+1} \to L_{s+1}^{\sigma^{q_s}({n})}, \quad
        \prod_i L^{q} \to \prod_i L^{q}, \quad T^{q}_{i} \to T^{q}_{i},\quad R^{q} \to R^{q}.
    \end{aligned}
\end{equation}

\subsection{Hamiltonian maps}

In this subsection, we show how the Hamiltonian terms are mapped back to themselves after three steps of gauging. Schematically, the process can be summarized as follows,
\begin{align*}
    H^{(0)} \xrightarrow[]{CC^{(1)} \circ \Gamma_1}
    H^{(1)} \xrightarrow[]{CC^{(0)} \circ \Gamma_0}
    H^{(2)} \xrightarrow[]{CC^{(2)} \circ \Gamma_2}
    H^{(0)} .
\end{align*}
As discussed in the main text, after each step of gauging one must apply a circuit of controlled-conjugation gates to restore the symmetry operators so that the subsequent gauging step can be carried out. Each local controlled-conjugation gate is defined by
\begin{equation}
    \begin{aligned}
        CC_{i,j} \ket{n_i, q_i, n_j, q_j}
        &= \ket{n_i, q_i, \sigma^{q_i}({n_j}), q_j}, \quad CC_{i,j}^{\dagger} \ket{n_i, q_i, n_j, q_j}
        &= \ket{n_i, q_i, \sigma^{\bar{q}_i}({n_j}), q_j}.
    \end{aligned}
\end{equation}
The global controlled-conjugation circuits $CC^{(1)}$ and $CC^{(0)}$ are illustrated in Fig.~\ref{fig:figure_1}(c).

After applying gauging on all the 1-triangles, and followed by applying the global controlled-conjugation gate $CC^{(1)}$, the local Hamiltonian terms become:

\raisebox{-0.5\height}{
\begin{minipage}{\textwidth}
\centering
\begin{tikzpicture}[scale=1, baseline=(current bounding box.center)]
\tikzset{
  hexbond/.style={line width=0.6pt},
  vdot/.style={circle, fill=black, inner sep=1.2pt}
}

\coordinate (A_1) at ( 0, 2*0.8660254);
\coordinate (A_2) at ( -1, 0);
\coordinate (A_3) at ( 1, 0);
\draw[hexbond] (A_1)--(A_2)--(A_3)--cycle;

\node at ($(A_1)+(0,0.3)$) {$L_1^{n}$};
\node at ($(A_3)+(0.5,0)$) {$L_2^{n}$};
\node at ($(A_2)+(-0.5,0)$) {$L_3^{n}$};

\node[myRed] at (0, 2*0.8660254/3) {$0$}; 
\end{tikzpicture}, 
\begin{tikzpicture}[scale=1, baseline=(current bounding box.center)]
\tikzset{
  hexbond/.style={line width=0.6pt},
  vdot/.style={circle, fill=black, inner sep=1.2pt}
}

\coordinate (A_1) at ( 0, 2*0.8660254);
\coordinate (A_2) at ( -1, 0);
\coordinate (A_3) at ( 1, 0);
\draw[hexbond] (A_1)--(A_2)--(A_3)--cycle;

\coordinate (A_4) at ( -2, 2*0.8660254);
\coordinate (A_5) at ( -3, 0);
\draw[hexbond] (A_4)--(A_2)--(A_5)--cycle;

\node at ($(A_1)+(0,0.2)$) {$X^q$};
\node at ($(A_2)+(0,-0.3)$) {$C^qX^q$};
\node at ($(A_3)+(0,-0.3)$) {$X^q$};
\node at ($(A_4)+(0,0.2)$) {$C^q$};
\node at ($(A_5)+(0,-0.3)$) {$C^q$};
\node[myGreen] at (0, 2*0.8660254/3) {$1$};
\end{tikzpicture},
\begin{tikzpicture}[scale=1, baseline=(current bounding box.center)]
\tikzset{
  hexbond/.style={line width=0.6pt},
  vdot/.style={circle, fill=black, inner sep=1.2pt}
}

\coordinate (A_1) at ( 0, 2*0.8660254);
\coordinate (A_2) at ( -1, 0);
\coordinate (A_3) at ( 1, 0);
\draw[hexbond] (A_1)--(A_2)--(A_3)--cycle;

\coordinate (A_4) at ( 2, 2*0.8660254);
\coordinate (A_5) at ( 3, 0);
\draw[hexbond] (A_4)--(A_3)--(A_5)--cycle;

\node at ($(A_1)+(0,0.2)$) {$L_1^{\sigma^{q_2 \bar{q}_4}({n})}$};
\node at ($(A_2)+(0,-0.3)$) {$L_3^{n}$};
\node at ($(A_3)+(0,-0.3)$) {$L_4^{\sigma^{q_5 \bar{q}_4}({n})}$};

\node at ($(A_1)+(0,-0.3)$) {\footnotesize $1$};
\node at ($(A_4)+(0,-0.3)$) {\footnotesize $2$};
\node at ($(A_2)+(0,0.3)$) {\footnotesize $3$};
\node at ($(A_3)+(0,0.3)$) {\footnotesize $4$};
\node at ($(A_5)+(0,0.3)$) {\footnotesize $5$};

\node[myGreen] at (0, 2*0.8660254/3) {$1$};
\end{tikzpicture},\\
\begin{tikzpicture}[scale=1, baseline=(current bounding box.center)]
\tikzset{
  hexbond/.style={line width=0.6pt},
  vdot/.style={circle, fill=black, inner sep=1.2pt}
}

\coordinate (A_1) at ( 0, 2*0.8660254);
\coordinate (A_2) at ( -1, 0);
\coordinate (A_3) at ( 1, 0);
\draw[hexbond] (A_1)--(A_2)--(A_3)--cycle;

\node at ($(A_1)+(0,0.2)$) {$R^{q}$};
\node at ($(A_2)+(-0.5,0)$) {$R^{q}$};
\node at ($(A_3)+(0.5,0)$) {$R^{q}$};
\node[myBlue] at (0, 2*0.8660254/3) {$2$};
\end{tikzpicture},
\begin{tikzpicture}[scale=1, baseline=(current bounding box.center)]
\tikzset{
  hexbond/.style={line width=0.6pt},
  vdot/.style={circle, fill=black, inner sep=1.2pt}
}

\coordinate (A_1) at ( 0, 2*0.8660254);
\coordinate (A_2) at ( -1, 0);
\coordinate (A_3) at ( 1, 0);
\draw[hexbond] (A_1)--(A_2)--(A_3)--cycle;

\coordinate (A_4) at ( -2, 2*0.8660254);
\coordinate (A_5) at ( 0, -2*0.8660254);
\coordinate (A_6) at ( 2, 2*0.8660254);
\draw[hexbond] (A_4)--(A_5)--(A_6)--cycle;

\coordinate (A_7) at ( 4, 2*0.8660254);
\coordinate (A_8) at ( 3, 0);
\coordinate (A_9) at ( 2, -2*0.8660254);
\draw[hexbond] (A_6)--(A_7)--(A_8)--cycle;
\draw[hexbond] (A_3)--(A_8)--(A_9)--cycle;
\draw[hexbond] (A_5)--(A_9);

\node at ($(A_4)+(0,-0.3)$) {\footnotesize $1$};
\node at ($(A_1)+(0,-0.3)$) {\footnotesize $2$};
\node at ($(A_6)+(0,-0.3)$) {\footnotesize $3$};
\node at ($(A_2)+(0,-0.3)$) {\footnotesize $6$};
\node at ($(A_3)+(0,-0.3)$) {\footnotesize $4$};
\node at ($(A_5)+(0,0.3)$) {\footnotesize $5$};
\node at ($(A_7)+(0,-0.3)$) {\footnotesize $7$};
\node at ($(A_8)+(0,-0.3)$) {\footnotesize $8$};
\node at ($(A_9)+(0,0.3)$) {\footnotesize $9$};

\node at ($(A_4)+(0,0.3)$) {$Z_1^{\widehat{\sigma^{q_2}(n) \dagger}}$};
\node at ($(A_1)+(0,0.3)$) {$Z^{\widehat{\sigma^{q_2}(n)}}_2$};
\node at ($(A_6)+(0,0.3)$) {$Z_3^{\widehat{\sigma^{q_8 \bar{q}_9 q_4}(n) \dagger}}$};
\node at ($(A_3)+(1,0.3)$) {$Z_4^{\widehat{\sigma^{q_8 \bar{q}_9 q_4}(n)}}$};
\node at ($(A_5)+(0,-0.3)$) {$Z_5^{\widehat{\sigma^{q_4}(n) \dagger}}$};
\node at ($(A_2)+(-0.8,0)$) {$Z_6^{\widehat{\sigma^{q_4}(n)}}$};

\node[myBlue] at (0, 2*0.8660254/3) {$2$}; 
\end{tikzpicture},
\end{minipage}
}
where $C^q |n_i ,q_i\rangle = |\sigma^q(n_i), q_i \rangle$, and $X^q |n_i, q_i \rangle = |n_i, q q_i\rangle$.

Then, we apply the gauging map on all the 0-triangles, and followed by applying the global controlled-conjugation gate $CC^{(0)}$, the local Hamiltonian terms become:

\raisebox{-0.5\height}{
\begin{minipage}{0.95\textwidth}
\centering
\begin{tikzpicture}[scale=1, baseline=(current bounding box.center)]
\tikzset{
  hexbond/.style={line width=0.6pt},
  vdot/.style={circle, fill=black, inner sep=1.2pt}
}

\coordinate (A_1) at ( 0, 2*0.8660254);
\coordinate (A_2) at ( -1, 0);
\coordinate (A_3) at ( 1, 0);
\draw[hexbond] (A_1)--(A_2)--(A_3)--cycle;

\coordinate (A_4) at ( -2, 2*0.8660254);
\coordinate (A_5) at ( 0, -2*0.8660254);
\coordinate (A_6) at ( 2, 2*0.8660254);
\draw[hexbond] (A_4)--(A_5)--(A_6)--cycle;

\node at ($(A_4)+(0,-0.3)$) {\footnotesize $1$};
\node at ($(A_1)+(0,-0.3)$) {\footnotesize $2$};
\node at ($(A_6)+(0,-0.3)$) {\footnotesize $3$};
\node at ($(A_2)+(0,-0.3)$) {\footnotesize $6$};
\node at ($(A_3)+(0,-0.3)$) {\footnotesize $4$};
\node at ($(A_5)+(0,0.3)$) {\footnotesize $5$};

\node at ($(A_1)+(0,0.3)$) {$L_2^{\sigma^{\bar{q}_3 q_4}({n})}$};
\node at ($(A_3)+(0.5,0)$) {$L_4^{n}$};
\node at ($(A_2)+(-0.8,0)$) {$L_6^{\sigma^{\bar{q}_6 q_5}({n})}$};

\node[myRed] at (0, 2*0.8660254/3) {$0$}; 
\end{tikzpicture},
\begin{tikzpicture}[scale=1, baseline=(current bounding box.center)]
\tikzset{
  hexbond/.style={line width=0.6pt},
  vdot/.style={circle, fill=black, inner sep=1.2pt}
}

\coordinate (A_1) at ( 0, 2*0.8660254);
\coordinate (A_2) at ( -1, 0);
\coordinate (A_3) at ( 1, 0);
\draw[hexbond] (A_1)--(A_2)--(A_3)--cycle;

\node at ($(A_1)+(0,0.2)$) {$X^q$};
\node at ($(A_2)+(-0.4,0)$) {$X^q$};
\node at ($(A_3)+(0.4,0)$) {$X^q$};
\node[myGreen] at (0, 2*0.8660254/3) {$1$};
\end{tikzpicture}, \begin{tikzpicture}[scale=1, baseline=(current bounding box.center)]
\tikzset{
  hexbond/.style={line width=0.6pt},
  vdot/.style={circle, fill=black, inner sep=1.2pt}
}

\coordinate (A_1) at ( 0, 2*0.8660254);
\coordinate (A_2) at ( -1, 0);
\coordinate (A_3) at ( 1, 0);
\draw[hexbond] (A_1)--(A_2)--(A_3)--cycle;

\node at ($(A_1)+(0,0.2)$) {$L^{n}$};
\node at ($(A_2)+(-0.4,0)$) {$L^{n}$};
\node at ($(A_3)+(0.4,0)$) {$L^{n}$};
\node[myBlue] at (0, 2*0.8660254/3) {$2$};
\end{tikzpicture}, \\
\begin{tikzpicture}[scale=1, baseline=(current bounding box.center)]
\tikzset{
  hexbond/.style={line width=0.6pt},
  vdot/.style={circle, fill=black, inner sep=1.2pt}
}

\coordinate (A_1) at ( 0, 2*0.8660254);
\coordinate (A_2) at ( -1, 0);
\coordinate (A_3) at ( 1, 0);
\draw[hexbond] (A_1)--(A_2)--(A_3)--cycle;

\node at ($(A_1)+(0,0.2)$) {$C^{\bar{q}}R^{q}$};
\node at ($(A_2)+(-0.4,-0.2)$) {$C^{\bar{q}}R^{q}$};
\node at ($(A_3)+(0.4,-0.2)$) {$C^{\bar{q}}R^{q}$};
\node[myBlue] at (0, 2*0.8660254/3) {$2$};
\end{tikzpicture},
\begin{tikzpicture}[scale=1, baseline=(current bounding box.center)]
\tikzset{
  hexbond/.style={line width=0.6pt},
  vdot/.style={circle, fill=black, inner sep=1.2pt}
}

\coordinate (A_1) at ( 0, 2*0.8660254);
\coordinate (A_2) at ( -1, 0);
\coordinate (A_3) at ( 1, 0);
\draw[hexbond] (A_1)--(A_2)--(A_3)--cycle;

\coordinate (A_4) at ( -2, 2*0.8660254);
\coordinate (A_5) at ( 0, -2*0.8660254);
\coordinate (A_6) at ( 2, 2*0.8660254);
\draw[hexbond] (A_4)--(A_5)--(A_6)--cycle;

\node at ($(A_4)+(0,-0.3)$) {\footnotesize $1$};
\node at ($(A_1)+(0,-0.3)$) {\footnotesize $2$};
\node at ($(A_6)+(0,-0.3)$) {\footnotesize $3$};
\node at ($(A_2)+(0,-0.3)$) {\footnotesize $6$};
\node at ($(A_3)+(0,-0.3)$) {\footnotesize $4$};
\node at ($(A_5)+(0,0.3)$) {\footnotesize $5$};

\node at ($(A_4)+(0,0.3)$) {$Z_1^{\widehat{\sigma^{\bar{q}_2}(n) \dagger}}$};
\node at ($(A_1)+(0,0.3)$) {$Z^{\widehat{\sigma^{\bar{q}_2}(n)}}_2$};
\node at ($(A_6)+(0,0.3)$) {$Z_3^{\widehat{\sigma^{\bar{q}_4}(n) \dagger}}$};
\node at ($(A_3)+(0.6,0)$) {$Z_4^{\widehat{\sigma^{\bar{q}_4}(n)}}$};
\node at ($(A_5)+(0,-0.3)$) {$Z_5^{\widehat{\sigma^{\bar{q}_6}(n) \dagger}}$};
\node at ($(A_2)+(-0.8,0)$) {$Z_6^{\widehat{\sigma^{\bar{q}_6}(n)}}$};

\node[myGreen] at (0, 2*0.8660254/3) {$1$}; 
\end{tikzpicture}.
\end{minipage}
}

Finally, we apply the gauging map on all the $2$-triangles, followed by applying the global on-site controlled-conjugation gate $CC^{(2)} = \prod_i CC_{i,i}$. The set generated by local Hamiltonian terms then maps back to its original form.

\subsection{Ribbon operator maps} \label{sec:general_ribbon_gauging}

In this subsection, we derive the mappings between different logical states under the sequence of finite-depth local unitary gates discussed in the previous subsections. Consider the deformed lattice with periodic boundary condition in Fig.~\ref{fig:deformed_lattice}, where we relabel the qudits for convenience. The orientation rule of the hexagonal lattice is defined as follows: for vertices inside the blue triangles (2-triangles), all connected edges point toward the vertex, while for vertices inside the green triangles (1-triangles), all connected edges point outward from the vertex. 

In general, the ribbon operator $F^{h,g}_{\xi}$ on the hexagonal lattice is defined as follows~\cite{kitaev2003fault},
\begin{center}
\begin{tikzpicture}[scale=1.2]
\tikzset{
  hexbond/.style={line width=0.6pt},
  vdot/.style={circle, fill=black, inner sep=1.2pt}
}



\coordinate (A_U)  at ( 0.0000000,  1.0000000);
\coordinate (A_UR) at ( 0.8660254,  0.5000000);
\coordinate (A_LR) at ( 0.8660254, -0.5000000);
\coordinate (A_D)  at ( 0.0000000, -1.0000000);
\coordinate (A_LL) at (-0.8660254, -0.5000000);
\coordinate (A_UL) at (-0.8660254,  0.5000000);
\draw[hexbond] (A_U)--(A_UR)--(A_LR)--(A_D)--(A_LL)--(A_UL)--cycle;

\coordinate (B_U)  at ( 1.7320508+0.0000000,  0.0000000+1.0000000);
\coordinate (B_UR) at ( 1.7320508+0.8660254,  0.0000000+0.5000000);
\coordinate (B_LR) at ( 1.7320508+0.8660254,  0.0000000-0.5000000);
\coordinate (B_D)  at ( 1.7320508+0.0000000,  0.0000000-1.0000000);
\coordinate (B_LL) at ( 1.7320508-0.8660254,  0.0000000-0.5000000);
\coordinate (B_UL) at ( 1.7320508-0.8660254,  0.0000000+0.5000000);
\draw[hexbond] (B_U)--(B_UR)--(B_LR)--(B_D)--(B_LL)--(B_UL)--cycle;

\coordinate (C_U)  at ( 3.4641016+0.0000000,  0.0000000+1.0000000);
\coordinate (C_UR) at ( 3.4641016+0.8660254,  0.0000000+0.5000000);
\coordinate (C_LR) at ( 3.4641016+0.8660254,  0.0000000-0.5000000);
\coordinate (C_D)  at ( 3.4641016+0.0000000,  0.0000000-1.0000000);
\coordinate (C_LL) at ( 3.4641016-0.8660254,  0.0000000-0.5000000);
\coordinate (C_UL) at ( 3.4641016-0.8660254,  0.0000000+0.5000000);
\draw[hexbond] (C_U)--(C_UR)--(C_LR)--(C_D)--(C_LL)--(C_UL)--cycle;

\coordinate (D_U)  at ( 5.1961524+0.0000000,  0.0000000+1.0000000);
\coordinate (D_UR) at ( 5.1961524+0.8660254,  0.0000000+0.5000000);
\coordinate (D_LR) at ( 5.1961524+0.8660254,  0.0000000-0.5000000);
\coordinate (D_D)  at ( 5.1961524+0.0000000,  0.0000000-1.0000000);
\coordinate (D_LL) at ( 5.1961524-0.8660254,  0.0000000-0.5000000);
\coordinate (D_UL) at ( 5.1961524-0.8660254,  0.0000000+0.5000000);
\draw[hexbond] (D_U)--(D_UR)--(D_LR)--(D_D)--(D_LL)--(D_UL)--cycle;

\coordinate (E_U)  at ( 6.9282032+0.0000000,  0.0000000+1.0000000);
\coordinate (E_UR) at ( 6.9282032+0.8660254,  0.0000000+0.5000000);
\coordinate (E_LR) at ( 6.9282032+0.8660254,  0.0000000-0.5000000);
\coordinate (E_D)  at ( 6.9282032+0.0000000,  0.0000000-1.0000000);
\coordinate (E_LL) at ( 6.9282032-0.8660254,  0.0000000-0.5000000);
\coordinate (E_UL) at ( 6.9282032-0.8660254,  0.0000000+0.5000000);
\draw[hexbond] (E_U)--(E_UR)--(E_LR)--(E_D)--(E_LL)--(E_UL)--cycle;


\coordinate (F_U)  at ( 0.8660254+0.0000000,  1.5+1.0000000);
\coordinate (F_UR) at ( 0.8660254+0.8660254,  1.5+0.5000000);
\coordinate (F_LR) at ( 0.8660254+0.8660254,  1.5-0.5000000);
\coordinate (F_D)  at ( 0.8660254+0.0000000,  1.5-1.0000000);
\coordinate (F_LL) at ( 0.8660254-0.8660254,  1.5-0.5000000);
\coordinate (F_UL) at ( 0.8660254-0.8660254,  1.5+0.5000000);
\draw[hexbond] (F_U)--(F_UR)--(F_LR)--(F_D)--(F_LL)--(F_UL)--cycle;

\coordinate (G_U)  at ( 2.5980762+0.0000000,  1.5+1.0000000);
\coordinate (G_UR) at ( 2.5980762+0.8660254,  1.5+0.5000000);
\coordinate (G_LR) at ( 2.5980762+0.8660254,  1.5-0.5000000);
\coordinate (G_D)  at ( 2.5980762+0.0000000,  1.5-1.0000000);
\coordinate (G_LL) at ( 2.5980762-0.8660254,  1.5-0.5000000);
\coordinate (G_UL) at ( 2.5980762-0.8660254,  1.5+0.5000000);
\draw[hexbond] (G_U)--(G_UR)--(G_LR)--(G_D)--(G_LL)--(G_UL)--cycle;

\coordinate (H_U)  at ( 4.3301270+0.0000000,  1.5+1.0000000);
\coordinate (H_UR) at ( 4.3301270+0.8660254,  1.5+0.5000000);
\coordinate (H_LR) at ( 4.3301270+0.8660254,  1.5-0.5000000);
\coordinate (H_D)  at ( 4.3301270+0.0000000,  1.5-1.0000000);
\coordinate (H_LL) at ( 4.3301270-0.8660254,  1.5-0.5000000);
\coordinate (H_UL) at ( 4.3301270-0.8660254,  1.5+0.5000000);
\draw[hexbond] (H_U)--(H_UR)--(H_LR)--(H_D)--(H_LL)--(H_UL)--cycle;

\coordinate (I_U)  at ( 6.0621778+0.0000000,  1.5+1.0000000);
\coordinate (I_UR) at ( 6.0621778+0.8660254,  1.5+0.5000000);
\coordinate (I_LR) at ( 6.0621778+0.8660254,  1.5-0.5000000);
\coordinate (I_D)  at ( 6.0621778+0.0000000,  1.5-1.0000000);
\coordinate (I_LL) at ( 6.0621778-0.8660254,  1.5-0.5000000);
\coordinate (I_UL) at ( 6.0621778-0.8660254,  1.5+0.5000000);
\draw[hexbond] (I_U)--(I_UR)--(I_LR)--(I_D)--(I_LL)--(I_UL)--cycle;

\coordinate (J_U)  at ( 7.7942286+0.0000000,  1.5+1.0000000);
\coordinate (J_UR) at ( 7.7942286+0.8660254,  1.5+0.5000000);
\coordinate (J_LR) at ( 7.7942286+0.8660254,  1.5-0.5000000);
\coordinate (J_D)  at ( 7.7942286+0.0000000,  1.5-1.0000000);
\coordinate (J_LL) at ( 7.7942286-0.8660254,  1.5-0.5000000);
\coordinate (J_UL) at ( 7.7942286-0.8660254,  1.5+0.5000000);
\draw[hexbond] (J_U)--(J_UR)--(J_LR)--(J_D)--(J_LL)--(J_UL)--cycle;

\draw[gray] ($(A_UL)!0.5!(A_LL)$)--($(E_UR)!0.5!(E_LR)$);
\fill[gray,nearly transparent] ($(A_UL)!0.5!(A_LL)$)--($(E_UR)!0.5!(E_LR)$)--(E_UR)--(E_U)--(D_UR)--(D_U)--(C_UR)--(C_U)--(B_UR)--(B_U)--(A_UR)--(A_U)--(A_UL)--($(A_UL)!0.5!(A_LL)$);

\draw[-<] (A_U)--($(A_U)!0.5!(A_UL)$);
\draw[-<] (A_U)--($(A_U)!0.5!(A_UR)$);
\draw[-<] (B_U)--($(B_U)!0.5!(B_UL)$);
\draw[-<] (B_U)--($(B_U)!0.5!(B_UR)$);
\draw[-<] (C_U)--($(C_U)!0.5!(C_UL)$);
\draw[-<] (C_U)--($(C_U)!0.5!(C_UR)$);
\draw[-<] (D_U)--($(D_U)!0.5!(D_UL)$);
\draw[-<] (D_U)--($(D_U)!0.5!(D_UR)$);
\draw[-<] (E_U)--($(E_U)!0.5!(E_UL)$);
\draw[-<] (E_U)--($(E_U)!0.5!(E_UR)$);

\draw[-<] (A_LL)--($(A_LL)!0.5!(A_UL)$);
\draw[-<] (B_LL)--($(B_LL)!0.5!(B_UL)$);
\draw[-<] (C_LL)--($(C_LL)!0.5!(C_UL)$);
\draw[-<] (D_LL)--($(D_LL)!0.5!(D_UL)$);
\draw[-<] (E_LL)--($(E_LL)!0.5!(E_UL)$);
\draw[-<] (E_LR)--($(E_LR)!0.5!(E_UR)$);

\draw[dotted] (A_UL)--($(A_U)!0.5!(A_D)$);
\draw[dotted] (A_U)--($(A_U)!0.5!(A_D)$);
\draw[dotted] (A_UR)--($(A_U)!0.5!(A_D)$);
\draw[dotted] (B_UL)--($(B_U)!0.5!(B_D)$);
\draw[dotted] (B_U)--($(B_U)!0.5!(B_D)$);
\draw[dotted] (B_UR)--($(B_U)!0.5!(B_D)$);
\draw[dotted] (C_UL)--($(C_U)!0.5!(C_D)$);
\draw[dotted] (C_U)--($(C_U)!0.5!(C_D)$);
\draw[dotted] (C_UR)--($(C_U)!0.5!(C_D)$);
\draw[dotted] (D_UL)--($(D_U)!0.5!(D_D)$);
\draw[dotted] (D_U)--($(D_U)!0.5!(D_D)$);
\draw[dotted] (D_UR)--($(D_U)!0.5!(D_D)$);
\draw[dotted] (E_UL)--($(E_U)!0.5!(E_D)$);
\draw[dotted] (E_U)--($(E_U)!0.5!(E_D)$);
\draw[dotted] (E_UR)--($(E_U)!0.5!(E_D)$);

\node at ($(A_U)!0.5!(A_UL)+(0,0.3)$) {$|x_n\rangle$};
\node at ($(A_U)!0.5!(A_UR)+(0,0.3)$) {$|x_1\rangle$};
\node at ($(B_U)!0.5!(B_UL)+(0,0.3)$) {$|x_2\rangle$};
\node at ($(B_U)!0.5!(B_UR)+(0,0.3)$) {$|x_3\rangle$};
\node at ($(C_U)!0.5!(C_UL)+(0,0.3)$) {$|x_4\rangle$};
\node at ($(C_U)!0.5!(C_UR)+(0,0.3)$) {$|x_5\rangle$};
\node at ($(D_U)!0.5!(D_UL)+(0,0.3)$) {$|x_6\rangle$};
\node at ($(D_U)!0.5!(D_UR)+(0,0.3)$) {$|x_7\rangle$};
\node at ($(E_U)!0.5!(E_UL)+(0,0.3)$) {$|x_8\rangle$};
\node at ($(E_U)!0.5!(E_UR)+(0,0.3)$) {$|x_9\rangle$};

\node at ($(A_UL)!0.5!(A_LL)+(-0.3,-0.2)$) {$|y_n\rangle$};
\node at ($(B_UL)!0.5!(B_LL)+(-0.3,-0.2)$) {$|y_2\rangle$};
\node at ($(C_UL)!0.5!(C_LL)+(-0.3,-0.2)$) {$|y_4\rangle$};
\node at ($(D_UL)!0.5!(D_LL)+(-0.3,-0.2)$) {$|y_6\rangle$};
\node at ($(E_UL)!0.5!(E_LL)+(-0.3,-0.2)$) {$|y_8\rangle$};

\node at ($(A_U) + (-1.5, 0)$) {$F^{h,g}_{\xi}$};

\def\dy{-3.5}

\coordinate (A2_U)  at ($(A_U)+(0,\dy)$);
\coordinate (A2_UR) at ($(A_UR)+(0,\dy)$);
\coordinate (A2_LR) at ($(A_LR)+(0,\dy)$);
\coordinate (A2_D)  at ($(A_D)+(0,\dy)$);
\coordinate (A2_LL) at ($(A_LL)+(0,\dy)$);
\coordinate (A2_UL) at ($(A_UL)+(0,\dy)$);
\draw[hexbond] (A2_U)--(A2_UR)--(A2_LR)--(A2_D)--(A2_LL)--(A2_UL)--cycle;

\coordinate (B2_U)  at ($(B_U)+(0,\dy)$);
\coordinate (B2_UR) at ($(B_UR)+(0,\dy)$);
\coordinate (B2_LR) at ($(B_LR)+(0,\dy)$);
\coordinate (B2_D)  at ($(B_D)+(0,\dy)$);
\coordinate (B2_LL) at ($(B_LL)+(0,\dy)$);
\coordinate (B2_UL) at ($(B_UL)+(0,\dy)$);
\draw[hexbond] (B2_U)--(B2_UR)--(B2_LR)--(B2_D)--(B2_LL)--(B2_UL)--cycle;

\coordinate (C2_U)  at ($(C_U)+(0,\dy)$);
\coordinate (C2_UR) at ($(C_UR)+(0,\dy)$);
\coordinate (C2_LR) at ($(C_LR)+(0,\dy)$);
\coordinate (C2_D)  at ($(C_D)+(0,\dy)$);
\coordinate (C2_LL) at ($(C_LL)+(0,\dy)$);
\coordinate (C2_UL) at ($(C_UL)+(0,\dy)$);
\draw[hexbond] (C2_U)--(C2_UR)--(C2_LR)--(C2_D)--(C2_LL)--(C2_UL)--cycle;

\coordinate (D2_U)  at ($(D_U)+(0,\dy)$);
\coordinate (D2_UR) at ($(D_UR)+(0,\dy)$);
\coordinate (D2_LR) at ($(D_LR)+(0,\dy)$);
\coordinate (D2_D)  at ($(D_D)+(0,\dy)$);
\coordinate (D2_LL) at ($(D_LL)+(0,\dy)$);
\coordinate (D2_UL) at ($(D_UL)+(0,\dy)$);
\draw[hexbond] (D2_U)--(D2_UR)--(D2_LR)--(D2_D)--(D2_LL)--(D2_UL)--cycle;

\coordinate (E2_U)  at ($(E_U)+(0,\dy)$);
\coordinate (E2_UR) at ($(E_UR)+(0,\dy)$);
\coordinate (E2_LR) at ($(E_LR)+(0,\dy)$);
\coordinate (E2_D)  at ($(E_D)+(0,\dy)$);
\coordinate (E2_LL) at ($(E_LL)+(0,\dy)$);
\coordinate (E2_UL) at ($(E_UL)+(0,\dy)$);
\draw[hexbond] (E2_U)--(E2_UR)--(E2_LR)--(E2_D)--(E2_LL)--(E2_UL)--cycle;

\coordinate (F2_U)  at ($(F_U)+(0,\dy)$);
\coordinate (F2_UR) at ($(F_UR)+(0,\dy)$);
\coordinate (F2_LR) at ($(F_LR)+(0,\dy)$);
\coordinate (F2_D)  at ($(F_D)+(0,\dy)$);
\coordinate (F2_LL) at ($(F_LL)+(0,\dy)$);
\coordinate (F2_UL) at ($(F_UL)+(0,\dy)$);
\draw[hexbond] (F2_U)--(F2_UR)--(F2_LR)--(F2_D)--(F2_LL)--(F2_UL)--cycle;

\coordinate (G2_U)  at ($(G_U)+(0,\dy)$);
\coordinate (G2_UR) at ($(G_UR)+(0,\dy)$);
\coordinate (G2_LR) at ($(G_LR)+(0,\dy)$);
\coordinate (G2_D)  at ($(G_D)+(0,\dy)$);
\coordinate (G2_LL) at ($(G_LL)+(0,\dy)$);
\coordinate (G2_UL) at ($(G_UL)+(0,\dy)$);
\draw[hexbond] (G2_U)--(G2_UR)--(G2_LR)--(G2_D)--(G2_LL)--(G2_UL)--cycle;

\coordinate (H2_U)  at ($(H_U)+(0,\dy)$);
\coordinate (H2_UR) at ($(H_UR)+(0,\dy)$);
\coordinate (H2_LR) at ($(H_LR)+(0,\dy)$);
\coordinate (H2_D)  at ($(H_D)+(0,\dy)$);
\coordinate (H2_LL) at ($(H_LL)+(0,\dy)$);
\coordinate (H2_UL) at ($(H_UL)+(0,\dy)$);
\draw[hexbond] (H2_U)--(H2_UR)--(H2_LR)--(H2_D)--(H2_LL)--(H2_UL)--cycle;

\coordinate (I2_U)  at ($(I_U)+(0,\dy)$);
\coordinate (I2_UR) at ($(I_UR)+(0,\dy)$);
\coordinate (I2_LR) at ($(I_LR)+(0,\dy)$);
\coordinate (I2_D)  at ($(I_D)+(0,\dy)$);
\coordinate (I2_LL) at ($(I_LL)+(0,\dy)$);
\coordinate (I2_UL) at ($(I_UL)+(0,\dy)$);
\draw[hexbond] (I2_U)--(I2_UR)--(I2_LR)--(I2_D)--(I2_LL)--(I2_UL)--cycle;

\coordinate (J2_U)  at ($(J_U)+(0,\dy)$);
\coordinate (J2_UR) at ($(J_UR)+(0,\dy)$);
\coordinate (J2_LR) at ($(J_LR)+(0,\dy)$);
\coordinate (J2_D)  at ($(J_D)+(0,\dy)$);
\coordinate (J2_LL) at ($(J_LL)+(0,\dy)$);
\coordinate (J2_UL) at ($(J_UL)+(0,\dy)$);
\draw[hexbond] (J2_U)--(J2_UR)--(J2_LR)--(J2_D)--(J2_LL)--(J2_UL)--cycle;

\draw[gray] ($(A2_UL)!0.5!(A2_LL)$)--($(E2_UR)!0.5!(E2_LR)$);
\fill[gray,nearly transparent]
  ($(A2_UL)!0.5!(A2_LL)$)--($(E2_UR)!0.5!(E2_LR)$)--
  (E2_UR)--(E2_U)--(D2_UR)--(D2_U)--(C2_UR)--(C2_U)--(B2_UR)--(B2_U)--(A2_UR)--(A2_U)--(A2_UL)--($(A2_UL)!0.5!(A2_LL)$);

\draw[-<] (A2_U)--($(A2_U)!0.5!(A2_UL)$);
\draw[-<] (A2_U)--($(A2_U)!0.5!(A2_UR)$);
\draw[-<] (B2_U)--($(B2_U)!0.5!(B2_UL)$);
\draw[-<] (B2_U)--($(B2_U)!0.5!(B2_UR)$);
\draw[-<] (C2_U)--($(C2_U)!0.5!(C2_UL)$);
\draw[-<] (C2_U)--($(C2_U)!0.5!(C2_UR)$);
\draw[-<] (D2_U)--($(D2_U)!0.5!(D2_UL)$);
\draw[-<] (D2_U)--($(D2_U)!0.5!(D2_UR)$);
\draw[-<] (E2_U)--($(E2_U)!0.5!(E2_UL)$);
\draw[-<] (E2_U)--($(E2_U)!0.5!(E2_UR)$);

\draw[-<] (A2_LL)--($(A2_LL)!0.5!(A2_UL)$);
\draw[-<] (B2_LL)--($(B2_LL)!0.5!(B2_UL)$);
\draw[-<] (C2_LL)--($(C2_LL)!0.5!(C2_UL)$);
\draw[-<] (D2_LL)--($(D2_LL)!0.5!(D2_UL)$);
\draw[-<] (E2_LL)--($(E2_LL)!0.5!(E2_UL)$);
\draw[-<] (E2_LR)--($(E2_LR)!0.5!(E2_UR)$);

\draw[dotted] (A2_UL)--($(A2_U)!0.5!(A2_D)$);
\draw[dotted] (A2_U)--($(A2_U)!0.5!(A2_D)$);
\draw[dotted] (A2_UR)--($(A2_U)!0.5!(A2_D)$);

\draw[dotted] (B2_UL)--($(B2_U)!0.5!(B2_D)$);
\draw[dotted] (B2_U)--($(B2_U)!0.5!(B2_D)$);
\draw[dotted] (B2_UR)--($(B2_U)!0.5!(B2_D)$);

\draw[dotted] (C2_UL)--($(C2_U)!0.5!(C2_D)$);
\draw[dotted] (C2_U)--($(C2_U)!0.5!(C2_D)$);
\draw[dotted] (C2_UR)--($(C2_U)!0.5!(C2_D)$);

\draw[dotted] (D2_UL)--($(D2_U)!0.5!(D2_D)$);
\draw[dotted] (D2_U)--($(D2_U)!0.5!(D2_D)$);
\draw[dotted] (D2_UR)--($(D2_U)!0.5!(D2_D)$);

\draw[dotted] (E2_UL)--($(E2_U)!0.5!(E2_D)$);
\draw[dotted] (E2_U)--($(E2_U)!0.5!(E2_D)$);
\draw[dotted] (E2_UR)--($(E2_U)!0.5!(E2_D)$);

\node at ($(A2_U)!0.5!(A2_UL)+(0,0.3)$) {$|x_n\rangle$};
\node at ($(A2_U)!0.5!(A2_UR)+(0,0.3)$) {$|x_1\rangle$};
\node at ($(B2_U)!0.5!(B2_UL)+(0,0.3)$) {$|x_2\rangle$};
\node at ($(B2_U)!0.5!(B2_UR)+(0,0.3)$) {$|x_3\rangle$};
\node at ($(C2_U)!0.5!(C2_UL)+(0,0.3)$) {$|x_4\rangle$};
\node at ($(C2_U)!0.5!(C2_UR)+(0,0.3)$) {$|x_5\rangle$};
\node at ($(D2_U)!0.5!(D2_UL)+(0,0.3)$) {$|x_6\rangle$};
\node at ($(D2_U)!0.5!(D2_UR)+(0,0.3)$) {$|x_7\rangle$};
\node at ($(E2_U)!0.5!(E2_UL)+(0,0.3)$) {$|x_8\rangle$};
\node at ($(E2_U)!0.5!(E2_UR)+(0,0.3)$) {$|x_9\rangle$};

\node at ($(A2_UL)!0.5!(A2_LL)+(-0.5,-0.2)$) {$|y_n h_n^{-1}\rangle$};
\node at ($(B2_UL)!0.5!(B2_LL)+(-0.5,-0.2)$) {$|y_2 h_2^{-1}\rangle$};
\node at ($(C2_UL)!0.5!(C2_LL)+(-0.5,-0.2)$) {$|y_4 h_4^{-1}\rangle$};
\node at ($(D2_UL)!0.5!(D2_LL)+(-0.5,-0.2)$) {$|y_6 h_{6}^{-1}\rangle$};
\node at ($(E2_UL)!0.5!(E2_LL)+(-0.5,-0.2)$) {$|y_8 h_8^{-1}\rangle$};

\node at ($(A2_U) + (-3, 0)$) {$= \delta_{x_1 x_2^{-1} \cdots x_{n-1} x_{n}^{-1}, g}$};

\end{tikzpicture}
\end{center}
where $h_{2i} = (x_1 x_2^{-1} x_3 x_4^{-1} \cdots x_{2i-1})^{-1} h  (x_1 x_2^{-1} x_3 x_4^{-1} \cdots x_{2i-1})$.

Anyon ribbon operators are given by the linear combinations of the ribbon operators. For $\mathcal{D}(G)$ quantum double model, general anyonic ribbon operators can be written as follows~\cite{kitaev2003fault,bombin2008family,bravyi2022adaptive}:
\begin{align}
    F^{(C,R); (\mathbf{u},\mathbf{v})}_{\xi} := \frac{d_{\pi}}{|E(C)|} \sum_{k \in E(C)} \left(\rho_{\pi}^{-1}(k) \right)_{j,j'} F_{\xi}^{(c_i^{-1}, p_i k p_{i'}^{-1})},
\end{align}
in which $E(C)$ is the centralizer group of an element $r_C \in C$, $d_{\pi}$ is the dimension of the irreducible representation of the centralizer group. Here $\mathbf{u}= (i,j)$ and $\mathbf{v} = (i', j')$, where $i, i' \in \{1, \ldots, |C| \}$ and $j, j' \in \{1,\ldots, d_{\pi} \}$ are the indices for the elements of the conjugacy class and the matrix row/column indices respectively. $\rho_{\pi}(k)$ is the $d_{\pi}$ dimensional irreducible representation of $k \in E(C)$. Lastly, we choose $\{p_i\}_{i=1}^{|C|} \in G$ such that $c_i = p_i r_C p_i^{-1} \in C$.

Equivalently, ribbon operators can also be written in the form of generalized Pauli $X$ and Pauli $Z$ operators. Consider a ribbon operator winding around a closed loop (from qudit $1$ to qudit $n$). The pure charge anyons labeled by $([e], \pi_n)$, where $\pi_n$ is an irreducible representation of $G$, correspond to the orbit of an $N$-representation under the conjugation action of $Q$. The ribbon operator associated with such a pure charge can be written as
\begin{equation}
    \begin{aligned}
        F_{\xi}^{([e], \pi_n)} &\propto P_{\prod_i q_i = 1} \left( \sum_{q \in Q} \left(Z_n Z_1^{\dagger}\right)^{\widehat{\sigma^{q}(n)}} \left(Z_2 Z_3^{\dagger}\right)^{\widehat{\sigma^{q_2 \bar{q}_1 q}(n)}} \left(Z_4 Z_5^{\dagger}\right)^{\widehat{\sigma^{q_4 \bar{q}_3 q_2 \bar{q}_1 q}(n)}} \cdots\right). \label{eq:general_electric_ribbon}
    \end{aligned}
\end{equation}
Here $P_{\prod_i q_i = 1}$ is the projector onto the zero $Q$-flux subspace. Since this projector remains invariant throughout the entire process, we focus only on the $\mathbb{Z}_n$ operators. After performing gauging on the 1-triangles, we obtain
\begin{equation}
    \begin{aligned}
        \left(Z_n Z_1^{\dagger}\right)^{\widehat{\sigma^{q}(n)}} \left(Z_2 Z_3^{\dagger}\right)^{\widehat{\sigma^{q_2 \bar{q}_1 q}(n)}} \left(Z_4 Z_5^{\dagger}\right)^{\widehat{\sigma^{q_4 \bar{q}_3 q_2 \bar{q}_1 q}(n)}} \cdots
        \xrightarrow[]{\Gamma_1} L_1^{\sigma^{q}({n})} L_3^{\sigma^{q_2 \bar{q}_1 q}({n})} L_5^{\sigma^{q_4 \bar{q}_3 q_2 \bar{q}_1 q}({n})} \cdots.
    \end{aligned}
\end{equation}
We then apply the global controlled-conjugation gate shown in Fig.~\ref{fig:figure_1}(c), after which the operator becomes
\begin{align}
    L_1^{\sigma^{q_{2'} \bar{q}_1 q}({n})} L_3^{\sigma^{q_{4'} \bar{q}_3 q_2 \bar{q}_1 q}({n})} L_5^{\sigma^{q_{6'} \bar{q}_5 q_4 \bar{q}_3 q_2 \bar{q}_1 q}({n})} \cdots.
\end{align}

Next, we apply gauging on the 0-triangles, yielding
\begin{align}
    \left(Z_{n'}^{\dagger} Z_1\right)^{\widehat{\sigma^{q_{2'} \bar{q}_1 q}(n)}} \left(Z_{2'}^{\dagger} Z_3\right)^{\widehat{\sigma^{q_{4'} \bar{q}_3 q_2 \bar{q}_1 q}(n)}} \left(Z_{4'}^{\dagger} Z_5\right)^{\widehat{\sigma^{q_{6'} \bar{q}_5 q_4 \bar{q}_3 q_2 \bar{q}_1 q}(n)}} \cdots.
\end{align}
Applying the global controlled-conjugation gate $CC^{(0)}$, we obtain
\begin{align}
   \left(Z_{n'}\right)^{\widehat{\sigma^{\bar{q}_n}(n) \dagger}}
    \left(Z_1\right)^{\widehat{\sigma^{\bar{q}_1 q}(n)}} \left(Z_{2'}\right)^{\widehat{\sigma^{\bar{q}_1 q}(n) \dagger}} \left(Z_3\right)^{\widehat{\sigma^{\bar{q}_3 q_2 \bar{q}_1 q}(n)}} \left(Z_{4'}\right)^{\widehat{\sigma^{\bar{q}_3 q_2 \bar{q}_1 q}(n) \dagger}} \cdots
\end{align}
Rearranging the above expression gives
\begin{align}
    \left(Z_1 Z_{2'}^{\dagger}\right)^{\widehat{\sigma^{\bar{q}_1 q}(n)}} \left(Z_3 Z_{4'}^{\dagger}\right)^{\widehat{\sigma^{\bar{q}_3 q_2 \bar{q}_1 q}(n)}} \cdots
\end{align}
We then perform gauging on the 2-triangles, which maps the operator to
\begin{align}
    L^{\sigma^{\bar{q}_1 q}({n})}_{2'} L^{\sigma^{\bar{q}_3 q_2 \bar{q}_1 q}({n})}_{4'} L^{\sigma^{\bar{q}_5 q_4 \bar{q}_3 q_2 \bar{q}_1 q}({n})}_{6'} \cdots.
\end{align}

Finally, we apply a global on-site controlled-conjugation gate, resulting in
\begin{align}
     L^{\sigma^{q_{2'} \bar{q}_1 q}({n})}_{2'} L^{\sigma^{q_{4'} \bar{q}_3 q_2 \bar{q}_1 q}({n})}_{4'} L^{\sigma^{q_{6'} \bar{q}_5 q_4 \bar{q}_3 q_2 \bar{q}_1 q}({n})}_{6'} \cdots.
\end{align}
By summing over all $q \in Q$, we generate all elements in the conjugacy class of $n$. The resulting operator is therefore a magnetic ribbon operator labeled by the conjugacy class $[n]$, and trivial irreducible representation. This process realizes an anyon permutation corresponding to a generalized electric-magnetic duality in non-Abelian quantum double models. Specifically, we have
\begin{align}
    F^{([e],\pi_n)}_{\xi} \;\longrightarrow\; F^{([n],1)}_{\xi} \;\longrightarrow\; F^{([e],\pi_n)}_{\xi'},
\end{align}
where $\xi' \sim \xi$ and differs from $\xi$ by a translation of one lattice spacing.

\section{Example: generalized electric-magnetic duality in $D_4$ quantum double} \label{sec:D4_gauging}

As an illustrative example, in this section we study an anyon permutation in the $D_4$ quantum double model. It is well known that the anyon theory of the $D_4$ quantum double is equivalent to the type-III $\mathbb{Z}_2^3$ twisted quantum double. Therefore, for convenience, we adopt the anyon labeling of the twisted quantum double model. A dictionary between the anyon labels in these two conventions is listed in Table~\ref{tab:dictionary}. In this appendix, we realize the following anyon permutation:
\begin{equation}
    \begin{aligned}
        m_r \mapsto m_{rg}, \quad m_g \mapsto m_g, \quad m_b \mapsto m_{gb}, \quad m_{rb} \mapsto m_{rb}, \quad
        e_r \mapsto e_r, \quad e_g \mapsto e_{rgb}, \quad e_b \mapsto e_b. \label{eq:anyon_permutation}
    \end{aligned}
\end{equation}
This anyon permutation belongs to class~I and corresponds to gauging the $\mathbb{Z}_2 \times \mathbb{Z}_2$ subgroup generated by $(r^2,1)$ and $(1,s)$. This permutation is the same as that studied in Ref.~\cite{kobayashi2025clifford}. In this work, the authors showed that, when applied to a type-III $\mathbb{Z}_2^3$ twisted quantum double with a specific boundary condition, this permutation implements a logical $T$ gate.

\begin{table}
\begin{tabular}{||c | c | c | c ||} 
 \hline
 $\mathcal{D}(D_4)$ & Centralizers & Characters of irreps of centralizers & $\mathcal{D}^{\alpha}(\mathbb{Z}_2^3)$ \\ [0.5ex] 
 \hline\hline
 $([e], 1)$ & $\langle r, s\rangle$ & trivial & 1 \\
 \hline
 $([e], \mathbf{1}_s)$ & $\langle r, s\rangle$ & $\chi([s]) = \chi([rs]) = -1,\ \chi([e]) = \chi([r]) = \chi([r^2]) = 1$ & $e_{rg}$ \\
 \hline
 $([e], \mathbf{1}_r)$ & $\langle r, s\rangle$ & $\chi([r]) = \chi([rs]) = -1,\ \chi([e]) = \chi([r^2]) = \chi([s]) = 1$ & $e_{r}$ \\ 
 \hline
 $([e], \mathbf{1}_r\mathbf{1}_s)$ & $\langle r, s\rangle$ & $\chi([s]) = \chi([r]) = -1,\ \chi([e]) = \chi([r^2]) = \chi([rs]) = 1$ & $e_{g}$ \\
 \hline
 $([e], \pi)$ & $\langle r, s\rangle$ & $\chi([e])=2,\ \chi([r^2]) = -2,\ \chi([r]) = \chi([s]) = \chi([rs]) = 0$ & $m_{b}$\\
 \hline
 $([r^2], 1)$ & $\langle r, s\rangle$ & trivial & $e_{rgb}$\\
 \hline
 $([r^2], \mathbf{1}_s)$ & $\langle r, s\rangle$ & $\chi([s]) = \chi([rs]) = -1,\ \chi([e]) = \chi([r]) = \chi([r^2]) = 1$ & $e_b$ \\
 \hline
 $([r^2], \mathbf{1}_r)$ & $\langle r, s\rangle$ & $\chi([r]) = \chi([rs]) = -1,\ \chi([e]) = \chi([r^2]) = \chi([s]) = 1$ & $e_{gb}$ \\
 \hline
 $([r^2], \mathbf{1}_r \mathbf{1}_s)$ & $\langle r, s\rangle$ & $\chi([s]) = \chi([r]) = -1\ \chi([e]) = \chi([r^2]) = \chi([rs]) = 1$ & $e_{rb}$ \\
 \hline
 $([r^2], \pi)$ & $\langle r, s\rangle$ & $\chi([e])=2,\ \chi([r^2]) = -2,\ \chi([r]) = \chi([s]) = \chi([rs]) = 0$ & $f_b$\\
 \hline
 $([r], 1)$ & $\langle r\rangle$ & trivial & $m_{rg}$ \\
 \hline
 $([r], \omega)$ & $\langle r\rangle$ & $\chi(r^j) = e^{\frac{j \pi i}{2}}$ & $s_{rgb}$ \\
 \hline
 $([r], \omega^2)$ & $\langle r\rangle$ & $\chi(r^j) = e^{j \pi i}$ & $f_{rg}$ \\
 \hline
 $([r], \omega^3)$ & $\langle r\rangle$ & $\chi(r^j) = e^{-\frac{j \pi i}{2}}$ & $\bar{s}_{rgb}$\\
 \hline
 $([s], (1,1))$ & $\langle r^2, s\rangle$ & trivial & $m_{gb}$ \\
 \hline
 $([s], (-1,1))$ & $\langle r^2, s\rangle$ & $\chi(r^2) = \chi(r^2s) = -1,\ \chi(e) = \chi(s) = 1$ & $m_g$ \\
 \hline
 $([s], (1,-1))$ & $\langle r^2, s\rangle$ & $\chi(s) = \chi(r^2s) = -1,\ \chi(e) = \chi(r^2) = 1$ & $f_g$ \\
 \hline
 $([s], (-1,-1))$ & $\langle r^2, s\rangle$ & $\chi(r^2) = \chi(s) = -1,\ \chi(e) = \chi(r^2s) = 1$ & $f_{gb}$ \\
 \hline
 $([rs], (1,1))$ & $\langle r^2, s\rangle$ & trivial & $m_{RB}$ \\
 \hline
 $([rs], (-1,1))$ & $\langle r^2, s\rangle$ & $\chi(r^2) = \chi(r^2s) = -1,\ \chi(e) = \chi(s) = 1$ & $m_r$ \\
 \hline
 $([rs], (1,-1))$ & $\langle r^2, s\rangle$ & $\chi(s) = \chi(r^2s) = -1\, \chi(e) = \chi(r^2) = 1$ & $f_{r}$ \\
 \hline
 $([rs], (-1,-1))$ & $\langle r^2, s\rangle$ & $\chi(r^2) = \chi(s) = -1,\ \chi(e) = \chi(r^2s) = 1$ & $f_{rb}$ \\
 \hline
\end{tabular}
\caption{A dictionary between anyons in the $D_4$ quantum double and those in the type-III $\mathbb{Z}_2^3$ twisted quantum double. To simplify the notation, in the third column we use $\chi([g])$ to denote the character of the irreducible representation corresponding to the conjugacy class of $g$, and we write $\chi(g)$ when the associating centralizer group is Abelian.}
\label{tab:dictionary}
\end{table}

Consider a one-dimensional spin chain with local Hilbert space spanned by $\{|g\rangle \mid g \in D_4\}$, where $r, s \in D_4$ satisfy $r^4 = 1$, $s^2 = 1$, and $srsr = 1$. We denote group elements by pairs $(r^i, s^j)$, with $i \in \{0,1,2,3\}$ and $j \in \{0,1\}$. Consider the following group extension,
\begin{align}
    1 \longrightarrow \mathbb{Z}_2 \times \mathbb{Z}_2 \longrightarrow D_4 \longrightarrow \mathbb{Z}_2 \longrightarrow 1,
\end{align}
where we choose the normal subgroup $\mathbb{Z}_2 \times \mathbb{Z}_2$ to be generated by $(r^2, 1)$ and $(1, s)$. The corresponding 2-cocycle is trivial. Therefore, the local Hilbert space of the $D_4$ spin chain can be equivalently represented by three-qubit states. For simplicity, we denote these states as $|i, j, k\rangle$, corresponding to the group element $r^{2i} s^j (rs)^k$. We define the following local operators:
\begin{equation}
    \begin{aligned}
        &L^{r^2} |i, j, k\rangle = |i+1, j, k\rangle, \quad L^{s} | i, j , k\rangle = |i, j+1, k\rangle, \quad L^{rs} |i, j, k\rangle = |i+j, j, k+1\rangle, \\
        &R^{r^2} |i, j, k\rangle = |i+1, j, k\rangle, \quad R^{s} |i, j, k\rangle = |i+k, j+1, k\rangle \quad R^{rs} |i, j, k\rangle = |i, j, k+1\rangle,   \\
        &\widetilde{Z}^2 | i, j, k\rangle = (-1)^k |i, j, k\rangle \quad Z |i, j, k\rangle = (-1)^{j+k}|i, j, k\rangle, \quad Z^C |i, j, k\rangle = (-1)^i |i, j, k\rangle.
    \end{aligned}
\end{equation}

When gauging the normal subgroup $N = \mathbb{Z}_2 \times \mathbb{Z}_2$, the resulting emergent symmetry is isomorphic to $\mathbb{Z}_2 \times \mathbb{Z}_2$. To realize the gauging map as a self-duality, we further choose an embedding from $\hat{N}$ into $N$. Specifically, we take the embedding $(\hat{r}^2, 1) \mapsto (1, s)$ and $(1, \hat{s}) \mapsto (r^2, 1)$. This choice of embedding fixes the bicharacter appearing in Eq.~\eqref{eq:general_gauging}. The resulting gauging map is therefore given by
\begin{equation}
    \Gamma: |\{i_s, j_s, k_s\} \rangle \to \sum_{\{p_s, q_s\}} (-1)^{p_s (i_s - i_{s-1}) + q_s (j_s - j_{s-1})} |\{q_s, p_s, k_s\}\rangle, \label{eq:D4_gauging_1}
\end{equation}
where $i_s, j_s, k_s, q_s, p_s \in \{0, 1\}$. Under this gauging map, the operator mappings are given as follows.
\begin{equation}
    \begin{aligned}
        &L_s^{r^2} \to Z_s \widetilde{Z}_s^2 Z_{s+1} \widetilde{Z}_{s+1}^2 \to L_{s+1}^{r^2}, \quad L^{s}_s \to Z^C_s Z^C_{s+1} \to L^{s}_{s+1}, \quad \prod_{s} L^{rs}_s \to \prod_{s} L^{rs}_s,\quad R^{rs} \to R^{rs} \\
        &R_s^{r^2} \to Z_s \widetilde{Z}_s^2 Z_{s+1} \widetilde{Z}_{s+1}^2 \to R_{s+1}^{r^2}, \quad R^{s}_s \to (Z_s \widetilde{Z}_s^2 Z_{s+1} \widetilde{Z}_{s+1}^2)^{k_s} (Z^C_s Z^C_{s+1}) \to (L^{r^2}_{s+1})^{k_s} L^{s}_{s+1}.
    \end{aligned}
\end{equation}

\subsection{Hamiltonian maps}

The Hamiltonian of the $D_4$ quantum double on the triangular lattice we studied is defined as follows. The flux free terms are defined on the 0-triangles. We have

\raisebox{-0.5\height}{
\begin{minipage}{\textwidth}
\centering
$B^{(0)}_1$ = \begin{tikzpicture}[scale=1, baseline=(current bounding box.center)]
\tikzset{
  hexbond/.style={line width=0.6pt},
  vdot/.style={circle, fill=black, inner sep=1.2pt}
}

\coordinate (A_1) at ( 0, 2*0.8660254);
\coordinate (A_2) at ( -1, 0);
\coordinate (A_3) at ( 1, 0);
\draw[hexbond] (A_1)--(A_2)--(A_3)--cycle;

\coordinate (A_4) at ( -2, 2*0.8660254);
\coordinate (A_5) at ( 0, -2*0.8660254);
\coordinate (A_6) at ( 2, 2*0.8660254);
\draw[hexbond] (A_4)--(A_5)--(A_6)--cycle;

\node at ($(A_4)+(0,-0.3)$) {\footnotesize $1$};
\node at ($(A_1)+(0,-0.3)$) {\footnotesize $2$};
\node at ($(A_6)+(0,-0.3)$) {\footnotesize $3$};
\node at ($(A_2)+(0,-0.3)$) {\footnotesize $6$};
\node at ($(A_3)+(0,-0.3)$) {\footnotesize $4$};
\node at ($(A_5)+(0,0.3)$) {\footnotesize $5$};

\node at ($(A_4)+(0,0.3)$) {\footnotesize $(\widetilde{Z}^2_2 Z_2)^{k_2 + k_3}$};
\node at ($(A_1)+(0,0.3)$) {\footnotesize $(\widetilde{Z}^2_2 Z_2)^{k_2 + k_3}$};
\node at ($(A_5)+(0,-0.3)$) {\footnotesize $(\widetilde{Z}^2_5 Z_5)^{k_4 + k_5}$};
\node at ($(A_2)+(-0.9,0)$) {\footnotesize $(\widetilde{Z}^2_6 Z_6)^{k_4 + k_5}$};

\node[myRed] at (0, 2*0.8660254/3) {$0$}; 
\end{tikzpicture} $\times Z_1^C Z_2^C \cdots Z_6^C$, \\ $B^{(0)}_2 = (\widetilde{Z}^2_1 Z_1) (\widetilde{Z}^2_2 Z_2) \cdots (\widetilde{Z}^2_6 Z_6)$, \quad $B^{(0)}_3 = \widetilde{Z}^2_1 \widetilde{Z}^2_2 \cdots \widetilde{Z}^2_6$.
\end{minipage}
}
Since $B^{(0)}_3$ is always invariant under the gauging maps, we omit this term in the future for convenience. On each 1-triangle and 2-triangle, we define the following terms.

\raisebox{-0.5\height}{
\begin{minipage}{0.95\textwidth}
\centering
$A^{(1)}_{1}=$\begin{tikzpicture}[scale=1, baseline=(current bounding box.center)]
\tikzset{
  hexbond/.style={line width=0.6pt},
  vdot/.style={circle, fill=black, inner sep=1.2pt}
}

\coordinate (A_1) at ( 0, 2*0.8660254);
\coordinate (A_2) at ( -1, 0);
\coordinate (A_3) at ( 1, 0);
\draw[hexbond] (A_1)--(A_2)--(A_3)--cycle;

\node at ($(A_1)+(0,0.2)$) {$L^{r^2}$};
\node at ($(A_2)+(-0.5,0)$) {$L^{r^2}$};
\node at ($(A_3)+(0.5,0)$) {$L^{r^2}$};
\node[myGreen] at (0, 2*0.8660254/3) {$1$};
\end{tikzpicture},
\quad
$A^{(1)}_{2}=$\begin{tikzpicture}[scale=1, baseline=(current bounding box.center)]
\tikzset{
  hexbond/.style={line width=0.6pt},
  vdot/.style={circle, fill=black, inner sep=1.2pt}
}

\coordinate (A_1) at ( 0, 2*0.8660254);
\coordinate (A_2) at ( -1, 0);
\coordinate (A_3) at ( 1, 0);
\draw[hexbond] (A_1)--(A_2)--(A_3)--cycle;

\node at ($(A_1)+(0,0.2)$) {$L^{s}$};
\node at ($(A_2)+(-0.4,0)$) {$L^{s}$};
\node at ($(A_3)+(0.4,0)$) {$L^{s}$};
\node[myGreen] at (0, 2*0.8660254/3) {$1$};
\end{tikzpicture}, \quad
$A^{(1)}_{3}=$\begin{tikzpicture}[scale=1, baseline=(current bounding box.center)]
\tikzset{
  hexbond/.style={line width=0.6pt},
  vdot/.style={circle, fill=black, inner sep=1.2pt}
}

\coordinate (A_1) at ( 0, 2*0.8660254);
\coordinate (A_2) at ( -1, 0);
\coordinate (A_3) at ( 1, 0);
\draw[hexbond] (A_1)--(A_2)--(A_3)--cycle;

\node at ($(A_1)+(0,0.2)$) {$L^{rs}$};
\node at ($(A_2)+(-0.4,0)$) {$L^{rs}$};
\node at ($(A_3)+(0.4,0)$) {$L^{rs}$};
\node[myGreen] at (0, 2*0.8660254/3) {$1$};
\end{tikzpicture}, \\

$A^{(2)}_{1}=$\begin{tikzpicture}[scale=1, baseline=(current bounding box.center)]
\tikzset{
  hexbond/.style={line width=0.6pt},
  vdot/.style={circle, fill=black, inner sep=1.2pt}
}

\coordinate (A_1) at ( 0, 2*0.8660254);
\coordinate (A_2) at ( -1, 0);
\coordinate (A_3) at ( 1, 0);
\draw[hexbond] (A_1)--(A_2)--(A_3)--cycle;

\node at ($(A_1)+(0,0.2)$) {$R^{r^2}$};
\node at ($(A_2)+(-0.5,0)$) {$R^{r^2}$};
\node at ($(A_3)+(0.5,0)$) {$R^{r^2}$};
\node[myBlue] at (0, 2*0.8660254/3) {$2$};
\end{tikzpicture},
\quad
$A^{(2)}_{2}=$\begin{tikzpicture}[scale=1, baseline=(current bounding box.center)]
\tikzset{
  hexbond/.style={line width=0.6pt},
  vdot/.style={circle, fill=black, inner sep=1.2pt}
}

\coordinate (A_1) at ( 0, 2*0.8660254);
\coordinate (A_2) at ( -1, 0);
\coordinate (A_3) at ( 1, 0);
\draw[hexbond] (A_1)--(A_2)--(A_3)--cycle;

\node at ($(A_1)+(0,0.2)$) {$R^{s}$};
\node at ($(A_2)+(-0.4,0)$) {$R^{s}$};
\node at ($(A_3)+(0.4,0)$) {$R^{s}$};
\node[myBlue] at (0, 2*0.8660254/3) {$2$};
\end{tikzpicture}, \quad
$A^{(2)}_{3}=$\begin{tikzpicture}[scale=1, baseline=(current bounding box.center)]
\tikzset{
  hexbond/.style={line width=0.6pt},
  vdot/.style={circle, fill=black, inner sep=1.2pt}
}

\coordinate (A_1) at ( 0, 2*0.8660254);
\coordinate (A_2) at ( -1, 0);
\coordinate (A_3) at ( 1, 0);
\draw[hexbond] (A_1)--(A_2)--(A_3)--cycle;

\node at ($(A_1)+(0,0.2)$) {$R^{rs}$};
\node at ($(A_2)+(-0.4,0)$) {$R^{rs}$};
\node at ($(A_3)+(0.4,0)$) {$R^{rs}$};
\node[myBlue] at (0, 2*0.8660254/3) {$2$};
\end{tikzpicture}.
\end{minipage}
}

After applying the gauging map on all the 1-triangles, followed by applying the global controlled-conjugation circuit displayed in Fig.~\ref{fig:figure_1}(c), the star terms of the Hamiltonian become the following:

\raisebox{-0.5\height}{
\begin{minipage}{0.95\textwidth}
\centering
\begin{tikzpicture}[scale=1, baseline=(current bounding box.center)]
\tikzset{
  hexbond/.style={line width=0.6pt},
  vdot/.style={circle, fill=black, inner sep=1.2pt}
}

\coordinate (A_1) at ( 0, 2*0.8660254);
\coordinate (A_2) at ( -1, 0);
\coordinate (A_3) at ( 1, 0);
\draw[hexbond] (A_1)--(A_2)--(A_3)--cycle;

\node at ($(A_1)+(0,0.2)$) {$L^{s}$};
\node at ($(A_2)+(-0.4,0)$) {$L^{s}$};
\node at ($(A_3)+(0.4,0)$) {$L^{s}$};
\node[myRed] at (0, 2*0.8660254/3) {$0$};
\end{tikzpicture},
\begin{tikzpicture}[scale=1, baseline=(current bounding box.center)]
\tikzset{
  hexbond/.style={line width=0.6pt},
  vdot/.style={circle, fill=black, inner sep=1.2pt}
}

\coordinate (A_1) at ( 0, 2*0.8660254);
\coordinate (A_2) at ( -1, 0);
\coordinate (A_3) at ( 1, 0);
\draw[hexbond] (A_1)--(A_2)--(A_3)--cycle;

\node at ($(A_1)+(0,0.2)$) {$L^{r^2}$};
\node at ($(A_2)+(-0.5,0)$) {$L^{r^2}$};
\node at ($(A_3)+(0.5,0)$) {$L^{r^2}$};
\node[myRed] at (0, 2*0.8660254/3) {$0$};
\end{tikzpicture},
\begin{tikzpicture}[scale=1, baseline=(current bounding box.center)]
\tikzset{
  hexbond/.style={line width=0.6pt},
  vdot/.style={circle, fill=black, inner sep=1.2pt}
}

\coordinate (A_1) at ( 0, 2*0.8660254);
\coordinate (A_2) at ( -1, 0);
\coordinate (A_3) at ( 1, 0);
\draw[hexbond] (A_1)--(A_2)--(A_3)--cycle;

\node at ($(A_1)+(0,0.2)$) {$L^{r^2}$};
\node at ($(A_2)+(-0.5,0)$) {$L^{r^2}$};
\node at ($(A_3)+(0.5,0)$) {$L^{r^2}$};
\node[myGreen] at (0, 2*0.8660254/3) {$1$};
\end{tikzpicture},\\
\begin{tikzpicture}[scale=1, baseline=(current bounding box.center)]
\tikzset{
  hexbond/.style={line width=0.6pt},
  vdot/.style={circle, fill=black, inner sep=1.2pt}
}

\coordinate (A_1) at ( 0, 2*0.8660254);
\coordinate (A_2) at ( -1, 0);
\coordinate (A_3) at ( 1, 0);
\draw[hexbond] (A_1)--(A_2)--(A_3)--cycle;

\coordinate (A_4) at ( 2, 2*0.8660254);
\coordinate (A_5) at ( 3, 0);
\draw[hexbond] (A_4)--(A_3)--(A_5)--cycle;

\node at ($(A_1)+(0,0.2)$) {\small $L_1^{s} \left(L^{r^2}_1\right)^{k_2 + k_4}$};
\node at ($(A_2)+(0,-0.5)$) {\small $L_3^{s}$};
\node at ($(A_3)+(0,-0.5)$) {\small $L_4^{s}\left(L^{r^2}_4\right)^{k_5 + k_4}$};

\node at ($(A_1)+(0,-0.3)$) {\footnotesize $1$};
\node at ($(A_4)+(0,-0.3)$) {\footnotesize $2$};
\node at ($(A_2)+(0,0.3)$) {\footnotesize $3$};
\node at ($(A_3)+(0,0.3)$) {\footnotesize $4$};
\node at ($(A_5)+(0,0.3)$) {\footnotesize $5$};

\node[myGreen] at (0, 2*0.8660254/3) {$1$};
\end{tikzpicture},
\begin{tikzpicture}[scale=1, baseline=(current bounding box.center)]
\tikzset{
  hexbond/.style={line width=0.6pt},
  vdot/.style={circle, fill=black, inner sep=1.2pt}
}

\coordinate (A_1) at ( 0, 2*0.8660254);
\coordinate (A_2) at ( -1, 0);
\coordinate (A_3) at ( 1, 0);
\draw[hexbond] (A_1)--(A_2)--(A_3)--cycle;

\coordinate (A_4) at ( -2, 2*0.8660254);
\coordinate (A_5) at ( -3, 0);
\draw[hexbond] (A_4)--(A_2)--(A_5)--cycle;

\node at ($(A_1)+(0,0.2)$) {$R^{rs}$};
\node at ($(A_2)+(0,-0.3)$) {$C^{rs} R^{rs}$};
\node at ($(A_3)+(0,-0.3)$) {$R^{rs}$};
\node at ($(A_4)+(0,0.2)$) {$C^{rs}$};
\node at ($(A_5)+(0,-0.3)$) {$C^{rs}$};
\node[myGreen] at (0, 2*0.8660254/3) {$1$};
\end{tikzpicture},
\begin{tikzpicture}[scale=1, baseline=(current bounding box.center)]
\tikzset{
  hexbond/.style={line width=0.6pt},
  vdot/.style={circle, fill=black, inner sep=1.2pt}
}

\coordinate (A_1) at ( 0, 2*0.8660254);
\coordinate (A_2) at ( -1, 0);
\coordinate (A_3) at ( 1, 0);
\draw[hexbond] (A_1)--(A_2)--(A_3)--cycle;

\node at ($(A_1)+(0,0.2)$) {$R^{rs}$};
\node at ($(A_2)+(-0.4,0)$) {$R^{rs}$};
\node at ($(A_3)+(0.4,0)$) {$R^{rs}$};
\node[myBlue] at (0, 2*0.8660254/3) {$2$};
\end{tikzpicture},
\end{minipage}
}
where $C^{rs} |i, j, k\rangle = |i+j, j, k\rangle$. The flux-free terms are generated by the following:

\raisebox{-0.5\height}{
\begin{minipage}{0.95\textwidth}
\centering
\begin{tikzpicture}[scale=0.85, baseline=(current bounding box.center)]
\tikzset{
  hexbond/.style={line width=0.6pt},
  vdot/.style={circle, fill=black, inner sep=1.2pt}
}

\coordinate (A_1) at ( 0, 2*0.8660254);
\coordinate (A_2) at ( -1, 0);
\coordinate (A_3) at ( 1, 0);
\draw[hexbond] (A_1)--(A_2)--(A_3)--cycle;

\coordinate (A_4) at ( -2, 2*0.8660254);
\coordinate (A_5) at ( 0, -2*0.8660254);
\coordinate (A_6) at ( 2, 2*0.8660254);
\draw[hexbond] (A_4)--(A_5)--(A_6)--cycle;

\node at ($(A_4)+(0,-0.3)$) {\footnotesize $1$};
\node at ($(A_1)+(0,-0.3)$) {\footnotesize $2$};
\node at ($(A_6)+(0,-0.3)$) {\footnotesize $3$};
\node at ($(A_2)+(0,-0.3)$) {\footnotesize $6$};
\node at ($(A_3)+(0,-0.3)$) {\footnotesize $4$};
\node at ($(A_5)+(0,0.3)$) {\footnotesize $5$};

\node at ($(A_4)+(0,0.3)$) {$\widetilde{Z}^2_1 Z_1$};
\node at ($(A_1)+(0,0.3)$) {$\widetilde{Z}^2_2 Z_2$};
\node at ($(A_6)+(0,0.3)$) {$\widetilde{Z}^2_3 Z_3$};
\node at ($(A_3)+(0.5,0)$) {$\widetilde{Z}^2_4 Z_4$};
\node at ($(A_5)+(0,-0.3)$) {$\widetilde{Z}^2_5 Z_5$};
\node at ($(A_2)+(-0.5,0)$) {$\widetilde{Z}^2_6 Z_6$};

\node[myBlue] at (0, 2*0.8660254/3) {$2$}; 
\end{tikzpicture},
\begin{tikzpicture}[scale=0.85, baseline=(current bounding box.center)]
\tikzset{
  hexbond/.style={line width=0.6pt},
  vdot/.style={circle, fill=black, inner sep=1.2pt}
}

\coordinate (A_1) at ( 0, 2*0.8660254);
\coordinate (A_2) at ( -1, 0);
\coordinate (A_3) at ( 1, 0);
\draw[hexbond] (A_1)--(A_2)--(A_3)--cycle;

\coordinate (A_4) at ( -2, 2*0.8660254);
\coordinate (A_5) at ( 0, -2*0.8660254);
\coordinate (A_6) at ( 2, 2*0.8660254);
\draw[hexbond] (A_4)--(A_5)--(A_6)--cycle;

\node at ($(A_4)+(0,-0.3)$) {\footnotesize $1$};
\node at ($(A_1)+(0,-0.3)$) {\footnotesize $2$};
\node at ($(A_6)+(0,-0.3)$) {\footnotesize $3$};
\node at ($(A_2)+(0,-0.3)$) {\footnotesize $6$};
\node at ($(A_3)+(0,-0.3)$) {\footnotesize $4$};
\node at ($(A_5)+(0,0.3)$) {\footnotesize $5$};

\node at ($(A_4)+(0,0.3)$) {$Z_1^C$};
\node at ($(A_1)+(0,0.3)$) {$Z_2^C$};
\node at ($(A_6)+(0,0.3)$) {$Z_3^C$};
\node at ($(A_3)+(0.5,0)$) {$Z_4^C$};
\node at ($(A_5)+(0,-0.3)$) {$Z_5^C$};
\node at ($(A_2)+(-0.5,0)$) {$Z_6^C$};

\node[myBlue] at (0, 2*0.8660254/3) {$2$}; 
\end{tikzpicture} $\times$ \begin{tikzpicture}[scale=0.85, baseline=(current bounding box.center)]
\tikzset{
  hexbond/.style={line width=0.6pt},
  vdot/.style={circle, fill=black, inner sep=1.2pt}
}

\coordinate (A_1) at ( 0, 2*0.8660254);
\coordinate (A_2) at ( -1, 0);
\coordinate (A_3) at ( 1, 0);
\draw[hexbond] (A_1)--(A_2)--(A_3)--cycle;

\coordinate (A_4) at ( -2, 2*0.8660254);
\coordinate (A_5) at ( 0, -2*0.8660254);
\coordinate (A_6) at ( 2, 2*0.8660254);
\draw[hexbond] (A_4)--(A_5)--(A_6)--cycle;

\coordinate (A_7) at ( 4, 2*0.8660254);
\coordinate (A_8) at ( 3, 0);
\coordinate (A_9) at ( 2, -2*0.8660254);
\draw[hexbond] (A_6)--(A_7)--(A_8)--cycle;
\draw[hexbond] (A_3)--(A_8)--(A_9)--cycle;
\draw[hexbond] (A_5)--(A_9);

\node at ($(A_4)+(0,-0.3)$) {\footnotesize $1$};
\node at ($(A_1)+(0,-0.3)$) {\footnotesize $2$};
\node at ($(A_6)+(0,-0.3)$) {\footnotesize $3$};
\node at ($(A_2)+(0,-0.3)$) {\footnotesize $6$};
\node at ($(A_3)+(0,-0.3)$) {\footnotesize $4$};
\node at ($(A_5)+(0,0.3)$) {\footnotesize $5$};
\node at ($(A_7)+(0,-0.3)$) {\footnotesize $7$};
\node at ($(A_8)+(0,-0.3)$) {\footnotesize $8$};
\node at ($(A_9)+(0,0.3)$) {\footnotesize $9$};

\node at ($(A_4)+(0,0.3)$) {$(\widetilde{Z}_1^2 Z_1)^{k_2}$};
\node at ($(A_1)+(0,0.3)$) {$(\widetilde{Z}_2^2 Z_2)^{k_2}$};
\node at ($(A_6)+(0.6,0.3)$) {$(\widetilde{Z}_3^2 Z_3)^{k_4 + k_8 + k_9}$};
\node at ($(A_3)+(1.4,0.3)$) {$(\widetilde{Z}_4^2 Z_4)^{k_4 + k_8 + k_9}$};
\node at ($(A_5)+(0,-0.3)$) {$(\widetilde{Z}_5^2 Z_5)^{k_4}$};
\node at ($(A_2)+(-0.8,0)$) {$(\widetilde{Z}_6^2 Z_6)^{k_4}$};

\node[myBlue] at (0, 2*0.8660254/3) {$2$}; 
\end{tikzpicture}.
\end{minipage}
}

Then, we applying the gauging map on all the 0-triangles, followed by applying the global controlled-conjugation $CC^{(0)}$. The Hamiltonian terms we get are the following:

\raisebox{-0.5\height}{
\begin{minipage}{0.95\textwidth}
\centering
\begin{tikzpicture}[scale=1, baseline=(current bounding box.center)]
\tikzset{
  hexbond/.style={line width=0.6pt},
  vdot/.style={circle, fill=black, inner sep=1.2pt}
}

\coordinate (A_1) at ( 0, 2*0.8660254);
\coordinate (A_2) at ( -1, 0);
\coordinate (A_3) at ( 1, 0);
\draw[hexbond] (A_1)--(A_2)--(A_3)--cycle;

\coordinate (A_4) at ( -2, 2*0.8660254);
\coordinate (A_5) at ( 0, -2*0.8660254);
\coordinate (A_6) at ( 2, 2*0.8660254);
\draw[hexbond] (A_4)--(A_5)--(A_6)--cycle;

\node at ($(A_4)+(0,-0.3)$) {\footnotesize $1$};
\node at ($(A_1)+(0,-0.3)$) {\footnotesize $2$};
\node at ($(A_6)+(0,-0.3)$) {\footnotesize $3$};
\node at ($(A_2)+(0,-0.3)$) {\footnotesize $6$};
\node at ($(A_3)+(0,-0.3)$) {\footnotesize $4$};
\node at ($(A_5)+(0,0.3)$) {\footnotesize $5$};

\node at ($(A_1)+(0.5,0.4)$) {$L_2^{s} \left(L^{r^2}_2\right)^{k_3 + k_4}$};
\node at ($(A_3)+(0.5,0)$) {$L_4^{s}$};
\node at ($(A_2)+(-1.8,0)$) {$L_6^{s} \left(L^{r^2}_6\right)^{k_5 + k_6}$};

\node[myRed] at (0, 2*0.8660254/3) {$0$}; 
\end{tikzpicture},
\begin{tikzpicture}[scale=1, baseline=(current bounding box.center)]
\tikzset{
  hexbond/.style={line width=0.6pt},
  vdot/.style={circle, fill=black, inner sep=1.2pt}
}

\coordinate (A_1) at ( 0, 2*0.8660254);
\coordinate (A_2) at ( -1, 0);
\coordinate (A_3) at ( 1, 0);
\draw[hexbond] (A_1)--(A_2)--(A_3)--cycle;

\node at ($(A_1)+(0,0.2)$) {$L^{r^2}$};
\node at ($(A_2)+(-0.5,0)$) {$L^{r^2}$};
\node at ($(A_3)+(0.5,0)$) {$L^{r^2}$};
\node[myRed] at (0, 2*0.8660254/3) {$0$};
\end{tikzpicture},
\begin{tikzpicture}[scale=1, baseline=(current bounding box.center)]
\tikzset{
  hexbond/.style={line width=0.6pt},
  vdot/.style={circle, fill=black, inner sep=1.2pt}
}

\coordinate (A_1) at ( 0, 2*0.8660254);
\coordinate (A_2) at ( -1, 0);
\coordinate (A_3) at ( 1, 0);
\draw[hexbond] (A_1)--(A_2)--(A_3)--cycle;

\coordinate (A_4) at ( -2, 2*0.8660254);
\coordinate (A_5) at ( 0, -2*0.8660254);
\coordinate (A_6) at ( 2, 2*0.8660254);
\draw[hexbond] (A_4)--(A_5)--(A_6)--cycle;

\node at ($(A_4)+(0,-0.3)$) {\footnotesize $1$};
\node at ($(A_1)+(0,-0.3)$) {\footnotesize $2$};
\node at ($(A_6)+(0,-0.3)$) {\footnotesize $3$};
\node at ($(A_2)+(0,-0.3)$) {\footnotesize $6$};
\node at ($(A_3)+(0,-0.3)$) {\footnotesize $4$};
\node at ($(A_5)+(0,0.3)$) {\footnotesize $5$};

\node at ($(A_4)+(0,0.3)$) {$\widetilde{Z}^2_1 Z_1$};
\node at ($(A_1)+(0,0.3)$) {$\widetilde{Z}^2_2 Z_2$};
\node at ($(A_6)+(0,0.3)$) {$\widetilde{Z}^2_3 Z_3$};
\node at ($(A_3)+(0.5,0)$) {$\widetilde{Z}^2_4 Z_4$};
\node at ($(A_5)+(0,-0.3)$) {$\widetilde{Z}^2_5 Z_5$};
\node at ($(A_2)+(-0.8,0)$) {$\widetilde{Z}^2_6 Z_6$};

\node[myGreen] at (0, 2*0.8660254/3) {$1$}; 
\end{tikzpicture},\\
\begin{tikzpicture}[scale=1, baseline=(current bounding box.center)]
\tikzset{
  hexbond/.style={line width=0.6pt},
  vdot/.style={circle, fill=black, inner sep=1.2pt}
}

\coordinate (A_1) at ( 0, 2*0.8660254);
\coordinate (A_2) at ( -1, 0);
\coordinate (A_3) at ( 1, 0);
\draw[hexbond] (A_1)--(A_2)--(A_3)--cycle;

\node at ($(A_1)+(0,0.2)$) {$R^{rs}$};
\node at ($(A_2)+(-0.4,0)$) {$R^{rs}$};
\node at ($(A_3)+(0.4,0)$) {$R^{rs}$};
\node[myGreen] at (0, 2*0.8660254/3) {$1$};
\end{tikzpicture},
\begin{tikzpicture}[scale=1, baseline=(current bounding box.center)]
\tikzset{
  hexbond/.style={line width=0.6pt},
  vdot/.style={circle, fill=black, inner sep=1.2pt}
}

\coordinate (A_1) at ( 0, 2*0.8660254);
\coordinate (A_2) at ( -1, 0);
\coordinate (A_3) at ( 1, 0);
\draw[hexbond] (A_1)--(A_2)--(A_3)--cycle;

\coordinate (A_4) at ( -2, 2*0.8660254);
\coordinate (A_5) at ( 0, -2*0.8660254);
\coordinate (A_6) at ( 2, 2*0.8660254);
\draw[hexbond] (A_4)--(A_5)--(A_6)--cycle;

\node at ($(A_4)+(0,-0.3)$) {\footnotesize $1$};
\node at ($(A_1)+(0,-0.3)$) {\footnotesize $2$};
\node at ($(A_6)+(0,-0.3)$) {\footnotesize $3$};
\node at ($(A_2)+(0,-0.3)$) {\footnotesize $6$};
\node at ($(A_3)+(0,-0.3)$) {\footnotesize $4$};
\node at ($(A_5)+(0,0.3)$) {\footnotesize $5$};

\node at ($(A_4)+(0,0.3)$) {$Z^C_1$};
\node at ($(A_1)+(0,0.3)$) {$Z^C_2$};
\node at ($(A_6)+(0,0.3)$) {$Z^C_3$};
\node at ($(A_3)+(0.5,0)$) {$Z^C_4$};
\node at ($(A_5)+(0,-0.3)$) {$Z^C_5$};
\node at ($(A_2)+(-0.8,0)$) {$Z^C_6$};

\node[myGreen] at (0, 2*0.8660254/3) {$1$}; 
\end{tikzpicture} $\times$ 
\begin{tikzpicture}[scale=1, baseline=(current bounding box.center)]
\tikzset{
  hexbond/.style={line width=0.6pt},
  vdot/.style={circle, fill=black, inner sep=1.2pt}
}

\coordinate (A_1) at ( 0, 2*0.8660254);
\coordinate (A_2) at ( -1, 0);
\coordinate (A_3) at ( 1, 0);
\draw[hexbond] (A_1)--(A_2)--(A_3)--cycle;

\coordinate (A_4) at ( -2, 2*0.8660254);
\coordinate (A_5) at ( 0, -2*0.8660254);
\coordinate (A_6) at ( 2, 2*0.8660254);
\draw[hexbond] (A_4)--(A_5)--(A_6)--cycle;

\node at ($(A_4)+(0,-0.3)$) {\footnotesize $1$};
\node at ($(A_1)+(0,-0.3)$) {\footnotesize $2$};
\node at ($(A_6)+(0,-0.3)$) {\footnotesize $3$};
\node at ($(A_2)+(0,-0.3)$) {\footnotesize $6$};
\node at ($(A_3)+(0,-0.3)$) {\footnotesize $4$};
\node at ($(A_5)+(0,0.3)$) {\footnotesize $5$};

\node at ($(A_4)+(0,0.3)$) {$(\widetilde{Z}^2_1 Z_1)^{k_2}$};
\node at ($(A_1)+(0,0.3)$) {$(\widetilde{Z}^2_2 Z_2)^{k_2}$};
\node at ($(A_6)+(0,0.3)$) {$(\widetilde{Z}^2_3 Z_3)^{k_4}$};
\node at ($(A_3)+(0.8,0)$) {$(\widetilde{Z}^2_4 Z_4)^{k_4}$};
\node at ($(A_5)+(0,-0.3)$) {$(\widetilde{Z}^2_5 Z_5)^{k_6}$};
\node at ($(A_2)+(-0.8,0)$) {$(\widetilde{Z}^2_6 Z_6)^{k_6}$};

\node[myGreen] at (0, 2*0.8660254/3) {$1$}; 
\end{tikzpicture},\\
\begin{tikzpicture}[scale=1, baseline=(current bounding box.center)]
\tikzset{
  hexbond/.style={line width=0.6pt},
  vdot/.style={circle, fill=black, inner sep=1.2pt}
}

\coordinate (A_1) at ( 0, 2*0.8660254);
\coordinate (A_2) at ( -1, 0);
\coordinate (A_3) at ( 1, 0);
\draw[hexbond] (A_1)--(A_2)--(A_3)--cycle;

\node at ($(A_1)+(0,0.2)$) {$L^{r^2}$};
\node at ($(A_2)+(-0.5,0)$) {$L^{r^2}$};
\node at ($(A_3)+(0.5,0)$) {$L^{r^2}$};
\node[myBlue] at (0, 2*0.8660254/3) {$2$};
\end{tikzpicture},
\begin{tikzpicture}[scale=1, baseline=(current bounding box.center)]
\tikzset{
  hexbond/.style={line width=0.6pt},
  vdot/.style={circle, fill=black, inner sep=1.2pt}
}

\coordinate (A_1) at ( 0, 2*0.8660254);
\coordinate (A_2) at ( -1, 0);
\coordinate (A_3) at ( 1, 0);
\draw[hexbond] (A_1)--(A_2)--(A_3)--cycle;

\node at ($(A_1)+(0,0.2)$) {$L^{s}$};
\node at ($(A_2)+(-0.4,0)$) {$L^{s}$};
\node at ($(A_3)+(0.4,0)$) {$L^{s}$};
\node[myBlue] at (0, 2*0.8660254/3) {$2$};
\end{tikzpicture},
\begin{tikzpicture}[scale=1, baseline=(current bounding box.center)]
\tikzset{
  hexbond/.style={line width=0.6pt},
  vdot/.style={circle, fill=black, inner sep=1.2pt}
}

\coordinate (A_1) at ( 0, 2*0.8660254);
\coordinate (A_2) at ( -1, 0);
\coordinate (A_3) at ( 1, 0);
\draw[hexbond] (A_1)--(A_2)--(A_3)--cycle;

\node at ($(A_1)+(0,0.2)$) {$L^{rs}$};
\node at ($(A_2)+(-0.4,-0.2)$) {$L^{rs}$};
\node at ($(A_3)+(0.4,-0.2)$) {$L^{rs}$};
\node[myBlue] at (0, 2*0.8660254/3) {$2$};
\end{tikzpicture}.
\end{minipage}
}

Finally, we can apply the gauging map on all the 2-triangles, followed by a transversal controlled-conjugation gate $CC^{(2)} = \prod_{i} CC_{i,i}$. The Hamiltonian goes back to its original form.

\subsection{Ribbon operators maps}

In this subsection, we study the mappings between ribbon operators under the gauging map discussed in the previous subsection. We explicitly show how the logical ribbon operators transform under the anyon permutation circuit constructed in the previous subsection. In particular, we work out the ribbon operator transformations associated with the following anyon permutations:
\begin{align}
    e_g \mapsto e_{rgb}, \quad m_g \mapsto m_{gb}.
\end{align}
These two anyon permutations serve as representative examples of anyons with quantum dimension $1$ and quantum dimension $2$, respectively. All other ribbon operator transformations associated with the anyon permutations in Eq.~\eqref{eq:anyon_permutation} can be derived in a similar manner.

First, let's consider the mapping between the ribbon operator of the anyon $e_g$ and the ribbon operator of the anyon $e_{rgb}$. Consider the lattice in Fig.~\ref{fig:deformed_lattice}, the $e_g$ ribbon operator is given by the following:
\begin{equation}
    \begin{aligned}
        F^{e_g}_{\xi} \propto (1 + \widetilde{Z}^2_1 \widetilde{Z}^2_2 \cdots \widetilde{Z}^2_n) (\widetilde{Z}^2_1 Z_1) (\widetilde{Z}^2_2 Z_2) \cdots (\widetilde{Z}^2_n Z_n)
    \end{aligned}
\end{equation}
Since the projector is invariant under the gauging maps, we only need to take care of the rest parts. After applying the gauging map on all the 1-triangles, followed with the global controlled-conjugation gate. We have,
\begin{equation}
    \begin{aligned}
        (\widetilde{Z}^2_1 Z_1) (\widetilde{Z}^2_2 Z_2) \cdots (\widetilde{Z}^2_n Z_n) \to L^{r^2}_1 L^{r^2}_3 \cdots L^{r^2}_{n-1}.
    \end{aligned}
\end{equation}

Then, we apply the gauging map on all the 0-triangles, followed with the global controlled-conjugation gate. We have,
\begin{equation}
    \begin{aligned}
        L^{r^2}_1 L^{r^2}_3 \cdots L^{r^2}_{n-1} \to (\widetilde{Z}^2_{n'} Z_{n'}) (\widetilde{Z}^2_1 Z_1) (\widetilde{Z}^2_{2'} Z_{2'}) (\widetilde{Z}^2_3 Z_3) \cdots (\widetilde{Z}^2_{n-1} Z_{n-1}).
    \end{aligned}
\end{equation}

Finally, we apply the gauging map on all the 2-triangles, followed with the transversal controlled-conjugation gate. We have
\begin{equation}
    \begin{aligned}
        (\widetilde{Z}^2_{n'} Z_{n'}) (\widetilde{Z}^2_1 Z_1) (\widetilde{Z}^2_{2'} Z_{2'}) (\widetilde{Z}^2_3 Z_3) \cdots (\widetilde{Z}^2_{n-1} Z_{n-1}) \to L^{r^2}_{2'} L^{r^2}_{4'} \cdots L^{r^2}_{n'}.
    \end{aligned}
\end{equation}

The final ribbon operator we get is the following.
\begin{equation}
    \begin{aligned}
        F^{e_{rgb}} \propto (1 + \widetilde{Z}^2_1 \widetilde{Z}^2_2 \cdots \widetilde{Z}^2_n) L^{r^2}_{2'} L^{r^2}_{4'} \cdots L^{r^2}_{n'}.
    \end{aligned}
\end{equation}

Therefore, the above circuit realizes the anyon permutation $e_g \mapsto e_{rgb}$.

Second, let's consider the mapping between the ribbon operator of the anyon $m_b$ and the ribbon operator of the anyon $m_{gb}$.

Consider the lattice in Fig.~\ref{fig:deformed_lattice}, the $m_b$ ribbon operator can be written as follows.
\begin{equation}
    \begin{aligned}
        F^{m_b}_{\xi} \propto (1 + \widetilde{Z}^2_1 \widetilde{Z}^2_2 \cdots \widetilde{Z}^2_n) \left(Z^C_1 Z^C_2 (\widetilde{Z}^2_2 Z_2)^{k_1 + k_2} Z^C_3 (\widetilde{Z}^2_3 Z_3)^{k_1 + k_2} \cdots + Z^C_1 (\widetilde{Z}^2_1 Z_1) Z^C_2 (\widetilde{Z}^2_2 Z_2)^{k_1 + k_2 + 1} + \cdots \right)
    \end{aligned}
\end{equation}
The projector is invariant under the above circuit. Therefore, we can mainly focus on the second part. After applying gauging map on all the 1-triangles and the controlled-conjugation gate, we have
\begin{equation}
    \begin{aligned}
        &\quad \ \left( Z^C_1 Z^C_2 \cdots Z^C_n\right)(\widetilde{Z}^2_2 Z_2)^{k_2 + k_1} (\widetilde{Z}^2_3 Z_3)^{k_2 + k_1} (\widetilde{Z}^2_4 Z_4)^{k_4 + k_3 + k_2 + k_1} \cdots \\
        &\to L^{s}_1 (L^{r^2}_1)^{k_1 + k_{2'}} L^{s}_3 (L^{r^2}_3)^{k_1 + k_2 + k_3 + k_{4'}} L^{s}_5 (L^{r^2}_5)^{k_1 + k_2 + \cdots + k_5 + k_{6'}} \cdots.
    \end{aligned}
\end{equation}

Then, we apply the gauging map on all the 0-triangles and the following controlled-conjugation circuit. We get
\begin{equation}
    \begin{aligned}
        &\quad\ L^{s}_1 (L^{r^2}_1)^{k_1 + k_{2'}} L^{s}_3 (L^{r^2}_3)^{k_1 + k_2 + k_3 + k_{4'}} L^{s}_5 (L^{r^2}_5)^{k_1 + k_2 + \cdots + k_5 + k_{6'}} \cdots\\
        &\to Z^C_{n'} (\widetilde{Z}^2_{n'} Z_{n'})^{k_n} Z^C_{1} (\widetilde{Z}^2_{1} Z_{1})^{k_1} Z^C_{2} (\widetilde{Z}^2_{2} Z_{2})^{k_1} Z^C_{3} (\widetilde{Z}^2_{3} Z_{3})^{k_1 + k_2 + k_3} \cdots.
    \end{aligned}
\end{equation}

Finally, we apply the gauging map on all the 2-triangles and the following transversal controlled-conjugation gate. We get the following operator.
\begin{align}
    L^{s}_{2'} (L^{r^2}_{2'})^{k_1 + k_{2'}} L^{s}_{4'} (L^{r^2}_{4'})^{k_1 + k_2 + k_3 + k_{4'}} \cdots .
\end{align}

For the other term, we can similarly get the following result.
\begin{align}
    L^{s}_{2'} (L^{r^2}_{2'})^{k_1 + k_{2'} + 1} L^{s}_{4'} (L^{r^2}_{4'})^{k_1 + k_2 + k_3 + k_{4'} + 1} \cdots .
\end{align}

Combining them together, we get the ribbon operator for the $m_{gb}$ anyon. Therefore, the above circuit realizes the anyon permutation $m_b \mapsto m_{gb}$.

\section{Example: Anyon permutation from twisted gauging in $\mathbb{Z}_4 \times \mathbb{Z}_4$ quantum double model} \label{sec:twisted_gauging}

In this appendix, we study the anyon permutation associated with twisted gauging. Consider the group $\mathbb{Z}_4 \times \mathbb{Z}_4$. It has the following group extension.
\begin{align}
    1 \longrightarrow \mathbb{Z}_2 \times \mathbb{Z}_2 \longrightarrow \mathbb{Z}_4 \times \mathbb{Z}_4 \longrightarrow \mathbb{Z}_2 \times \mathbb{Z}_2 \to 1,
\end{align}
with a nontrivial 2-cocycle $[\omega] \in  H^2(\mathbb{Z}_2 \times \mathbb{Z}_2, \mathbb{Z}_2 \times \mathbb{Z}_2)$, such that
\begin{align}
    \omega(q, q) = n,\quad \omega (q', q') = n',
\end{align}
where $q, q' \in Q = \mathbb{Z}_2 \times \mathbb{Z}_2$ and $n, n' \in N = \mathbb{Z}_2 \times \mathbb{Z}_2$. Consider the following twisted gauging map:
\begin{align}
    \Gamma^{\alpha}: |\{n_s, n_s', q_s, q_s'\}\rangle \mapsto \sum_{m_s, m_s'} (-1)^{m_s (n_{s-1}' + n_s') + m_{s-1}'(n_{s-1} + n_s) + n_s' (n_s + n_{s+1}) + m_s (m_{s-1}' + m_s')} |\{m_s, m_s', q_s, q_s'\}\rangle,
\end{align}
where $\alpha$ denotes the nontrivial twisting gauging map. In this example, we choose the following 2-cocycle:
\begin{align}
    \alpha((n_1, n'_1, q_1, q'_1),(n_2, n'_2, q_2, q'_2)) = (-1)^{n_1 n_2'}.
\end{align}
We note that this twisted gauging has a different mixing with lattice translation than the generalized gauging map defined in the main text. We define the local operators:
\begin{equation}
    \begin{aligned}
        X_s | n_s, n_s', q_s, q_s' \rangle = | n_s + q_s, n_s', q_s+1, q_s' \rangle&, \quad X_s' | n_s, n_s', q_s, q_s' \rangle = | n_s, n_s'+q_s', q_s, q_s'+1 \rangle, \\
        X_s^C | n_s, n_s', q_s, q_s' \rangle = | n_s + 1, n_s', q_s, q_s' \rangle&, \quad X_s'^C | n_s, n_s', q_s, q_s' \rangle = | n_s, n_s'+1, q_s, q_s' \rangle, \\
        Z_s | n_s, n_s', q_s, q_s' \rangle = (-1)^{q_s} | n_s, n_s', q_s, q_s' \rangle&, \quad Z_s' | n_s, n_s', q_s, q_s' \rangle = (-1)^{q_s'} | n_s, n_s', q_s, q_s' \rangle, \\
        Z_s^C | n_s, n_s', q_s, q_s' \rangle = (-1)^{n_s} | n_s, n_s', q_s, q_s' \rangle&, \quad Z_s'^C | n_s, n_s', q_s, q_s' \rangle = (-1)^{n_s'} | n_s, n_s', q_s, q_s' \rangle.
    \end{aligned}
\end{equation}

Under the gauging map, the above operators map to the following.
\begin{equation}
    \begin{aligned}
        &X_s^C \mapsto X_s^C, \quad X_s'^C \mapsto X_s'^C,\quad X_s \mapsto X_s,\quad X_s' \mapsto X_s', \quad Z_s \mapsto Z_s, \quad Z_s' \mapsto Z_s'\\
        & Z_s^C Z_{s+1}^C \mapsto Z_s^C X_{s}'^C Z_{s+1}^C,\quad Z_s'^C Z_{s+1}'^C \mapsto Z_s'^C X_{s+1}^C Z_{s+1}'^C. \label{eq:twisted_gauging_operator_maps}
    \end{aligned}
\end{equation}

\subsection{Hamiltonian maps}

To define the Hamiltonian on the triangular lattice shown in Fig.~\ref{fig:figure_1}(a), we assign four qubits to each vertex, where two qubits together represent a single $\mathbb{Z}_4$ degree of freedom. Among these, one qubit corresponds to the normal subgroup degree of freedom, while the other corresponds to the quotient subgroup degree of freedom. A $\mathbb{Z}_4 \times \mathbb{Z}_4$ quantum double ground state is the simultaneous $+1$-eigenstate of all the following terms, where the operator $W_0\equiv (Z_1 Z_2 + Z_3 Z_4 + Z_5 Z_6 -1)/2$:

\raisebox{-0.5\height}{
\begin{minipage}{0.95\textwidth}
\centering
\begin{tikzpicture}[scale=0.9, baseline=(current bounding box.center)]
\tikzset{
  hexbond/.style={line width=0.6pt},
  vdot/.style={circle, fill=black, inner sep=1.2pt}
}

\coordinate (A_1) at ( 0, 2*0.8660254);
\coordinate (A_2) at ( -1, 0);
\coordinate (A_3) at ( 1, 0);
\draw[hexbond] (A_1)--(A_2)--(A_3)--cycle;

\node at ($(A_1)+(0,0.2)$) {$X^C$};
\node at ($(A_2)+(-0.4,0)$) {$X^C$};
\node at ($(A_3)+(0.4,0)$) {$X^C$};
\node[myGreen] at (-0.2, 2*0.8660254/3) {$1$};
\node at (0, 2*0.8660254/3-0.1) {$,$};
\node[myBlue] at (0.2, 2*0.8660254/3) {$2$};
\end{tikzpicture},
\begin{tikzpicture}[scale=0.9, baseline=(current bounding box.center)]
\tikzset{
  hexbond/.style={line width=0.6pt},
  vdot/.style={circle, fill=black, inner sep=1.2pt}
}

\coordinate (A_1) at ( 0, 2*0.8660254);
\coordinate (A_2) at ( -1, 0);
\coordinate (A_3) at ( 1, 0);
\draw[hexbond] (A_1)--(A_2)--(A_3)--cycle;

\coordinate (A_4) at ( -2, 2*0.8660254);
\coordinate (A_5) at ( 0, -2*0.8660254);
\coordinate (A_6) at ( 2, 2*0.8660254);
\draw[hexbond] (A_4)--(A_5)--(A_6)--cycle;

\node at ($(A_1)+(0,0.2)$) {$Z^C$};
\node at ($(A_2)+(-0.4,0)$) {$Z^C$};
\node at ($(A_3)+(0.4,0)$) {$Z^C$};
\node at ($(A_4)+(0,0.2)$) {$Z^C$};
\node at ($(A_5)+(0,-0.2)$) {$Z^C$};
\node at ($(A_6)+(0,0.2)$) {$Z^C$};
\node[myRed] at (0, 2*0.8660254/3) {$0$};

\node at ($(A_4)+(0,-0.3)$) {\footnotesize $1$};
\node at ($(A_1)+(0,-0.3)$) {\footnotesize $2$};
\node at ($(A_6)+(0,-0.3)$) {\footnotesize $3$};
\node at ($(A_2)+(0,-0.3)$) {\footnotesize $6$};
\node at ($(A_3)+(0,-0.3)$) {\footnotesize $4$};
\node at ($(A_5)+(0,0.3)$) {\footnotesize $5$};

\end{tikzpicture} $\times\ W_0 $, \\
\begin{tikzpicture}[scale=0.9, baseline=(current bounding box.center)]
\tikzset{
  hexbond/.style={line width=0.6pt},
  vdot/.style={circle, fill=black, inner sep=1.2pt}
}

\coordinate (A_1) at ( 0, 2*0.8660254);
\coordinate (A_2) at ( -1, 0);
\coordinate (A_3) at ( 1, 0);
\draw[hexbond] (A_1)--(A_2)--(A_3)--cycle;

\node at ($(A_1)+(0,0.2)$) {$X'^C$};
\node at ($(A_2)+(-0.4,0)$) {$X'^C$};
\node at ($(A_3)+(0.4,0)$) {$X'^C$};
\node[myGreen] at (-0.2, 2*0.8660254/3) {$1$};
\node at (0, 2*0.8660254/3-0.1) {$,$};
\node[myBlue] at (0.2, 2*0.8660254/3) {$2$};
\end{tikzpicture},
\begin{tikzpicture}[scale=0.9, baseline=(current bounding box.center)]
\tikzset{
  hexbond/.style={line width=0.6pt},
  vdot/.style={circle, fill=black, inner sep=1.2pt}
}

\coordinate (A_1) at ( 0, 2*0.8660254);
\coordinate (A_2) at ( -1, 0);
\coordinate (A_3) at ( 1, 0);
\draw[hexbond] (A_1)--(A_2)--(A_3)--cycle;

\coordinate (A_4) at ( -2, 2*0.8660254);
\coordinate (A_5) at ( 0, -2*0.8660254);
\coordinate (A_6) at ( 2, 2*0.8660254);
\draw[hexbond] (A_4)--(A_5)--(A_6)--cycle;

\node at ($(A_1)+(0,0.2)$) {$Z'^C$};
\node at ($(A_2)+(-0.4,0)$) {$Z'^C$};
\node at ($(A_3)+(0.4,0)$) {$Z'^C$};
\node at ($(A_4)+(0,0.2)$) {$Z'^C$};
\node at ($(A_5)+(0,-0.2)$) {$Z'^C$};
\node at ($(A_6)+(0,0.2)$) {$Z'^C$};
\node[myRed] at (0, 2*0.8660254/3) {$0$};

\node at ($(A_4)+(0,-0.3)$) {\footnotesize $1$};
\node at ($(A_1)+(0,-0.3)$) {\footnotesize $2$};
\node at ($(A_6)+(0,-0.3)$) {\footnotesize $3$};
\node at ($(A_2)+(0,-0.3)$) {\footnotesize $6$};
\node at ($(A_3)+(0,-0.3)$) {\footnotesize $4$};
\node at ($(A_5)+(0,0.3)$) {\footnotesize $5$};

\end{tikzpicture} $\times\ W_0 $ ,\\
\begin{tikzpicture}[scale=0.9, baseline=(current bounding box.center)]
\tikzset{
  hexbond/.style={line width=0.6pt},
  vdot/.style={circle, fill=black, inner sep=1.2pt}
}

\coordinate (A_1) at ( 0, 2*0.8660254);
\coordinate (A_2) at ( -1, 0);
\coordinate (A_3) at ( 1, 0);
\draw[hexbond] (A_1)--(A_2)--(A_3)--cycle;

\node at ($(A_1)+(0,0.2)$) {$X$};
\node at ($(A_2)+(-0.4,0)$) {$X$};
\node at ($(A_3)+(0.4,0)$) {$X$};
\node[myGreen] at (-0.2, 2*0.8660254/3) {$1$};
\node at (0, 2*0.8660254/3-0.1) {$,$};
\node[myBlue] at (0.2, 2*0.8660254/3) {$2$};
\end{tikzpicture},
\begin{tikzpicture}[scale=0.9, baseline=(current bounding box.center)]
\tikzset{
  hexbond/.style={line width=0.6pt},
  vdot/.style={circle, fill=black, inner sep=1.2pt}
}

\coordinate (A_1) at ( 0, 2*0.8660254);
\coordinate (A_2) at ( -1, 0);
\coordinate (A_3) at ( 1, 0);
\draw[hexbond] (A_1)--(A_2)--(A_3)--cycle;

\coordinate (A_4) at ( -2, 2*0.8660254);
\coordinate (A_5) at ( 0, -2*0.8660254);
\coordinate (A_6) at ( 2, 2*0.8660254);
\draw[hexbond] (A_4)--(A_5)--(A_6)--cycle;

\node at ($(A_1)+(0,0.2)$) {$Z$};
\node at ($(A_2)+(-0.4,0)$) {$Z$};
\node at ($(A_3)+(0.4,0)$) {$Z$};
\node at ($(A_4)+(0,0.2)$) {$Z$};
\node at ($(A_5)+(0,-0.2)$) {$Z$};
\node at ($(A_6)+(0,0.2)$) {$Z$};
\node[myRed] at (0, 2*0.8660254/3) {$0$}; 
\end{tikzpicture},
\begin{tikzpicture}[scale=0.9, baseline=(current bounding box.center)]
\tikzset{
  hexbond/.style={line width=0.6pt},
  vdot/.style={circle, fill=black, inner sep=1.2pt}
}

\coordinate (A_1) at ( 0, 2*0.8660254);
\coordinate (A_2) at ( -1, 0);
\coordinate (A_3) at ( 1, 0);
\draw[hexbond] (A_1)--(A_2)--(A_3)--cycle;

\node at ($(A_1)+(0,0.2)$) {$X'$};
\node at ($(A_2)+(-0.4,0)$) {$X'$};
\node at ($(A_3)+(0.4,0)$) {$X'$};
\node[myGreen] at (-0.2, 2*0.8660254/3) {$1$};
\node at (0, 2*0.8660254/3-0.1) {$,$};
\node[myBlue] at (0.2, 2*0.8660254/3) {$2$};
\end{tikzpicture},
\begin{tikzpicture}[scale=0.9, baseline=(current bounding box.center)]
\tikzset{
  hexbond/.style={line width=0.6pt},
  vdot/.style={circle, fill=black, inner sep=1.2pt}
}

\coordinate (A_1) at ( 0, 2*0.8660254);
\coordinate (A_2) at ( -1, 0);
\coordinate (A_3) at ( 1, 0);
\draw[hexbond] (A_1)--(A_2)--(A_3)--cycle;

\coordinate (A_4) at ( -2, 2*0.8660254);
\coordinate (A_5) at ( 0, -2*0.8660254);
\coordinate (A_6) at ( 2, 2*0.8660254);
\draw[hexbond] (A_4)--(A_5)--(A_6)--cycle;

\node at ($(A_1)+(0,0.2)$) {$Z'$};
\node at ($(A_2)+(-0.4,0)$) {$Z'$};
\node at ($(A_3)+(0.4,0)$) {$Z'$};
\node at ($(A_4)+(0,0.2)$) {$Z'$};
\node at ($(A_5)+(0,-0.2)$) {$Z'$};
\node at ($(A_6)+(0,0.2)$) {$Z'$};
\node[myRed] at (0, 2*0.8660254/3) {$0$}; 
\end{tikzpicture}.
\end{minipage}
}

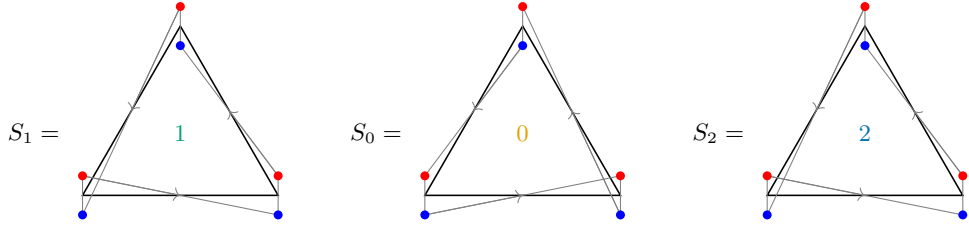
\begin{figure}
    \centering
    \begin{tikzpicture}[scale=1.3,baseline=(current bounding box.center)]
    \tikzset{
  hexbond/.style={line width=0.6pt},
  vdot/.style={circle, fill=black, inner sep=1.2pt}
}
    \coordinate (A_1) at ( 0, 2*0.8660254);
    \coordinate (A_2) at ( -1, 0);
    \coordinate (A_3) at ( 1, 0);
    \draw[hexbond] (A_1)--(A_2)--(A_3)--cycle;

    \coordinate (A_11) at ( 0, 2*0.8660254+0.2);
    \coordinate (A_12) at ( 0, 2*0.8660254-0.2);    
    \coordinate (A_21) at ( -1, 0.2);
    \coordinate (A_22) at ( -1, -0.2); 
    \coordinate (A_31) at ( 1, 0.2);
    \coordinate (A_32) at ( 1, -0.2);

    \coordinate (A) at (0, 0.8660254/2+0.2);
    \coordinate (A0) at (-1.5, 0.8660254/2+0.2);

    \draw[gray] (A_11)--(A_22)--(A_21)--(A_32)--(A_31)--(A_12)--cycle;
    \draw[->,gray] (A_11)--($(A_11)!0.5!(A_22)$);
    \draw[->,gray] (A_21)--($(A_21)!0.5!(A_32)$);
    \draw[->,gray] (A_31)--($(A_31)!0.5!(A_12)$);

    \node[vdot,red] at (A_11) {};
    \node[vdot,blue] at (A_12) {};
    \node[vdot,red] at (A_21) {};
    \node[vdot,blue] at (A_22) {};
    \node[vdot,red] at (A_31) {};
    \node[vdot,blue] at (A_32) {};
    \node[myGreen] at (A) {1};
    \node at (A0) {$S_1 = $};

    \def\dx{3.5}

    \coordinate (B_1) at ( 0+\dx, 2*0.8660254);
    \coordinate (B_2) at ( -1+\dx, 0);
    \coordinate (B_3) at ( 1+\dx, 0);
    \draw[hexbond] (B_1)--(B_2)--(B_3)--cycle;

    \coordinate (B_11) at ( 0+\dx, 2*0.8660254+0.2);
    \coordinate (B_12) at ( 0+\dx, 2*0.8660254-0.2);    
    \coordinate (B_21) at ( -1+\dx, 0.2);
    \coordinate (B_22) at ( -1+\dx, -0.2); 
    \coordinate (B_31) at ( 1+\dx, 0.2);
    \coordinate (B_32) at ( 1+\dx, -0.2);

    \coordinate (B) at (0+\dx, 0.8660254/2+0.2);
    \coordinate (B0) at (-1.5+\dx, 0.8660254/2+0.2);

    \draw[gray] (B_11)--(B_12)--(B_21)--(B_22)--(B_31)--(B_32)--cycle;
    \draw[->,gray] (B_12)--($(B_12)!0.5!(B_21)$);
    \draw[->,gray] (B_22)--($(B_22)!0.5!(B_31)$);
    \draw[->,gray] (B_32)--($(B_32)!0.5!(B_11)$);

    \node[vdot,red] at (B_11) {};
    \node[vdot,blue] at (B_12) {};
    \node[vdot,red] at (B_21) {};
    \node[vdot,blue] at (B_22) {};
    \node[vdot,red] at (B_31) {};
    \node[vdot,blue] at (B_32) {};
    \node[myRed] at (B) {0};
    \node at (B0) {$S_0 = $};

    \coordinate (C_1) at ( 0+2*\dx, 2*0.8660254);
    \coordinate (C_2) at ( -1+2*\dx, 0);
    \coordinate (C_3) at ( 1+2*\dx, 0);
    \draw[hexbond] (C_1)--(C_2)--(C_3)--cycle;

    \coordinate (C_11) at ( 0+2*\dx, 2*0.8660254+0.2);
    \coordinate (C_12) at ( 0+2*\dx, 2*0.8660254-0.2);    
    \coordinate (C_21) at ( -1+2*\dx, 0.2);
    \coordinate (C_22) at ( -1+2*\dx, -0.2); 
    \coordinate (C_31) at ( 1+2*\dx, 0.2);
    \coordinate (C_32) at ( 1+2*\dx, -0.2);

    \coordinate (C) at (0+2*\dx, 0.8660254/2+0.2);
    \coordinate (C0) at (-1.5+2*\dx, 0.8660254/2+0.2);

    \draw[gray] (C_11)--(C_22)--(C_21)--(C_32)--(C_31)--(C_12)--cycle;
    \draw[->,gray] (C_11)--($(C_11)!0.5!(C_22)$);
    \draw[->,gray] (C_21)--($(C_21)!0.5!(C_32)$);
    \draw[->,gray] (C_31)--($(C_31)!0.5!(C_12)$);

    \node[vdot,red] at (C_11) {};
    \node[vdot,blue] at (C_12) {};
    \node[vdot,red] at (C_21) {};
    \node[vdot,blue] at (C_22) {};
    \node[vdot,red] at (C_31) {};
    \node[vdot,blue] at (C_32) {};
    \node[myBlue] at (C) {2};
    \node at (C0) {$S_2 = $};
    
    \end{tikzpicture}
    \caption{Orientation of qudits on the three triangles. Each vertex hosts two qudits: the bottom (blue) qudit is associated with the first $\mathbb{Z}_4$ group labeled by $(n,q)$, while the top (red) qudit is associated with the second $\mathbb{Z}_4$ group labeled by $(n',q')$.}
    \label{fig:shift}
\end{figure}

To construct the anyon permutation circuit, we find it convenient in this case to combine the twisted gauging map with some special translation circuits $S_i$ defined in Fig.~\ref{fig:shift}, which realize nontrivial QCA in the $\mathbb{Z}_4\times \mathbb{Z}_4$ symmetric algebra in 1D. From the operator maps in Eq.~\eqref{eq:twisted_gauging_operator_maps}, we observe that only the degrees of freedom associated with the normal subgroup are transformed under the twisted gauging map, while the degrees of freedom associated with the quotient subgroup remain unchanged. Therefore, the degrees of freedom associated with the quotient subgroup are affected only by the sequence of $S_i$ circuits, which gives rise to an on-site swap between $q$ and $q'$. In the following, we focus on Hamiltonian terms corresponding to the degrees of freedom associated with the normal subgroup.

After applying twisted gauging and the translation operation $S_1$ on all the $1$-triangles, the generating set of Hamiltonian terms associated with the degrees of freedom associated with the normal subgroup is given as follows:

\raisebox{-0.5\height}{
\begin{minipage}{\textwidth}
\centering
\begin{tikzpicture}[scale=0.9, baseline=(current bounding box.center)]
\tikzset{
  hexbond/.style={line width=0.6pt},
  vdot/.style={circle, fill=black, inner sep=1.2pt}
}

\coordinate (A_1) at ( 0, 2*0.8660254);
\coordinate (A_2) at ( -1, 0);
\coordinate (A_3) at ( 1, 0);
\draw[hexbond] (A_1)--(A_2)--(A_3)--cycle;

\coordinate (A_7) at ( -1, 4*0.8660254);
\coordinate (A_8) at ( -2, -2*0.8660254);
\coordinate (A_9) at ( 3, 0);

\draw[hexbond] (A_1)--(A_7);
\draw[hexbond] (A_2)--(A_8);
\draw[hexbond] (A_3)--(A_9);

\draw[hexbond] (A_4)--(A_1)--(A_6)--(A_3)--(A_5)--(A_2)--cycle;

\node at ($(A_1)+(0.5,0.3)$) {$Z^C X'^C$};
\node at ($(A_2)+(-0.8,0.3)$) {$Z^C X'^C$};
\node at ($(A_3)+(0.8,0.3)$) {$Z^C X'^C$};
\node at ($(A_7)+(0,0.2)$) {$Z^C$};
\node at ($(A_8)+(0,-0.2)$) {$Z^C$};
\node at ($(A_9)+(0,0.3)$) {$Z^C$};
\node[myRed] at (0, 2*0.8660254/3) {$0$}; 

\node at ($(A_4)+(0,-0.3)$) {\footnotesize $1$};
\node at ($(A_1)+(0,-0.3)$) {\footnotesize $2$};
\node at ($(A_6)+(0,-0.3)$) {\footnotesize $3$};
\node at ($(A_2)+(0,-0.3)$) {\footnotesize $6$};
\node at ($(A_3)+(0,-0.3)$) {\footnotesize $4$};
\node at ($(A_5)+(0,0.3)$) {\footnotesize $5$};

\end{tikzpicture}$\times\ W_0$,\quad
\begin{tikzpicture}[scale=0.9, baseline=(current bounding box.center)]
\tikzset{
  hexbond/.style={line width=0.6pt},
  vdot/.style={circle, fill=black, inner sep=1.2pt}
}

\coordinate (A_1) at ( 0, 2*0.8660254);
\coordinate (A_2) at ( -1, 0);
\coordinate (A_3) at ( 1, 0);
\draw[hexbond] (A_1)--(A_2)--(A_3)--cycle;

\coordinate (A_4) at ( -2, 2*0.8660254);
\coordinate (A_5) at ( 0, -2*0.8660254);
\coordinate (A_6) at ( 2, 2*0.8660254);
\draw[hexbond] (A_4)--(A_5)--(A_6)--cycle;

\node at ($(A_1)+(0,0.3)$) {$Z'^C X^C$};
\node at ($(A_2)+(-0.7,0)$) {$Z'^C X^C$};
\node at ($(A_3)+(0.7,0)$) {$Z'^C X^C$};
\node at ($(A_4)+(0,0.3)$) {$Z'^C$};
\node at ($(A_5)+(0,-0.2)$) {$Z'^C$};
\node at ($(A_6)+(0,0.3)$) {$Z'^C$};
\node[myRed] at (0, 2*0.8660254/3) {$0$}; 

\node at ($(A_4)+(0,-0.3)$) {\footnotesize $1$};
\node at ($(A_1)+(0,-0.3)$) {\footnotesize $2$};
\node at ($(A_6)+(0,-0.3)$) {\footnotesize $3$};
\node at ($(A_2)+(0,-0.3)$) {\footnotesize $6$};
\node at ($(A_3)+(0,-0.3)$) {\footnotesize $4$};
\node at ($(A_5)+(0,0.3)$) {\footnotesize $5$};

\end{tikzpicture}$\times\ W_0 $,\quad \\
\begin{tikzpicture}[scale=0.9, baseline=(current bounding box.center)]
\tikzset{
  hexbond/.style={line width=0.6pt},
  vdot/.style={circle, fill=black, inner sep=1.2pt}
}

\coordinate (A_1) at ( 0, 2*0.8660254);
\coordinate (A_2) at ( -1, 0);
\coordinate (A_3) at ( 1, 0);
\draw[hexbond] (A_1)--(A_2)--(A_3)--cycle;

\node at ($(A_1)+(0,0.2)$) {$X^C$};
\node at ($(A_2)+(-0.4,0)$) {$X^C$};
\node at ($(A_3)+(0.4,0)$) {$X^C$};
\node[myGreen] at (0, 2*0.8660254/3) {$1$};
\end{tikzpicture},
\begin{tikzpicture}[scale=0.9, baseline=(current bounding box.center)]
\tikzset{
  hexbond/.style={line width=0.6pt},
  vdot/.style={circle, fill=black, inner sep=1.2pt}
}

\coordinate (A_1) at ( 0, 2*0.8660254);
\coordinate (A_2) at ( -1, 0);
\coordinate (A_3) at ( 1, 0);
\draw[hexbond] (A_1)--(A_2)--(A_3)--cycle;

\node at ($(A_1)+(0,0.2)$) {$X'^C$};
\node at ($(A_2)+(-0.4,0)$) {$X'^C$};
\node at ($(A_3)+(0.4,0)$) {$X'^C$};
\node[myGreen] at (0, 2*0.8660254/3) {$1$};
\end{tikzpicture},
\begin{tikzpicture}[scale=0.9, baseline=(current bounding box.center)]
\tikzset{
  hexbond/.style={line width=0.6pt},
  vdot/.style={circle, fill=black, inner sep=1.2pt}
}

\coordinate (A_1) at ( 0, 2*0.8660254);
\coordinate (A_2) at ( -1, 0);
\coordinate (A_3) at ( 1, 0);
\draw[hexbond] (A_1)--(A_2)--(A_3)--cycle;

\node at ($(A_1)+(0,0.2)$) {$X'^C$};
\node at ($(A_2)+(-0.4,0)$) {$X'^C$};
\node at ($(A_3)+(0.4,0)$) {$X'^C$};
\node[myBlue] at (0, 2*0.8660254/3) {$2$};
\end{tikzpicture},
\begin{tikzpicture}[scale=0.9, baseline=(current bounding box.center)]
\tikzset{
  hexbond/.style={line width=0.6pt},
  vdot/.style={circle, fill=black, inner sep=1.2pt}
}

\coordinate (A_1) at ( 0, 2*0.8660254);
\coordinate (A_2) at ( -1, 0);
\coordinate (A_3) at ( 1, 0);
\draw[hexbond] (A_1)--(A_2)--(A_3)--cycle;

\coordinate (A_4) at ( -2, 2*0.8660254);
\coordinate (A_5) at ( 0, -2*0.8660254);
\coordinate (A_6) at ( 2, 2*0.8660254);
\draw[hexbond] (A_4)--(A_5)--(A_6)--cycle;

\node at ($(A_4)+(0,0.2)$) {$X^C$};
\node at ($(A_5)+(0,-0.2)$) {$X^C$};
\node at ($(A_6)+(0,0.2)$) {$X^C$};
\node[myBlue] at (0, 2*0.8660254/3) {$2$}; 
\end{tikzpicture}.
\end{minipage}
}

Since the twisted gauging map acts trivially on the symmetry operators, the star terms do not get permuted between different triangles, in contrast to what we have in the untwisted gauging cases. Therefore, as there is no local symmetry operator defined on the $0$-triangles, we only apply the translation operation $S_0$ on the $0$-triangles. The resulting terms are given as follows:

\raisebox{-0.5\height}{
\begin{minipage}{\textwidth}
\centering
\begin{tikzpicture}[scale=0.9, baseline=(current bounding box.center)]
\tikzset{
  hexbond/.style={line width=0.6pt},
  vdot/.style={circle, fill=black, inner sep=1.2pt}
}

\coordinate (A_1) at ( 0, 2*0.8660254);
\coordinate (A_2) at ( -1, 0);
\coordinate (A_3) at ( 1, 0);
\draw[hexbond] (A_1)--(A_2)--(A_3)--cycle;

\coordinate (A_7) at ( 1, 4*0.8660254);
\coordinate (A_9) at ( 2, -2*0.8660254);
\coordinate (A_8) at ( -3, 0);

\draw[hexbond] (A_1)--(A_7);
\draw[hexbond] (A_2)--(A_8);
\draw[hexbond] (A_3)--(A_9);

\draw[hexbond] (A_4)--(A_1)--(A_6)--(A_3)--(A_5)--(A_2)--cycle;

\node at ($(A_1)+(-0.5,0.2)$) {$Z'^C X^C$};
\node at ($(A_2)+(-0.8,0.2)$) {$Z'^C X^C$};
\node at ($(A_3)+(0.8,0.2)$) {$Z'^C X^C$};
\node at ($(A_7)+(0,0.2)$) {$Z'^C$};
\node at ($(A_8)+(0,-0.2)$) {$Z'^C$};
\node at ($(A_9)+(0.3,0.2)$) {$Z'^C$};

\node at ($(A_4)+(0,-0.3)$) {\footnotesize $1$};
\node at ($(A_1)+(0,-0.3)$) {\footnotesize $2$};
\node at ($(A_6)+(0,-0.3)$) {\footnotesize $3$};
\node at ($(A_2)+(0,-0.3)$) {\footnotesize $6$};
\node at ($(A_3)+(0,-0.3)$) {\footnotesize $4$};
\node at ($(A_5)+(0,0.3)$) {\footnotesize $5$};

\node[myRed] at (0, 2*0.8660254/3) {$0$}; 
\end{tikzpicture}$\times\ W_0$,\quad 
\begin{tikzpicture}[scale=0.9, baseline=(current bounding box.center)]
\tikzset{
  hexbond/.style={line width=0.6pt},
  vdot/.style={circle, fill=black, inner sep=1.2pt}
}

\coordinate (A_1) at ( 0, 2*0.8660254);
\coordinate (A_2) at ( -1, 0);
\coordinate (A_3) at ( 1, 0);
\draw[hexbond] (A_1)--(A_2)--(A_3)--cycle;

\coordinate (A_4) at ( -2, 2*0.8660254);
\coordinate (A_5) at ( 0, -2*0.8660254);
\coordinate (A_6) at ( 2, 2*0.8660254);
\draw[hexbond] (A_4)--(A_5)--(A_6)--cycle;

\node at ($(A_1)+(0,0.2)$) {$Z^C X'^C$};
\node at ($(A_2)+(-0.7,0)$) {$Z^C X'^C$};
\node at ($(A_3)+(0.7,0)$) {$Z^C X'^C$};
\node at ($(A_4)+(0,0.2)$) {$Z^C$};
\node at ($(A_5)+(0,-0.2)$) {$Z^C$};
\node at ($(A_6)+(0,0.2)$) {$Z^C$};

\node at ($(A_4)+(0,-0.3)$) {\footnotesize $1$};
\node at ($(A_1)+(0,-0.3)$) {\footnotesize $2$};
\node at ($(A_6)+(0,-0.3)$) {\footnotesize $3$};
\node at ($(A_2)+(0,-0.3)$) {\footnotesize $6$};
\node at ($(A_3)+(0,-0.3)$) {\footnotesize $4$};
\node at ($(A_5)+(0,0.3)$) {\footnotesize $5$};

\node[myRed] at (0, 2*0.8660254/3) {$0$}; 
\end{tikzpicture}$\times\ W_0 $,\\
\begin{tikzpicture}[scale=0.9, baseline=(current bounding box.center)]
\tikzset{
  hexbond/.style={line width=0.6pt},
  vdot/.style={circle, fill=black, inner sep=1.2pt}
}

\coordinate (A_1) at ( 0, 2*0.8660254);
\coordinate (A_2) at ( -1, 0);
\coordinate (A_3) at ( 1, 0);
\draw[hexbond] (A_1)--(A_2)--(A_3)--cycle;

\node at ($(A_1)+(0,0.2)$) {$X^C$};
\node at ($(A_2)+(-0.4,0)$) {$X^C$};
\node at ($(A_3)+(0.4,0)$) {$X^C$};
\node[myGreen] at (0, 2*0.8660254/3) {$1$};
\end{tikzpicture},
\begin{tikzpicture}[scale=0.9, baseline=(current bounding box.center)]
\tikzset{
  hexbond/.style={line width=0.6pt},
  vdot/.style={circle, fill=black, inner sep=1.2pt}
}

\coordinate (A_1) at ( 0, 2*0.8660254);
\coordinate (A_2) at ( -1, 0);
\coordinate (A_3) at ( 1, 0);
\draw[hexbond] (A_1)--(A_2)--(A_3)--cycle;

\coordinate (A_4) at ( -2, 2*0.8660254);
\coordinate (A_5) at ( 0, -2*0.8660254);
\coordinate (A_6) at ( 2, 2*0.8660254);
\draw[hexbond] (A_4)--(A_5)--(A_6)--cycle;

\node at ($(A_4)+(0,0.2)$) {$X'^C$};
\node at ($(A_5)+(0,-0.2)$) {$X'^C$};
\node at ($(A_6)+(0,0.2)$) {$X'^C$};
\node[myGreen] at (0, 2*0.8660254/3) {$1$}; 
\end{tikzpicture},
\begin{tikzpicture}[scale=0.9, baseline=(current bounding box.center)]
\tikzset{
  hexbond/.style={line width=0.6pt},
  vdot/.style={circle, fill=black, inner sep=1.2pt}
}

\coordinate (A_1) at ( 0, 2*0.8660254);
\coordinate (A_2) at ( -1, 0);
\coordinate (A_3) at ( 1, 0);
\draw[hexbond] (A_1)--(A_2)--(A_3)--cycle;

\node at ($(A_1)+(0,0.2)$) {$X^C$};
\node at ($(A_2)+(-0.4,0)$) {$X^C$};
\node at ($(A_3)+(0.4,0)$) {$X^C$};
\node[myBlue] at (0, 2*0.8660254/3) {$2$};
\end{tikzpicture},
\begin{tikzpicture}[scale=0.9, baseline=(current bounding box.center)]
\tikzset{
  hexbond/.style={line width=0.6pt},
  vdot/.style={circle, fill=black, inner sep=1.2pt}
}

\coordinate (A_1) at ( 0, 2*0.8660254);
\coordinate (A_2) at ( -1, 0);
\coordinate (A_3) at ( 1, 0);
\draw[hexbond] (A_1)--(A_2)--(A_3)--cycle;

\node at ($(A_1)+(0,0.2)$) {$X'^C$};
\node at ($(A_2)+(-0.4,0)$) {$X'^C$};
\node at ($(A_3)+(0.4,0)$) {$X'^C$};
\node[myBlue] at (0, 2*0.8660254/3) {$2$};
\end{tikzpicture}.
\end{minipage}
}

Finally, we apply the twisted gauging map and the translation operation $S_2$ on all the $2$-triangles, after which the Hamiltonian terms return to their original forms. The above process can be schematically illustrated as follows:
\begin{align}
    H^{(0)} \xrightarrow[]{\overline{S_1} \circ \Gamma_1^{\alpha}} H^{(0)} \xrightarrow[]{\overline{S_0}} H^{(2)} \xrightarrow[]{\overline{S_2} \circ \Gamma_2^{\alpha}} H^{(0)}
\end{align}

\subsection{Ribbon operators maps}

Anyons of the $\mathbb{Z}_4 \times \mathbb{Z}_4$ quantum double model are generated by the following set $\{\widetilde{e}, \widetilde{m}, \widetilde{e}', \widetilde{m}'\}$, where we have $\widetilde{e}^4 = \widetilde{m}^4 = \widetilde{e}'^4 = \widetilde{m}'^4 = 1$, $S(\widetilde{e}, \widetilde{m}) = S(\widetilde{e}', \widetilde{m}') = i$, and $S(\widetilde{e}, \widetilde{m}') = S(\widetilde{e}', \widetilde{m}) = 1$.

Consider the lattice in Fig.~\ref{fig:deformed_lattice}, a closed $\widetilde{e}$ operator can be written as follows:
\begin{align}
    F^{\widetilde{e}} = Z^C_1 \sqrt{Z_1} Z^C_2 \sqrt{Z_2} \cdots Z^C_n \sqrt{Z_n}
\end{align}
After applying the twisted gauging map and the translation $S_1$ on all the $1$-triangles, the above operator becomes:
\begin{align}
     X^C_1 Z'^C_1 Z'^C_2 X^C_3 Z'^C_3 \cdots Z'^C_{n}\cdot \sqrt{Z'_1}\cdots \sqrt{Z'_n}.
\end{align}

Then we apply the translation $S_0$ on all the $0$-triangles, we get,
\begin{align}
    X'^C_{n'} Z^C_1 Z^C_2 X'^C_{2'} Z^C_3 \cdots Z^C_{n}\cdot \sqrt{Z_1}\cdots \sqrt{Z_n}.
\end{align}

Finally, we apply the twisted gauging map and the translation $S_2$ on all the $2$-triangles, we get,
\begin{align}
    \cdots Z'^C_1 X^C_{2} Z'^C_{2}  \cdots \cdot \sqrt{Z'_1}\cdots \sqrt{Z'_n},
\end{align}
which is the anyonic ribbon operator of the $\widetilde{e}'\widetilde{m}^2$ anyon. All other anyonic ribbon operator maps can be derived accordingly. The anyon permutation map of the above process is the following:
\begin{align}
    \widetilde{e} \mapsto \widetilde{e}'\widetilde{m}^2, \quad \widetilde{e}' \mapsto \widetilde{e}\widetilde{m}'^2,\quad \widetilde{m} \mapsto \widetilde{m}',\quad \widetilde{m}' \mapsto \widetilde{m}.
\end{align}

With a transversal SWAP gate between two $\mathbb{Z}_4$ degrees of freedom on each site, we eventually obtain the anyon permutation that is associated with the 1D twisted gauging map:
\begin{align}
    \widetilde{e} \mapsto \widetilde{e}\widetilde{m}'^2, \quad \widetilde{e}' \mapsto \widetilde{e}'\widetilde{m}^2,\quad \widetilde{m} \mapsto \widetilde{m},\quad \widetilde{m}' \mapsto \widetilde{m}'.
\end{align}

\section{Anyon permutation: class II}
\label{app:class_II}
To better illustrate the implementation of anyon permutations in class II, we use the $\mathbb{Z}_2 \times \mathbb{Z}_2$ quantum double model as an illustrative example. We also refer the reader to Ref.~\cite{warman2025transversal} for other examples.

The nontrivial 2-cocycle of the $\mathbb{Z}_2 \times \mathbb{Z}_2$ group is given as follows:
\begin{align}
    \alpha((i_1, j_1), (i_2, j_2)) = (-1)^{i_1 j_2},
\end{align}
where $(i,j)$ is the group element of $\mathbb{Z}_2 \times \mathbb{Z}_2$ with $i, j \in \{0,1\}$.

Following Eq.~\eqref{eq:phase_gate}, the corresponding phase gate can be written as follow:
\begin{align}
    U = \prod_{\mathrm{hexagon}} CZ_{1,6} CZ_{1,5} CZ_{6,5} CZ_{2,3} CZ_{2,4} CZ_{3,4}.
\end{align}
Here the $CZ_{a,b}$ gate is defined as follows.
\begin{align}
    CZ_{a,b} |(i_a, j_a), (i_b, j_b) \rangle = (-1)^{i_a j_b} |(i_a, j_a), (i_b, j_b) \rangle.
\end{align}
Pictorially, the gate on each plaquette can be illustrated as follows:
\begin{equation}
\begin{tikzpicture}[scale=0.8, baseline=(current bounding box.center)]
\tikzset{
  hexbond/.style={line width=0.6pt},
  vdot/.style={circle, fill=black, inner sep=1.2pt}
}

\coordinate (A_1) at ( -2, 0);
\coordinate (A_2) at ( -1, 1.732);
\coordinate (A_3) at ( 1, 1.732);
\coordinate (A_4) at ( 2, 0);
\coordinate (A_5) at ( 1, -1.732);
\coordinate (A_6) at ( -1, -1.732);

\draw[hexbond,postaction={decorate}, decoration={markings, mark=at position 0.5 with {\arrow{>}}}] (A_1)--(A_2);
\draw[hexbond,postaction={decorate}, decoration={markings, mark=at position 0.5 with {\arrow{>}}}] (A_3)--(A_2);
\draw[hexbond,postaction={decorate}, decoration={markings, mark=at position 0.5 with {\arrow{>}}}] (A_3)--(A_4);
\draw[hexbond,postaction={decorate}, decoration={markings, mark=at position 0.5 with {\arrow{>}}}] (A_5)--(A_4);
\draw[hexbond,postaction={decorate}, decoration={markings, mark=at position 0.5 with {\arrow{>}}}] (A_5)--(A_6);
\draw[hexbond,postaction={decorate}, decoration={markings, mark=at position 0.5 with {\arrow{>}}}] (A_1)--(A_6);

\draw[->,>=stealth',myGreen, thick, shorten <=2pt, shorten >=2pt]
  ($(A_1)!0.5!(A_2)$)
  .. controls +(0.6,-0.4) and +(0.6,0.4)
  .. ($(A_1)!0.5!(A_6)$);
  \draw[->,>=stealth',myGreen, thick, shorten <=2pt, shorten >=2pt]
  ($(A_1)!0.5!(A_2)$)
  .. controls +(0.6,-0.4) and +(0.2,0.6)
  .. ($(A_5)!0.5!(A_6)$);
  \draw[->,>=stealth',myGreen, thick, shorten <=2pt, shorten >=2pt]
  ($(A_1)!0.5!(A_6)$)
  .. controls +(0.6,0) and +(0,0.6)
  .. ($(A_5)!0.5!(A_6)$);

  \draw[->,>=stealth',myGreen, thick, shorten <=2pt, shorten >=2pt]
  ($(A_2)!0.5!(A_3)$)
  .. controls +(0,-0.5) and +(-0.5,-0.2)
  .. ($(A_3)!0.5!(A_4)$);
  \draw[->,>=stealth',myGreen, thick, shorten <=2pt, shorten >=2pt]
  ($(A_3)!0.5!(A_4)$)
  .. controls +(-0.3,-0.2) and +(-0.5,-0.2)
  .. ($(A_4)!0.5!(A_5)$);
  \draw[->,>=stealth',myGreen, thick, shorten <=2pt, shorten >=2pt]
  ($(A_2)!0.5!(A_3)$)
  .. controls +(0,-0.5) and +(-0.6,-0.2)
  .. ($(A_4)!0.5!(A_5)$);

\node at ($(A_1)+(-0.2+0.5,0+1.732/2+0.2)$) {$1$};
\node at ($(0,1.732+0.3)$) {$2$};
\node at ($(A_4)+(0.2-0.5,0+1.732/2+0.2)$) {$3$};
\node at ($(A_4)+(0.2-0.5,0-1.732/2-0.2)$) {$4$};
\node at ($(0,-1.732-0.3)$) {$5$};
\node at ($(A_1)+(-0.2+0.5,0-1.732/2-0.2)$) {$6$};
\end{tikzpicture}
\end{equation}
where the arrows represent the $CZ_{i,j}$ gates. The tail of each arrow is located at the control qubit, while the head of the arrow is located at the target qubit. Under this unitary gate, one can check the set of local stabilizers is invariant, while the anyon ribbon operators are mapped as follows:
\begin{align}
    m_1 \to m_1 e_2,\quad m_2 \to m_2 e_1,\quad e_1 \to e_1,\quad e_2 \to e_2.
\end{align}

\end{document}